\documentclass[12pt]{iopart}

\usepackage{iopams}
\usepackage{setstack}
\usepackage{graphicx} 

\usepackage{autopar}

\begin{document}

\title[Automatic parametrization of amplitudes]
{Towards a fully automated computation of RG-functions for the $3$-$d$ $O(N)$ vector model:\\ Parametrizing amplitudes
}

\author{
  Riccardo Guida$^1$
  and
  Paolo Ribeca$^2$
}
\address{
  $^1$Service de Physique Th\'eorique,
  CEA/DSM/SPhT \& CNRS/SPM/URA 2306,
  CEA/Saclay, F-91191 Gif-sur-Yvette C\'edex, France.
}
\address{
  $^2$Humboldt Universit\"at, Newtonstr. 15, D-12489 Berlin, Germany.
}

\begin{abstract}
Within the framework of field-theoretical description 
of second-order phase transitions via the $3$-dimensional $O(N)$ 
vector model, accurate predictions for critical exponents
can be obtained from (resummation of) the perturbative series of 
Renormalization-Group functions, 
which are in turn derived ---following Parisi's approach--- from the
expansions of appropriate field correlators evaluated 
at zero external momenta.

Such a technique was fully exploited $30$ years ago in two seminal works 
of Baker, Nickel, Green and Meiron (\cite{NickelI}-\cite{NickelII}), 
which lead to the knowledge of the $\beta$-function up to the $6$-loop level;
they succeeded in obtaining a precise numerical evaluation 
of all needed Feynman amplitudes in momentum space by lowering 
the dimensionalities of each integration with a cleverly arranged set 
of computational simplifications. 
In fact, extending this computation is not straightforward, due 
both to the factorial proliferation of relevant diagrams 
and the increasing dimensionality of their associated integrals; in any case,
this task can be reasonably carried on
only in the framework of an automated environment.

On the road towards the creation of such an environment, we here show  
how a strategy closely inspired by that of Nickel and coworkers
can be stated in algorithmic form, and successfully implemented 
on the computer. As an application, we plot the minimized distributions of
residual integrations for the sets of diagrams needed 
to obtain RG-functions to the full $7$-loop level; 
they represent a good evaluation of
the computational effort which will be required to improve 
the currently available estimates of critical exponents.
\end{abstract}


\eads{$^1$\mailto{guida@spht.saclay.cea.fr},
      $^2$\mailto{paolo@paoloribeca.net}}
\submitto{Journal of Statistical Mechanics: Theory and Experiment}
\vspace{20pt}
\begin{indented}
  \item[]\rm Preprint {\tt HU-EP-05/39}, {\tt SFB/CPP-05-64}, {\tt SPhT-T05/128}
\end{indented}
\maketitle

\section{Motivation}
A great achievement of the Renormalization-Group (RG) approach to critical
phenomena (introduced in \cite{Kadanoff:1966wm}, \cite{Wilson:1971bg}, 
\cite{Wilson:1971dh}, \cite{Wegner:1972my}, \cite{Wegner:1972xxx}, 
\cite{Wilson:1974jj})
is the realization that the universal behavior of many different
physical systems can be explained by the existence of a common
large-distance fixed point ---or {\em infrared}, or IR fixed point--- 
 of RG-equations;
these equations
describe the evolution of the effective Hamiltonians 
for such systems when the short-distance degrees of freedom 
---also called {\em ultraviolet}, or UV---
are increasingly summed up.
In particular, it turns out from this idea that  a quantitative description 
of physical universal quantites such as critical exponents 
and critical amplitudes 
can be obtained by considering the RG-evolution of
a euclidean quantum field theory appropriately chosen 
to stay in the same universality class of the systems to be described. 
In this perspective, second-order phase transitions of systems 
with $N$ phenomenological scalar components in the sense of Landau theory 
can be obtained by studying the IR fixed point(s) of the RG-equations for
quantum field theories of $N$ interacting scalar fields.
In the case of $O(N)$-symmetric interactions of $\left(\phi^2\right)^2$ type, the IR fixed point is 
known as {\em Wilson-Fisher fixed point} 
(see \cite{WilsonFisher}).
 
Following this line of thought, 
during the Cargese summer school in 1973, \cite{ParisiTwoLoops},
Parisi made the seminal observation that a quantitative description of the  
Wilson-Fisher fixed point can be obtained by analyzing the Callan-Symanzik 
RG-equations for a system of $3$-dimensional massive scalar bosons with an
$O(N)$-invariant interaction of $\left(\phi^2\right)^2$ type, 
the so-called $O(N)$ vector model 
(first published notes of this idea can be found in \cite{Parisi}).
Such a theory is a super-renormalizable one, i.e. the number of 
one-particle irreducible (OPI) diagrams with new primitive divergences 
is finite, and renormalization process is easier if compared 
to that required by other renormalizable theories. 

A price to pay for the simplicity of Parisi's approach 
is the fact that the value of the coupling is nonperturbatively large at the 
fixed point; thus, the 
perturbative series obtained in this framework have to be resummed 
by means of a suitable method (the one usually chosen being the 
so-called {\em Borel summation}\/). 
We do not dive here into the details of these resummation techniques; 
we only mention that for this field-theoretical model 
it has been rigorously proved in \cite{Magnen:1977ha} that
resummed perturbative series converge 
when their order increases; thus, the precision of the estimates obtained 
within Parisi's approach is only limited 
by our practical ability in generating and parametrizing 
the necessary amplitudes up to the desired order, and by the precision
we can reach when computing 
the high-dimensional integrals 
in terms of which amplitudes are expressed.
As an additional remark, one should bear in mind that,
to avoid instabilities in the resummation process 
at a given perturbative order, 
the numerical precision of each term of the series should be greater than 
that desired for the final result, and greater than that of 
subsequent terms as well.\footnote{Being 
the subject of critical phenomena so rich and complex, 
it is clearly impossible for us to give in this context 
much more than a very limited introduction to it;
however, the interested reader can easily find 
complete and pedagogical expositions in a variety of textbooks (see e.g.
\cite{LeBellac}, \cite{Zinn}, \cite{ParisiBook}, \cite{AmitMayorBook},  
\cite{Kleinert}; recent reviews are presented in \cite{BaBe} and 
\cite{PelissettoVicari}).
In these texts, many other subjects relevant to this article ---like
euclidean quantum field theories, theory of Renormalization-Group 
and resummation techniques--- are also extensively covered.}

The panorama of results in the computation of RG-functions 
for the $3$-dimensional $O(N)$ vector model is dominated by 
the work of B. G. Nickel and collaborators.
In 1976-1978 ---see \cite{NickelI} and \cite{NickelII}---, making use of a large and 
cleverly-arranged set of remarkable computational simplifications 
partially described in \cite{NickelDiagrams}, 
Baker, Nickel, Green and Meiron
were able to push the calculation from the $2$-loop- (see \cite{ParisiTwoLoops} and \cite{ZinnTwoLoops}) up to the
$6$-loop- level.
(See  also \cite{NickelPreprint}
 for a compilation of diagrams, weights and associated amplitude values;
 \cite{MuNi84} is an application of the same type of data to a different context.)
  In 1991 Murray and Nickel \cite{NickelPreprintII}
  completed the evaluation of an improved estimate of 
  the anomalous dimensions $\eta$ and $\eta_2$ including partial 
  $7$-loop results, but the full $7$-loop computation of  
  RG $\beta$-function and critical exponents
  is still an open problem.
It should be noticed that from a technical point of view
---and in particular when taking into account the lack of available 
computer power at the time---
 the $6$-loop calculation is already an outstanding masterpiece 
in its own right: in fact, 
it requires the high-precision numerical evaluation 
(with $8$-$10$ significant digits)
of a thousand diagrams, which were apparently 
parametrized by hand as integrals with up to $6$ residual dimensions, 
in the careful attempt of finding the most 
advantageous representation for each amplitude. 

As a matter of fact, a very conspicuous body of literature relies on
\cite{NickelII}, \cite{NickelPreprintII} 
and the compilation \cite{NickelPreprint} for the calculation 
of critical exponents as well as of other quantities
(see for example \cite{Guida:1998bx} 
and \cite{Calabrese:2002qi}, \cite{Calabrese:2002sz}, 
\cite{Kastening:2003iu}). However,
at least in our knowledge, no other group has been able 
to reproduce these computations independently so far:
this fact has been our main motivation for resuming the problem,
in the hope of being able to verify ---and possibly extend--- 
the results available at present.

The high number of diagrams which must be evaluated to complete the
calculation of the $7$th loop ($\sim 4000$, see \sref{GenerationOfGraphs}) 
makes it immediately clear that the only realistic
hope to push forward our knowledge  relies on finding a way of
automating computations by means of computer techniques; the goal of this paper is 
then to perform the first steps towards the build-up of such an automated 
environment, both for generation and parametrization of all needed 
Feynman amplitudes. Article structure is as follows:
\begin{itemize}
\item In \sref{ONVectorModel} we introduce all relevant information about 
the $3$-dimensional $O(N)$ vector model, together with the renormalization 
scheme and the RG-equations our computations are based on
\item In \sref{generation} 
we briefly discuss how to generate all needed Feynman diagrams, 
together with their combinatorial and $O(N)$ factors; 
after that, we analyze in great detail the set of analytic tricks 
which make possible an efficient parametrization of the associated amplitudes. 
A number of practical examples is given
\item In \sref{autopar} we provide a precise algorithmic formulation for each 
of the tasks required to actually find the set of optimal parametrizations 
at some given perturbative level
\item In \sref{ImplementationAndResults} 
we describe our computer implementation of such techniques, and 
the results we obtained so far. A critical discussion follows about 
how the set of chosen simplifications influences both the complexity 
of the parametrization code 
and the minimality of the resulting set of parametrizations.
\end{itemize}
   
\section{The $3$-dimensional $O(N)$ vector model}\label{ONVectorModel}

\subsection{Generalities}
The $3$-dimensional $O(N)$ vector model is described by
a classical (bare) action which is invariant under a transformation of
the scalar fields $\phi_\B^i$ ($i=1,\,\ldots\, ,\, N$) 
according to a fundamental representation of $O(N)$:
\beq
S_\B(\phi_\B) :=\int d^3x 
\left(
  \frac{1}{2}(\nabla \phi_\B)^2+{m^2_\B \over 2} \phi_\B^2
  +{\lambda_\B \over 4!}(\phi_\B^2)^2 
\right).
\eeq
The quantization of the model is obtained 
via functional integration, by considering the
(bare) generating functional,  
\beq
Z_\B (J_\B):=
\int [d\phi_\B] \exp \left(-S_\B(\phi_\B) +\int J_\B \phi_\B\right).
\eeq
An appropriate regularization ---parametrized by some 
ultraviolet cutoff denoted with $\Lambda$--- 
is necessary to give meaning
to functional integration. We assume that such a regularization exists,
but to maintain generality and simplicity of notations 
we will not show its details here.
As well known, perturbation theory for the bare model is plagued 
by ultraviolet divergences in the limit $\Lambda \to \infty$,
and must be complemented by renormalization; 
that amounts to a reparametrization
of bare (ultraviolet) quantities in terms of renormalized (infrared) 
ones.  
In this section we content ourselves with
briefly recalling only some essential points about the 
renormalization of the $O(N)$ vector model;
all other necessary definitions and some additional technical details 
about it, including a derivation of Callan-Symanzik equations in a generic
massive
renormalization scheme, are reported for reference in \ref{renormalization}.

Applying  standard power counting techniques to Feynman diagrams 
obtained from the perturbative expansion of  the $O(N)$ vector model, it is easy to
compute the superficial degree of divergence
$\omega$ for the Feynman amplitude $\AmplitudeD$ associated to
a connected \textit{one-particle irreducible}, or OPI, diagram $\Diagram$
with $\NExternal$ external legs, $\NVerticesII$ insertions of operator 
$\int \phi^2$ and $\NVerticesIV$ vertices with four legs. 
The obtained expression reads
\beq\label{omega}
\omega=3-\frac{\NExternal}{2}-\NVerticesIV-2 \NVerticesII;
\eeq
roughly speaking, $\omega$ parametrizes the behaviour of the amplitude 
in the limit
$\Lambda \to \infty$, where $\AmplitudeD\sim \Lambda^\omega$ 
(see \cite{ItzyksonZuber} 
for a more rigorous definition).
It turns out from \eref{omega} that the model is \emph{super-renormalizable}, i.e.
it has only a finite number of superficially divergent amplitudes.
When restricting to the conditions $\NExternal>0$ 
---as it is always the case in this article--- and $\NLoops>0$, 
amplitudes must satisfy $\NExternal=2$,
$\NVerticesII=0$ and $\NVerticesIV=1,2$ to have $\omega\ge0$ and to
possibly be superficially divergent; 
these constraints are verified
in our model only by the three graphs listed below:
\begin{eqnarray*}
\begin{array}{ccc}
\includegraphics[width=0.15\textwidth]{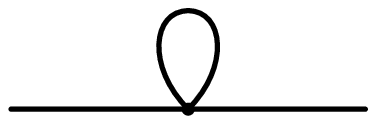},\qquad
&\includegraphics[width=0.15\textwidth]{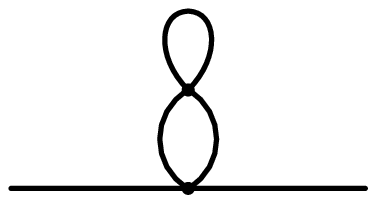},\qquad
&\includegraphics[width=0.15\textwidth]{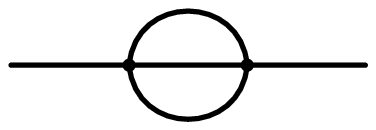},
\end{array}
\end{eqnarray*}
which we will call in the rest of this article ---from left to right--- 
\emph{``tadpole''}, \emph{``cactus''}, 
and \emph{``sunset''}, respectively.
It should also be
noticed that these divergent graphs require a mass renormalization,
but no divergent contribution affects the derivative 
of the two-point function w.r.t.\ $p^2$, i.e. wave-function renormalization 
is not needed to make the theory finite.

\subsection{Intermediate scheme}\label{IScheme}
From a practical point of view,
 the most effective way of proceeding is to first choose
an intermediate renormalization scheme, with the goal of making
the theory finite and the calculations as simple as possible;
only after that ---in the end--- one will switch 
to the final renormalization scheme
presented in section \ref{parisiScheme}.
More in detail, our own intermediate scheme 
(labeled with suffix $\I$ and referred to as \emph{``I-scheme''}) 
is defined by the following conditions:
\bea
\phi_\B=\phi_\I 
\\
\phi_\B^2=\left[\phi^2\right]_\I 
\\
\lambda_\B=m_\I g_\I \label{BIlambda}
\\
m_\B^2=m^2_\I+\delta m^2_\I
\label{BIm}
;\eea
the mass counterterm is in turn defined as:
\bea
\delta m^2_\I:=
-\left(
     \includegraphics[width=0.1\textwidth]{tadpole.eps}
      +\left. 
        \begin{minipage}[c]{0.1\textwidth}
        \includegraphics[width=\textwidth]{sunset.eps}
        \end{minipage}
        \right\vert_{\bo{p}=\bo{0}}
  \right).
  \label{BIdeltam}
\eea
From such definition it follows that in the I-scheme
the renormalized contribution of the ``tadpole''  \emph{fully} vanishes, 
while the renormalized contribution of the ``sunset'' vanishes 
at $\bo{p}=\bo{0}$. As a consequence, the renormalized contribution 
of the ``cactus'' diagram vanishes too, 
because the ``cactus'' amplitude 
factors in terms of a ``tadpole'' amplitude 
times a finite  term.   
(Please remark that in \eref{BIdeltam} we
did not write the explicit form of $\delta m^2_\I$, being it dependent
on the choice of regularization.)

As a first important property of our intermediate scheme, an inspection of 
counterterms listed in \eref{BIdeltam} 
readily shows the absence of potential overlapping divergent (sub)graphs. Thus,
the procedure of renormalization amounts in the I-scheme 
to mechanically replacing all the instances of divergent graphs with their 
renormalized counterparts when they occur as subgraphs embedded 
in larger graphs;
in practice, all needed operations can be performed by just putting ``tadpole"
subdiagrams to zero, and by replacing all occurrencies of the 
``sunset" with a renormalized version, obtained from the original one by 
subtracting the value of the diagram in zero.

This trivialization of renormalization is 
a fondamental assumption in the automated framework presented in this paper,
and its consistent practical advantages will be silently used everywhere 
throughout all our parametrization algorithms;
it should be noticed that this property 
---which is far from being always guaranteed--- depends in general both 
on the structure of primitive divergences of the model considered 
and on the type of subtractions
that are performed in a given renormalization scheme\footnote{See e.g.
\cite{Collins} for a pedagogical discussion about overlapping 
divergences and Zimmermann's forests formula.}.

Another remarkable property of the I-scheme is that 
 $\phi_\I$ and $[\phi^2]_\I$ 
do not renormalize, i.e.
$
 \zphi{\I}=\zphiphi{\I}=1
;$
 this fact implies the vanishing of corresponding anomalous dimensions 
 \beq
 \eta_\I(g_\I)=\eta_{2\I}(g_\I)=0
 \eeq
 (see \ref{renormalization} for general definitions of $Z$-terms 
and anomalous dimensions).
Triviality of $Z$-terms implies in turn the following set of relations, 
valid for $\NExternal>0$:
\beq\label{BIcorrelators}
\Gamma_{\I}^{(\NExternal,\NVerticesII)}\!\left(\Many{\bo{p}};\Many{\bo{q}};m^2_\I,g_\I\right)
=\Gamma_\B^{(\NExternal,\NVerticesII)}\!\left(
        \Many{\bo{p}};\Many{\bo{q}};
         m^2_\B=m^2_\I+\delta m^2_\I,\lambda_\B=m_\I g_\I\right)
.\eeq
Remark in the equation above
the use of $\Many{\bo{p}}$, $\Many{\bo{q}}$ to denote respectively
the collections $(\bo{p}_1,\,\ldots\, ,\,\bo{p}_\NExternal)$, 
$(\bo{q}_1,\,\ldots\, ,\,\bo{q}_\NVerticesII)$;
please consult \ref{renormalization} for more details on our notations.

The Callan-Symanzik operator in the I-scheme, $\mathcal{D}_\I$, 
is defined by specializing to such a scheme the general definition in \eref{RCallanSymanzikD}:
\beq
\mathcal{D}_\I := 
\left. m_\I \frac{d}{d m_\I} \right\vert_{\lambda_\B,\Lambda}.
\eeq
Applying $\mathcal{D}_\I$ to both sides of \eref{BIlambda} 
the expression for the $\beta$-function 
is easily derived:
\beq
\beta_\I(g_\I)=\mathcal{D}_\I[g_\I]=-g_\I
.\eeq
The \emph{exact} Callan-Symanzik equation in the I-scheme 
(assuming $\NExternal>0$) then reads:
\bea
\left[
 m_\I\frac{\de}{\de m_\I}
 -g_\I\frac{\de}{\de g_\I}
\right]
\Gamma_{\I}^{(\NExternal,\NVerticesII)}\!\left(\Many{\bo{p}};\Many{\bo{q}}\right)
=
m_\I^2 \sigma_\I(g_\I)\;
\Gamma_{\I}^{(\NExternal,\NVerticesII+1)}\!\left(\Many{\bo{p}};\Many{\bo{q}},\bo{q}_{\NVerticesII+1}=0\right)
\label{ICallanSymanzik}
\eea
where the $\sigma$-function is obtained by applying its definition 
\eref{RSigma} to \eref{BIm}, \eref{BIdeltam}:
\beq \sigma_\I \left( g_\I \right) 
= 2 + \frac{N + 2}{24 \pi} g_\I - \frac{N +2}{288 \pi^2} g_\I^2 
.\eeq
(It should be noticed that, regardeless of the dependence of $\delta m^2_\I$ 
on the cutoff $\Lambda$,
the expression of $\sigma_\I \left( g_\I \right)$ 
does not depend on the choice of regularization in the limit
$\Lambda \to\infty$, as it should be for all RG-functions
 in a well-behaved renormalization scheme.)
 
In spite of their apparent simplicity,
 Callan-Symanzik equations in the I-scheme are not at all trivial!
For instance, specializing to $(\NExternal,\NVerticesII)=(2,0)$ 
at $\bo{p}=\bo{0}$ and introducing the rescaled function
\beq
\widehat{\Gamma_\I^{( 2,0)}}\!\left( g_\I \right)
:=
\Gamma_\I^{( 2,0)}\!\left(\Many{\bo{0}}; m_\I, g_\I \right)/m^2_\I
\eeq
we obtain the exact relation
\beq
\left[2 - g_\I  \frac{\partial}{\partial g_\I} \right]
\widehat{\Gamma_\I^{( 2,0)}}\!\left( g_\I \right)
=\sigma_\I(g_\I)\;
\Gamma_{\I}^{(2,1)}\!\left(\Many{\bo{0}};\bo{q}=\bo{0};g_\I\right)
\label{CSIcheck}
\eeq
which can be used to link two quantities involved in the computation 
of RG-functions, see discussion below.

Before closing this section some comments are in order to motivate 
our own choice of the I-scheme as intermediate renormalization scheme.

It should be clear from the considerations presented above
---and in particular from the definition of mass counterterm 
in \eref{BIdeltam}--- that the main peculiarity of the I-scheme is 
that a minimal number of counterterms is introduced.
This property results in a minimal amount of renormalized diagrams and
turns out to be invaluable to simplify the automatization 
of the parametrization process, expecially in view of cost optimization
--- see \sref{autopar}.
This simplicity, together with the desire of testing existing results in
the most independent way as possible, mainly motivates our choice 
of the I-scheme, which is in fact quite different from the one employed in 
\cite{NickelPreprint}.

The price we have to pay for this conceptual simplification
is that in this scheme we may have more diagrams to evaluate than in other 
possible schemes; in our case, for example,  
the renormalized high-order contributions to 
$\Gamma^{\left( 2, 0\right)}_\I(\bo{p}=\bo{0})$
are not trivially vanishing --- apart from the ``sunset'' graph 
and all graphs with ``tadpole'' insertion(s).
In particular, one relevant difference w.r.t.\ the scheme 
used in \cite{NickelPreprint} is that in our I-scheme 
$2$-point connected OPI (sub-)graphs 
with external lines connecting to the same vertex 
(which we refer to as to \emph{generalized tadpoles}\footnote{They are called \emph{``Hartree-type self-energy insertions''} in \cite{NickelPreprint}.}) 
give in general rise to non-zero contributions, and must thus be evaluated.
However, this complication is not a dramatic problem for at least two reasons:
\begin{enumerate}
\item the expression of  $\Gamma^{\left( 2, 0\right)}_\I(\bo{p}=\bo{0})$
can actually be obtained 
from that of $\Gamma^{\left( 2, 1\right)}_\I(\bo{p}=\bo{q}=\bo{0})$  
by use of the exact Callan-Symanzik equation in I-scheme, \eref{CSIcheck}
\item anticipating \sref{parisiScheme}, we remark that 
in addition to that of $\Gamma^{\left( 2, 0\right)}_\I(\bo{p}=\bo{0})$
also the expressions for $\Gamma^{\left( 4, 0\right)}_\I$, 
$\Gamma^{\left( 2, 1\right)}_\I$ and 
${\partial \Gamma^{\left( 2, 0 \right)}_\I}/{\partial p^2}$ 
---all evaluated at zero external momenta---
are required to obtain the desired RG-functions in the massive scheme.
It turns out in practice that the actual complexity of the evaluation 
of $\Gamma^{\left( 2, 0\right)}_\I(\bo{p}=\bo{0})$
is relatively small if compared to the computational costs of the other 
three needed correlators just mentioned.
\end{enumerate}
Motivated by those considerations, 
we will include $\Gamma^{\left( 2, 0\right)}_\I$ in our cost analysis of 
\sref{ImplementationAndResults}, planning to use \eref{CSIcheck} 
as a consistency check on future results.

\subsection{Parisi's massive scheme}\label{parisiScheme}
The standard massive scheme ---suffix $\M$---, 
as introduced in {\cite{Parisi}},
 is defined by the following normalization conditions:
\bea
\left. \Gamma_\M^{(2,0)}(\bo{p},-\bo{p};m^2_\M,g_\M) \right\vert_{\bo{p}=\bo{0}} = m^2_\M
\label{MNormCond-I}
\\
\left. \frac{\de\Gamma_\M^{(2,0)}}{\de p^2}(\bo{p},-\bo{p};m^2_\M,g_\M) \right\vert_{\bo{p}=\bo{0}} = 1
\label{MNormCond-II}
\\
\left. \Gamma_\M^{(4,0)}(\Many{\bo{p}};m^2_\M,g_\M) \right\vert_{\Many{\bo{p}}=\bo{0}} = m_\M g_\M
\label{MNormCond-III}
\\
\left. \Gamma_\M^{(2,1)}(\bo{p},-\bo{p};\bo{q};m^2_\M,g_\M) \right\vert_{\bo{p}=\bo{q}=\bo{0}} =1.
\label{MNormCond-IV}
\eea
(Please remark that we are adopting here the same conventions 
introduced for bare correlators in equations~(\ref{on-fact-I}-\ref{on-fact-III}).)
The relation with bare quantities, in the case $\NExternal>0$,
 follows from \eref{GAR-GAB-II}:
\beq\label{BMcorrelators}
\Gamma_{\M}^{(\NExternal,\NVerticesII)}(\Many{\bo{p}};\Many{\bo{q}})
=
\left(\zphi{\M}\right)^{\NExternal/2} \left(\frac{\zphiphi{\M}}{\zphi{\M}}\right)^{\!\!\NVerticesII}
 \Gamma_\B^{(\NExternal,\NVerticesII)}(\Many{\bo{p}};\Many{\bo{q}}).
\eeq

Callan-Symanzik equations \eref{RCallanSymanzik}, 
and definitions of RG-functions as in equations (\ref{RBeta}-\ref{RSigma}),
hold in this scheme with the suffix replacement $\R \to \M$.
Specializing \eref{RCallanSymanzik} to \eref{MNormCond-I} we get
an additional relation specific to this renormalization scheme:
\beq
2-\eta_\M(g_\M) = \sigma_\M(g_\M).
\eeq
The other RG-functions are obtained as follows:
\begin{enumerate}
\item using \eref{BIcorrelators}, \eref{BMcorrelators}
and normalization conditions (\ref{MNormCond-I}-\ref{MNormCond-IV})
we obtain the relations
\bea\label{M-Z}
\zphi{\M}=\left(\left.\frac{\de \Gamma_{\I}^{(2,0)}}{\de p^2}(\bo{p},-\bo{p})
\right\vert_{\bo{p}=\bo{0}}\right)^{-1}
\\\label{M-ZII}
\zphiphi{\M}=\left(\Gamma_{\I}^{(2,1)}(\Many{\bo{0}};\bo{0})\right)^{-1}
\\\label{M-g}
g_\M=\frac{\Gamma_{\I}^{(4,0)}(\Many{\bo{0}})}{
\sqrt{\Gamma_{\I}^{(2,0)}(\Many{\bo{0}})
\left(\left.\displaystyle\frac{\de \Gamma_{\I}^{(2,0)}}{\de p^2}(\bo{p},-\bo{p})
\right\vert_{\bo{p}=\bo{0}}\right)^{3}}}
\\\label{M-mRatio}
\frac{m^2_\M}{m^2_\I}=
\frac{\Gamma_{\I}^{(2,0)}(\Many{\bo{0}})}{m^2_\I
\left(\left.\displaystyle\frac{\de \Gamma_{\I}^{(2,0)}}{\de p^2}(\bo{p},-\bo{p})
\right\vert_{\bo{p}=\bo{0}}\right)}
\eea
\item by definition (and use of chain rule) we observe that
\bea
\label{Mbeta}
\beta_\M(g_\M)
   =\mathcal{D}_\M[g_\M]
   =\mathcal{D}_\M[g_\I]\;\frac{\de g_\M}{\de g_\I}
\\
\eta_\M(g_\M)
   =\mathcal{D}_\M[\log \zphi{\M}]
   =\mathcal{D}_\M[g_\I]\;\frac{\de \log \zphi{\M}}{\de g_\I}
\\
\label{MetaII}
\eta_{2\M}(g_\M)
   =\mathcal{D}_\M[\log \zphiphi{\M}]
   =\mathcal{D}_\M[g_\I]\;\frac{\de \log \zphiphi{\M}}{\de g_\I}
\eea
\item the expression for $\mathcal{D}_\M[g_\I]$ is readily derived
by applying $\mathcal{D}_\M$ to both sides of relation
$\lambda_\B=m_\I g_\I$, and using the identity
\beq
\frac{\mathcal{D}_\M[m_\I]}{m_\I}
=1-\frac{1}{2}\;\mathcal{D}_\M\!\!\left[\log \frac{m^2_\M}{m^2_\I}\right]
=1-\frac{1}{2}\;\mathcal{D}_\M[g_\I]\;\frac{\de\log \frac{m^2_\M}{m^2_\I}}{\de g_\I}
.\eeq
\end{enumerate}
As a result we obtain:
\beq\label{DMgI}
\mathcal{D}_\M[g_\I]
={-g_\I}{\left(1-\frac{g_\I}{2}\frac{\de\log \frac{m^2_\M}{m^2_\I}}{\de g_\I}\right)}^{-1}
.\eeq
Equation \eref{DMgI} together with equations (\ref{M-Z}-\ref{M-mRatio}) 
and equations (\ref{Mbeta}-\ref{MetaII}) allow to determine 
RG-functions in the massive scheme from the knowledge of
$\Gamma_{\I}^{(2,0)}$, of its derivative w.r.t.\ $p^2$, of $\Gamma_{\I}^{(4,0)}$
and $\Gamma_{\I}^{(2,1)}$, all evaluated at zero external momenta.

To compute critical exponents one must first resum 
with some appropriate method (e.g., Borel summation) 
the divergent perturbative 
series obtained for $\beta_\M(g_\M)$, for anomalous dimensions 
$\eta_\M(g_\M)$, $\eta_{2\M}(g_\M)$,
and ---as a typical check--- for other series derived from anomalous dimensions by usual scaling and hyper-scaling
relations among exponents, 
like e.g. $\nu_\M(g_\M)=1/(2+\eta_{2\M}(g_\M)-\eta_{\M}(g_\M))$.
Critical exponents are then obtained by evaluating the resummed series at $g_\M=g_\M^*$,
the non trivial zero of $\beta_\M(g_\M)$, 
which corresponds to the critical region.
In spite of the fact that they can be obtained from scheme-dependent quantities,
exact critical exponents are universal numbers. 
(Notice nevertheless that the speed of convergence of
approximations obtained from resummation of 
perturbative series at fixed finite order might vary by choosing different 
renormalization schemes.)

\section{Setting up the problem}\label{generation}
\subsection{Flowchart}

As shown in previous section,
performing the calculation of RG-functions is tantamount to computing 
perturbative series up to the desired order for
correlators $\Gamma^{\left( 2, 0 \right)}_\I$, $\Gamma^{\left( 4, 0 \right)}_\I$, 
$\Gamma^{\left( 2, 1\right)}_\I$ 
and ${\partial \Gamma^{\left( 2, 0 \right)}_\I}/{\partial p^2}$,
all evaluated at zero external momenta.

\begin{figure}\label{TheRoadmap}
\begin{center}
\hfill\includegraphics[width=0.85\textwidth]{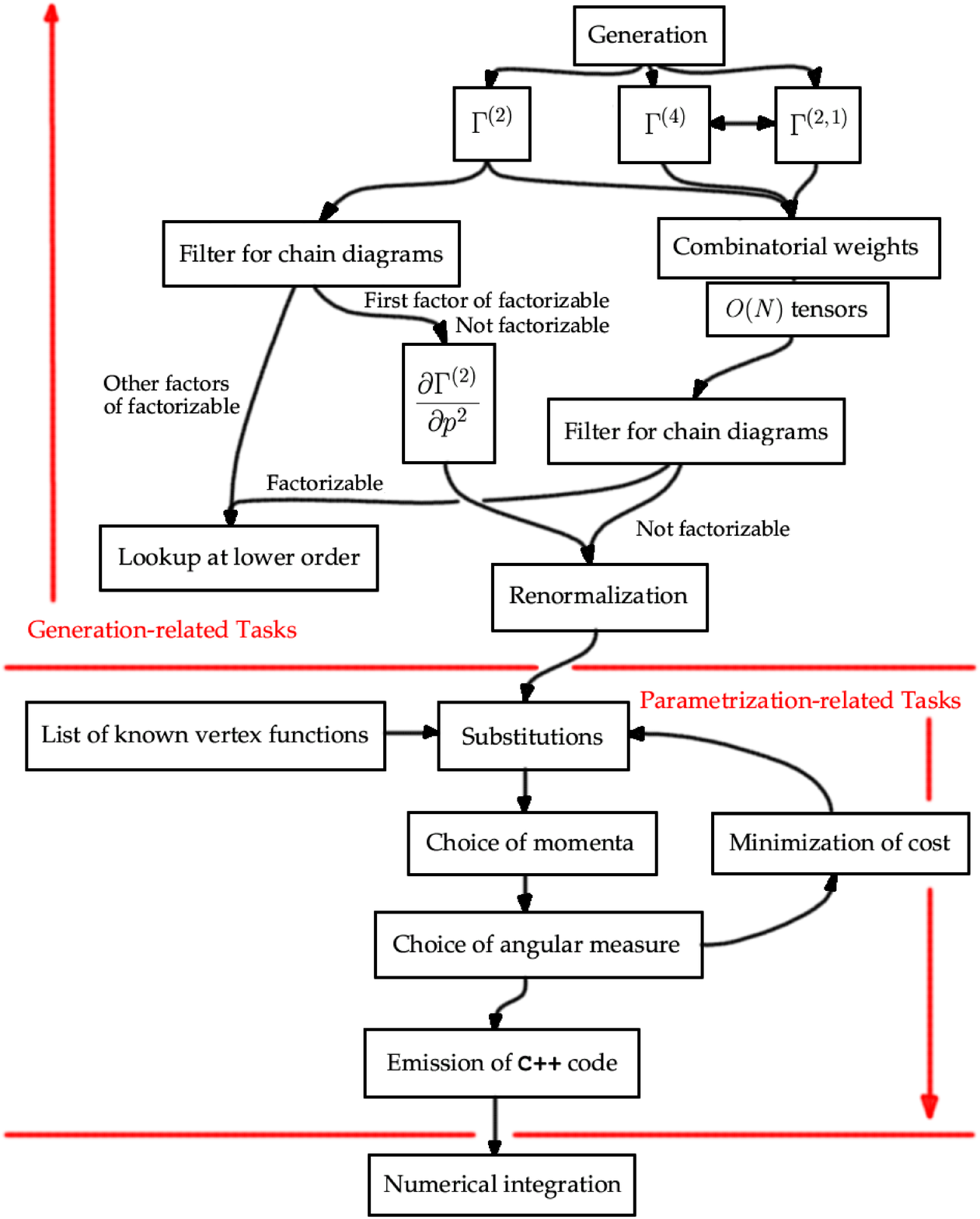}
\caption{Flowchart of our automatic framework. 
 Its details are clarified in sections \ref{generation}-\ref{autopar}.}
\end{center}
\end{figure}

In figure \ref{TheRoadmap} we present a flowchart where all the tasks 
contributing to the structure of our automatic framework are mentioned.
Quite schematically, we can identify three main steps:
the generation of all needed diagrams with their associated weights and $O(N)$
tensor structure, the parametrization of amplitudes associated to
diagrams and, finally, the numerical integration of generated code.

The problem of generating diagrams and combinatorial factors 
is computationally hard but standard, 
and will thus be addressed only briefly at the beginning 
of section \ref{generation}; the last part of section \ref{generation} 
is devoted to prepare the theoretical ground for section \ref{autopar},
where we present the algorithmic implementation of an efficient strategy
to find for each amplitude integral representations 
with low associated dimensionality.
As anticipated, in this paper we will not deal at all 
with the challenging problem posed by the numerical integration of
obtained parametrizations, leaving this task to future publications.

\subsection{Graphs, weights and tensors}
\subsubsection{Representation of graphs}
While many good computer packages for manipulating graphs do exist 
(see for instance \cite{GAP}), 
some of them being specifically tailored to applications in theoretical 
physics (one such example is \cite{FeynARTS}), 
we have anyway decided to re-implement  
our own graph tool from scratch. The main motivation for such an effort 
can be found in the fact that, since our entire parametrization chain 
is formulated in terms of graphs, we needed much more extended capabilities 
than those usually found in usual Feynman-graph manipulation tools; 
furthermore, we badly needed efficiency, because some of the stages of  
parametrization production are computationally very expensive 
(see \sref{DerivativeSection} and the end of \sref{Loop Momenta}). 

All these considerations brought us to the conclusion that only a new library,
written in terms of an efficient low-level programming language 
and providing a set of graph-theoretical operations specifically 
tailored to our needs, could solve our problems --- in the end, \texttt{C++}
has been our final choice of language.

Here are the basic design choices which have been formulated for our library:
\begin{itemize}
\item it allows graphs with {\em coloured vertices} and 
{\em coloured lines} to be described (in our situation, 
coloured vertices are used to model external lines and effective vertices 
---see \sref{SubstitutingVertices}---, while
coloured lines are needed i.e. for symmetries of 
arguments of effective vertices and cosine diagrams, 
see \sref{ChoiceOfAngularMeasure})
\item it is based on an {\em adjacency-matrix} representation 
(we briefly recall that for diagrams where only a single kind of lines 
is allowed, 
each entry of the adjacency matrix $\Adjacency_{v,v'}$ gives the
number of lines joining vertex $v$ and vertex $v'$; 
in more complicated situations $\Adjacency_{v,v'}$ encodes both the number 
and the colours of such lines)
\item it knows the concept of {\em canonical representative} 
for a graph. 
The basic idea here is the fact that the same graph 
can have in general different representations, 
the one obtained from the other by a relabeling of graph vertices;
since such a relabeling defines an equivalence relation, 
the canonical representative is ---as usual in similar situations--- 
the special representative picked up to label all the equivalent\footnote{Or, following standard graph terminology, {\em isomorphic}.} elements
belonging to the same class. 
It should be noticed that implementing the concept of canonical 
representative was particularly important in our context: in fact, 
the definition of the representative can be used to provide
an ordering relation for graphs, which is in turn essential 
to a good cooperation of our library with 
the data-types and the facilities offered by \texttt{C++} standard library
\cite{Stroustrup}.

However, while the knowledge of the canonical representative for a graph 
gives many advantages, it has the main drawback that its computation 
can turn out to be very expensive 
(in fact, this operation can imply the inspection of a number of graphs 
up to the factorial of the number of vertices, such being the 
maximal number of possible vertex relabelings).

On the other hand, 
many algorithms exist which allow to reduce the cost of finding 
the canonical representative (even if each algorithm known so far suffers 
from some ``difficult'' graphs); these algorithms usually consist 
in ``colouring'' vertices on the basis of some vertex property 
which is invariant by vertex relabeling, so to partition the vertex set 
into smaller subsets, and to reduce ---sometimes drastically--- 
the number of permutations to be checked.
Our personal choice among the many algorithms 
proposed in the literature (see for example \cite{SchmidtDruffel}, 
\cite{McKay} and \cite{Rosen} for an overview) has 
been to implement a twofold colouring, based both on leg partitions and on 
minimal distances (refer to \cite{SchmidtDruffel} 
for a description of the latter); these criteria, though being 
not as efficient for large graphs as those used in \cite{McKay}, 
are far less complicated 
from the point of view of software implementation, and nonetheless 
quite efficient for the small Feynman diagrams we need to cope with. 
\end{itemize}
Some of the graph-theoretical operations provided by our library 
are for example the identification of connected components, 
the computation of the order of automorphism group, 
the computation of spanning trees, 
the enumeration of all occurrences of a given subgraph in a larger graph 
(presented as an ordered sequence, that is as an {\em iterator} in 
\texttt{C++} terminology, see \cite{Stroustrup}), 
and the enumeration of all cycles, 
of all cycle bases and of all paths joining two vertices 
(in the form of iterators as well). All these capabilities will be essential  
to implement algorithms presented in \sref{autopar}.

\subsubsection{Generation of graphs}\label{GenerationOfGraphs}
The task of generating all inequivalent graphs which satisfy a
predefined set of topological conditions is known since a long time to be a
very hard one, due to the fact that all known generation algorithms 
are also plagued by the problem of isomorphic copies: 
diagrams corresponding to the 
same canonical representative are in general produced more than once, leading 
to a worser and worser inefficiency of the process as the number 
of vertices of the graph increases.
Various methods have been proposed in the literature;
here again, some packages are available (see for example \cite{qgraf}). 
In the spirit of \cite{Heap}, 
we generate the set of diagrams in a non-recursive way, 
taking advantage of our knowledge of the canonical representative 
to prune a large number of isomorphic copies out 
of the generated tree (see \cite{Nagle}).
The performance of such algorithm is very satisfactory, at least for the 
typical number of vertices we are interested in.
\begin{table}
\caption{\label{NumberOfDiagrams} Number of diagrams (OPI and without tadpoles) contributing to
the two-point and four-point function up to $8$ loops in perturbative order.}
\begin{indented}
\item[]\begin{tabular}{@{}ccccccccccccc}
\br
&\centre{9}{Incremental number}&\centre{3}{Total number}\\
\ns
&\crule{9}&\crule{3}\\
Loop number& 0&1&2&3&4&5&6&7&8 & 6&7&8\\
\br
$\Gamma^{(2,0)}_\I$& 0&0&1&2&6&19&75&317&1622 & 103&420&2042\\
\mr
$\Gamma^{(4,0)}_\I$& 1&1&2&8&27&129&660&3986&26540 & 828&4814&31354\\
\br
\end{tabular}
\end{indented}
\end{table}

In \tref{NumberOfDiagrams} we quote some results about the
number of diagrams needed to compute the two-point and four-point functions. 
The main facts we can deduce from these numbers are the following:
\begin{enumerate}
\item to complete the evaluation of the RG $\beta$-function at $7$-loop level 
$3986$ more diagrams must be evaluated just for the four-point function 
$\Gamma^{(4)}_\I$ alone
\item improving the computation of anomalous dimensions 
$\eta_{\M}$ and $\eta_{2\M}$ in terms of $g_{\I}$ to $8$ loops 
would require the evaluation in $\bo{p}=\bo{0}$ of both values 
and derivatives w.r.t.\ $p^2$ 
of the $1622$ diagrams contributing to $\Gamma^{(2,0)}_\I$,
plus ---eventually--- the evaluation of diagrams 
contributing to $\Gamma^{(2,1)}_\I(\bo{p}=\bo{q}=\bo{0})$.  
(For a more precise statement of the problem,
please refer to the discussion at the end of \sref{IScheme}.)
These diagrams have more lines and, accordingly, are more difficult 
to evaluate than those contributing to the seven-loop four-point function,
see \sref{ImplementationAndResults}.
\end{enumerate}
In any case, results in \tref{NumberOfDiagrams} state clearly that 
neither of the two tasks presented above can be performed 
without the help of an automated framework, able to supply 
the parametrization and the evaluation of all needed amplitudes.

\subsubsection{Symmetry factors}
When generating Feynman diagrams for the perturbative expansions of
correlators in quantum field theories, one is interested in
describing graphs whith unlabeled internal vertices, 
and unlabeled lines as well, because graphs differing by such relabelings 
contribute with the same amplitude; this causes the problem of computing
the {\em symmetry factor} for a given Feynman graph, 
which accounts for the multiplicity of
such identical contributions.

Practical computation of symmetry factors can be very difficult, since it 
requires the knowledge of the order 
of the automorphism group of the graph of interest; 
in our case, symmetry factors are derived directly from the adjacency-matrix
of the graph, following the directions given in \cite{RibecaPhD}.

\subsubsection{$O \left( N \right)$-factors} 
The computation of traces for $O(N)$ group is quite simple, and can be carried 
out in many ways; for example, tensors can be represented as graphs, 
and their contractions can be performed in full graphical form as well.
Since this point does not pose particular problems, we will not insist on it 
here.

\subsubsection{Consistency checks}
Of course, all possible precautions must be taken against the unpleasant 
possibility of accidentally omitting a graph, or computing a wrong 
symmetry/$O(N)$ factor. 
The consistency of our results has been carefully checked 
in many ways; among them we recall the fact that from 
{\em zero-dimensional field theory} some sum rules can be 
deduced, allowing to test both combinatorial and $O(N)$ factors 
(see \cite{Zinn} for a description of this technique). As an aside, we notice 
that the introduction of a mass counterterm in such a context allowed us 
to directly obtain sum rules for diagrams without tadpoles
(and an analogous solution has apparently been adopted 
in \cite{NickelPreprint}).

\subsection{Which parametrization to choose?}\label{WhichParametrization}
Ideally, in our approach to the computation of RG-functions the ``optimal'' 
parametrization for a given Feynman amplitude 
is the one minimizing the CPU-time 
which is needed for its numerical evaluation, up to the desired precision. 
In practice, many factors come into play while trying to quantify 
in a precise way the various contributions to such a computational cost; 
for example: 
\begin{itemize}
\item for a given amplitude, any choice of parametrization induces 
its own integral representation, with its associated specific dimensionality
\item an estimate of the number of evaluations of a given integrand needed 
to compute the value of a multidimensional integral within 
a specified numerical accuracy is in general 
not precisely known {\em a priori}, since it
depends in an unaccessible way on the regularity properties
of the integrand function itself

\item 
parametrizations of the same amplitude in terms of integrals of the same 
dimensionality may be obtained ---see \sref{SubstitutingVertices}---
by choosing in various ways the set of replaced effective-vertex functions; 
these functions may have in general very unequal evaluation timings

\item the same vertex function may possibly require a very different amount 
of floating-point operations when it is evaluated at different values 
of external momenta.
\end{itemize}
All these considerations suggest that we should not rely 
on a too much refined 
definition of the {\em cost function} whose minimum characterizes 
the optimal parametrization; on the other hand, the 
typical asymptotic behavior of the error estimate 
in deterministic multidimensional numerical integration,
\beq
\mathrm{error}\sim (\mathrm{integrand\ evaluations})^{-{\mathrm{constant}}/{\mathrm{dimensionality}}}\label{IntegrationCost}
,\eeq
suggests that it is the dimensionality of the integration which sets the basic 
scale of the number of required evaluations of the integrand.
Thus, we will discard all 
other possible definitions, basically sticking as our cost function 
to the dimensionality of the final integral needed to compute an amplitude 
--- even if for technical reasons the actual definition 
(given in \sref{autoparOverview} and used throughout the article) 
will need to be slightly more refined, its spirit remains the same.


Having in mind from now on the goal of minimizing the dimensionality 
of the integral representations of our amplitudes, let us evaluate 
such a dimensionality for four standard families of parametrizations: 
we will thus compute the basic cost of momentum representation ($\CostMom$),
that of position representation ($\CostPos$), and those of
Schwinger ($\CostSchw$) and Feynman ($\CostFey$) representations.
Such costs are here called \emph{basic} to
stress the fact they refer to the dimensionality of integrals as 
directly obtained from standard parametrization techniques, 
without considering the possible use of additional special simplifications  
(in the case of momentum representation, for instance, a more refined choice
of angular measure could be supplied; such a choice will indeed 
play an important role in lowering the number of residual integrations, 
as described in \sref{ChoiceOfAngularMeasure}).


Let us consider a connected OPI diagram $\Diagram$ 
with $\NExternal$ external lines, $\NVertices$ internal vertices 
($\NVertices=:\NVerticesIV+\NVerticesII$, 
$\NVerticesII$ and $\NVerticesIV$ being  
the number of vertices with $2$ and $4$ legs, respectively), 
$\NLoops$ loops and $\NLines$ internal lines.
The corresponding $d$-dimensional Feynman amplitude $\AmplitudeD$
evaluated at zero external momenta and unitary masses
can be schematically written in terms of
momentum, position, Schwinger and Feynman 
representations as (respectively): 
\bea
\label{parMom}
\AmplitudeD &\propto \int \prod_{l=1}^{\NLoops} d^d\ell_{l} \;\prod_{i=1}^{\NLines}\frac{1}{1+k^2_i}
\\
\label{parPos}
&\propto \int \prod_{v=1}^{\NVertices} d^d x_{v} 
  \; \delta^d(x_\NVertices) 
  \prod_{v=1}^{\NVertices} \;\prod_{v'=v+1}^{\NVertices}
  \left(\Propagator(x_v-x_{v'})\right)^{\Adjacency_{v,{v'}}}
\\
\label{parSchw}
 &\propto
 \int_{0}^{\infty} \prod_{i=1}^{\NLines} d\alpha_i 
 \; \frac{e^{-\sum_i \alpha_i}}{\left(Q_\NLoops(\Many{\alpha})\right)^{\frac{d}{2}}}
\\
\label{parFey}
 &\propto
 \int_{0}^{1} \prod_{i=1}^{\NLines} d\alpha_i 
 \; \delta(\sum_i \alpha_i-1)
 \; \left(Q_\NLoops(\Many{\alpha})\right)^{-\frac{d}{2}}
,\eea
where we denoted by $\ell_l$ the loop momenta,
by $k_i$ the momentum flowing in the $i$-th internal line 
---which can be expressed as an appropriate linear combination 
of loop momenta---, 
by $\Propagator(x)$ the propagator in position space, 
by $\Adjacency_{v,{v'}}$ the entries of the adjacency matrix 
---i.e., the number of lines connecting vertex $v$ to vertex $v'$---,
and by $Q_\NLoops(\Many{\alpha})$ an appropriate 
homogeneous function of $(\alpha_i)_{i=1}^{\NLines}$ of degree $\NLoops$
(see e.g.\ \cite{ItzyksonZuber} for more details).

Before moving to the evaluation of the basic costs 
for these integral representations,
a useful property of $SO(d)$-invariant integrands must be mentioned
(see \sref{ChoosingAxes} 
for a more detailed illustration in the case $d=3$).
When integrating a function over a collection of vectors
$v_1,\,\ldots\, ,\, v_K\in\mathbb{R}^d$,
 the dimensionality of the original integral 
can be reduced in a standard way 
if the integrand function depends \emph{only} on 
the scalar products among those vectors; 
the reduction is obtained 
by performing the trivial angular integrations which correspond 
to the invariance 
w.r.t.\ a global $SO(d)$ rotation of integration vectors 
$v_1,\,\ldots\, ,\, v_K$.
If $\mathrm{K}\geq d-1$, the number of such trivial integration is ${d(d-1)}/{2}$.
Due to vanishing external momenta we evaluate amplitudes at,
this property can be applied to
momentum representation \eref{parMom} as well as to position representation
\eref{parPos}.

With this property in mind, we are now ready to complete 
the evaluation of basic costs: 
the dimensionality of the integral required by each representation 
follows from previous 
expressions, by properly taking into account the delta functions in 
\eref{parPos}, \eref{parFey}, 
and by making use of the $SO(d)$-symmetry of \eref{parMom}, \eref{parPos} 
(corresponding estimates \eref{costMom}, \eref{costPos} 
hold for $\NLoops \geq d-1 $ and $\NVertices \geq d$, respectively):
\bea\label{costMom}
\CostMom&= d \NLoops -\frac{d(d-1)}{2}
\\\label{costPos}
\CostPos&= d \left(\NVertices-1\right) -\frac{d(d-1)}{2}
         = d \left(\NLoops+ \frac{\NExternal-3-d}{2}+\NVerticesII\right)
\\\label{costSchw}
\CostSchw&=\NLines =2 \NLoops+ \frac{\NExternal-4}{2}+\NVerticesII
\\\label{costFey}
\CostFey&=\NLines -1 =2 \NLoops+ \frac{\NExternal-4}{2}+\NVerticesII -1
;\eea
it should be noticed that the standard topological identities
\bea
\NVerticesIV&= \NLoops+ \frac{\NExternal-2}{2}
\\
\NLines &= 2\NLoops+ \frac{\NExternal-4}{2}+\NVerticesII
\eea   
have been used to eliminate $\NVerticesIV$ and $\NLines$ from formulas
(\ref{costPos}--\ref{costFey}). 

From expressions (\ref{costMom}-\ref{costFey}) it clearly appears that, 
when $d>2$ 
and the number of loops $\NLoops$ is large enough, 
$\CostFey < \CostMom \sim \CostPos$.
\begin{table}
\caption{\label{StandardCosts} Basic costs 
$(\CostMom,\CostPos,\CostSchw,\CostFey)$ 
---for momentum, position, Schwinger and Feynman representations 
respectively--- of a $\NLoops$-loop amplitude contributing 
to $\Gamma^{(\NExternal,\NVerticesII)}$ at $d=3$.}
\begin{indented}
 \item[]{
\begin{tabular}[ht]{@{}llll}
 \br
     & $\Gamma^{(2,0)}$&$\Gamma^{(4,0)}$&$\Gamma^{(2,1)}$ 
 \\
 \mr
 $\NLoops=6$ & $(15,12,11,10)$& $(15,15,12,11)$& $(15,15,12,11)$
 \\
 $\NLoops=7$ & $(18,15,13,12)$& $(18,18,14,13)$& $(18,18,14,13)$
 \\
 $\NLoops=8$ & $(21,18,15,14)$& $(21,21,16,15)$& $(21,21,16,15)$
 \\
 \br
\end{tabular}
 }
\end{indented}
\end{table}
%
In \tref{StandardCosts} we reproduce in detail the costs of integrating  
the $3$-dimensional amplitudes contributing at loop orders $\NLoops=6,7,8$ to 
$\Gamma^{(2,0)}$, $\Gamma^{(4,0)}$ and $\Gamma^{(2,1)}$.
From this analysis of standard representations, the apparent conclusion 
would be that ---in the case of interest--- Feynman representation considerably lowers 
the number of residual integrations if 
compared to momentum or position representations.

However, the seminal idea ---exposed in {\rm\cite{NickelDiagrams}}--- 
which allowed Baker, Nickel, Green and Meiron 
to obtain in {\rm\cite{NickelI}-\cite{NickelII}}
quite precise $6$-loop estimates of the RG-functions consisted, in fact,
in using a \emph{non-standard} representation of amplitudes in momentum space.
This new representation was obtained from the standard one by replacing 
in the amplitudes as many one-loop subintegrals as possible
--- corresponding to one-loop subdiagrams of the main diagram; 
in fact, since the analytic expression of one-loop correlators
for non-exceptional momenta is analytically known from the work of Melrose 
\cite{MelroseReduction}, 
a reduction of $3$ integrations in the basic cost readily follows from 
each substitution one can make in the original amplitude.

As a matter of fact, if we analyze in more detail the cost $\CostMomImpr$ 
of momentum representation when it is improved in the spirit of Nickel et al.,
we soon come to a surprising conclusion. Let us suppose
that we have been able to replace in an amplitude, say, 
$\NLoopsReplaced$ analytically known functions, 
each corresponding to a one-loop subdiagram. Assuming  
$\NLoops-\NLoopsReplaced\geq 2$, so that we can spare $3$ integrations 
thanks to the overall $SO(3)$-invariance of the amplitude, 
we then obtain the relation    
\beq
\CostMomImpr=3 \left( \NLoops-\NLoopsReplaced-1\right).
\eeq
Now, the requirement that the obtained parametrization have no more residual 
integrals than the standard Feynman parametrization is equivalent 
to the following condition:
\beq
\CostMomImpr\leq\CostFey \Leftrightarrow 3\NLoopsReplaced \geq \NLoops-\frac{\NExternal}{2}-\NVerticesII
\eeq
(notice that in the present context the lower bound on $\NLoopsReplaced$ is maximized by $\Gamma^{(2,0)}$).
More explicitly, when $\NLoops=7 (8)$ the condition $\CostMomImpr\leq\CostFey$  
reads  
$\NLoopsReplaced \geq 2 (3)$ for $\Gamma^{(2,0)}$ and 
$\NLoopsReplaced \geq 2 (2)$ for $\Gamma^{(2,1)}$ and $\Gamma^{(4,0)}$:
this means that if we are able to perform enough replacements the improved
momentum space representation will beat standard Feynman parametrization. 
This fact motivates the choice ---Nickel's as well as ours--- of this 
representation for amplitudes.

In addition, surprises are not yet over: it turns out that 
in this representation 
---complemented  with a convenient choice of renormalization scheme---
one can naturally take advantage of a whole set of powerful identities and properties.
For example, all the occurrences 
of the two-loop ``sunset'' diagram can be replaced by the analytic 
expression of its renormalized counterpart, with a gain of $6$ integrals; but 
much more can be done. 
The next sections are devoted to a systematic presentation of 
a set of additional simplifications and tricks
which are well suited to improve this framework.
In particular:
\begin{itemize}
\item the idea of replacing in the amplitudes 
some functions known analytically 
---which we will usually refer to as to
\emph{effective vertex functions}, or as to
{\em effective vertices} for short--- is very powerful, and can 
be generalized to effective vertices other than the ones which were 
presumably used by Nickel and coworkers. A detailed analysis of this 
method and its variations is carried out in \sref{EffectiveVerticesSection}.
\item other identities valid when computing amplitudes at zero external momenta 
can be used to achieve further reductions in the final dimensionality of 
some parametrizations; they are analyzed in \sref{ZeroExternalMomenta}
\item a good choice of {\em angular measures} ---that is, of the way 
of writing the parametrization of loop integration variables 
in spherical coordinates--- can lead to the analytic integration of a
conspicuous number of angles. However, being not explicitly tied to Nickel's 
framework this problem is in a sense a more general one, 
and requires some additional notation which will be introduced 
only at the beginning of \sref{autopar}; thus, we postpone
this issue to \sref{ChoiceOfAngularMeasure}
\item closely related to the last point, there is the possibility of 
factoring and computing analytically some easier one-dimensional integrals of 
chains of propagators; this subject is postponed as well 
to \sref{ChoiceOfAngularMeasure}.
\end{itemize}
However, it goes without saying that this approach to the parametrization 
of amplitudes also shows some drawbacks.

First of all, it must be stated that we do not know exactly which set of 
simplification rules has been used in \cite{NickelI}-\cite{NickelII} to lead 
to such a spectacular reduction of the complexity of the original problem, 
nor we do know whether 
such a set can be formulated in terms of simple algorithmic procedures. 
In fact, only a hint of the basics of effective-vertex technique is given 
by Nickel and coworkers in their published literature 
(see \cite{NickelPreprint} and \cite{NickelDiagrams}); we have thus tried, 
so to say, to ``reverse-engineer'' Nickel's results 
---and preprint \cite{NickelPreprint} in particular--- 
to deduce or re-invent the techniques we present here, 
selecting in the end the ones which seemed to us to be the most appropriate 
for automation; however, nothing prevents some (perhaps essential) 
ingredients of the original approach from being possibly very difficult 
to automate, and very difficult to state in algorithmic form.

Secondly, a related ---and much more fundamental--- 
problem posed by this otherwise appealing 
framework is the fact that its effectiveness cannot be 
proven {\emph{a priori}}: no theorem exists 
(at least in our knowledge) stating that at a given loop order, 
for a given set of available effective vertices, some minimal number of
replacements will be uniformly obtained for {\em all} graphs needed 
during the evaluation of some field-theoretical quantity; 
the risk exists that, due to stronger and stronger topological 
obstructions present in diagrams at higher orders, 
the set of effective vertices will prove inadequate to 
satisfactorily reduce some ``difficult'' graphs. As a matter of fact, a few 
such graphs requiring more residual integrations than their relatives do 
appear for some choices of effective vertices and sets of graphs 
to be parametrized. In the lack of a proof {\emph{a priori}}, 
the effectiveness of the method can be demonstrated 
only {\emph{a posteriori}} with an explicit inspection carried out 
diagram by diagram; considering the very large number of amplitudes, 
the large number of different possible replacements for a given diagram
and the tempting possibility of increasing the set of effective vertices, 
once again we arrive at the conclusion that such a complex optimization 
problem can be successfully analyzed only by means of an automatic framework, 
capable to handle the problem of minimizing the number of residual amplitudes 
fastly and more effectively than any human being. 
In addition, such an approach is the only one able to provide an evolutive 
set-up if new tricks are found or new parametrization strategies 
prove themselves to be necessary (one example of such a case could be 
an hypothetical $8$-loop computation, where one could wish to implement, 
for instance, a stage choosing the best parametrization among those given 
by both momentum and Feynman representations).  

Consequently, in \sref{autopar} we will show how to build a 
prototype framework
to optimize ---according to a reasonable choice of simplification rules---
the parametrization of a given amplitude, starting from a given
set of user-defined effective vertices.
As reported on in \sref{ImplementationAndResults} 
and \sref{epilogue}, the knowledge of a small number of
effective vertices already leads to very effective parametrizations 
for the cases $\NLoops=6$ and $\NLoops=7$.

\subsection{More on effective vertices}\label{EffectiveVerticesSection}
The idea of Nickel et al. of substituting effective vertices 
in a given amplitude, thus attempting to decrease the number of residual
integrations and find a cheaper parametrization, is very general. 
The underlying hypotheses are the line-locality of momentum representation 
 (that is, the mapping between lines of the diagram 
and products of corresponding propagators in the integrands),
integration over loop momenta,
the triviality of renormalization (absence of overlapping divergences, 
implying that one may replace a divergent block with its renormalized 
counterpart) and, of course, the knowledge of {\em zero-cost}\/ 
or ---more generally--- {\em low-cost}\/ expressions 
for effective vertex functions
describing subdiagrams of the original amplitude (by {\em zero-cost} 
expression we mean a subgraph which is known 
in terms of elementary functions and zero integrations, 
while with {\em low-cost} expression we indicate 
an equivalent integral representation of a subdiagram involving a 
smaller number of integrations than the one we started with).

Various classes of low-cost vertex functions are available;
they will be examined in the following sections.

\subsubsection{One-loop functions}
As already mentioned, analytic expressions in terms of elementary functions 
for one-loop functions in $d=3$ are known:
following our definition, we say that all such functions 
are zero-cost effective vertices.

More in detail, 
in \cite{MelroseReduction} 
one-loop correlators with $\NExternal\ge d+1$ legs ($d$ being the integer spacetime dimension)
are reduced to a linear combination of one-loop correlators with $\NExternal=d$ legs
(see also \cite{NickelDiagrams}).
Unfortunately, due to the presence of inverse powers of kinematical determinants
in the coefficients of the reductions, such formulas have a range 
of validity limited to non-exceptional momenta, 
i.e. to kinematical configurations such that the involved determinants are nonvanishing.
One possible way to patch this potentially catastrophic problem 
is to reject 
such exceptional points during  numerical integration:
this strategy ---suggested in \cite{NickelDiagrams}--- 
has the drawback of requiring
the compatibility of point-rejection with the chosen integration algorithm.
As a complementary alternative,
we developed in \cite{oneloop} a general theoretical framework 
to deal with reductions of one-loop correlators 
in all kinematical situations,
and we implemented it in an highly-optimized 
and robust \texttt{C++} library. 

\subsubsection{Functions in terms of spectral densities}\label{spectral}
All higher-loop amplitudes whose expressions can be simplified 
to reduce the number of residual integrals constitute potentially 
useful effective vertices.

A first concrete example of such a situation is obtained when considering
$\NLoops$-loop amplitudes $\AmplitudeD^{(2,0)}(p^2)$ contributing to $\Gamma_\I^{(2,0)}(\bo{p},-\bo{p})$: 
due to their analyticity in the complex $p^2$ plane cut at 
$p^2<-M^2_{\mathrm{threshold}}$,
these amplitudes possess a dispersive
one-dimensional integral representation, 
which ---when renormalization is not required--- has the standard unsubtracted form 
\beq 
\AmplitudeD^{(2,0)}(p^2):=\int \prod_{l=1}^{\NLoops} d^3\!\bell_{l}
 \;\prod_{i=1}^{\NLines}\frac{1}{k^2_i+m_i^2}
=\int_0^{\infty} \frac{d M^2}{p^2 + M^2}\;\AmplitudeDDensity\!\left(M\right)
\label{CauchyTheorem}
\eeq
where the density $\AmplitudeDDensity$ is related via Cauchy theorem 
to the discontinuity of the amplitude along the $p^2$-cut,
and can be obtained in terms of a sum over cuts of the diagram $\Diagram$ 
by use of  standard Cutkosky rules 
(see \cite{Remiddi:1982hn} for a useful introduction to such techniques).
\begin{table}
\caption{\label{SomeDensities} Analytic expressions for some densities
at $d=3$, adopting the normalizations of \eref{CauchyTheorem}.
In the table we used the notations $M_{12}:=m_1+m_2$ and 
$M_{123}:=m_1+m_2+m_3$; $\step(x)$ is the usual step function. 
 Please remark that in the case of ``sunset'' diagram a subtracted version of 
\eref{CauchyTheorem} must be employed, while  
the density for the ``double-triangle'' diagram is evaluated at $m_i=1$.
}
\begin{indented}
 \item[]{
\begin{tabular}[ht]{@{}cc}
 \br
    $\Diagram$ & $\AmplitudeDDensity (M)$
 \\
 \br
     \begin{minipage}[c]{0.1\textwidth}
     \includegraphics[width=\textwidth]{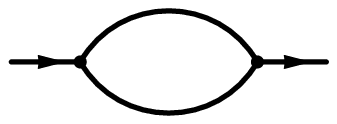}
     \end{minipage}
     & $\frac{\pi^2}{M} \;\step(M-M_{12})$
 \\
 \mr
     \begin{minipage}[c]{0.1\textwidth}
     \includegraphics[width=\textwidth]{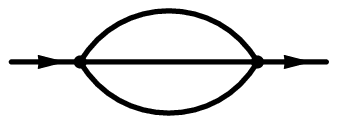}
     \end{minipage}
      &
     $2\pi^4 \left(1-\frac{M_{123}^{}}{M}\right) \;\step(M-M_{123})$
 \\
 \mr
 \begin{minipage}[c]{0.1\textwidth}
   \includegraphics[width=\textwidth]{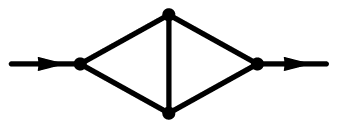}
 \end{minipage}
 &
 \begin{minipage}[c]{0.55\textwidth}
   \begin{center}
   $
   \frac{2\pi^4}{M^2\sqrt{M^2-3}}
   \;\log{\left(
       \frac{\left[6-M^2+M\sqrt{M^2-3}^{^{}}\right]^2}{9(M^2-4)}
     \right)}
   \;\step(M-2)
   $\\
   $
   +\frac{4\pi^4}{M^2\sqrt{M^2-3}}
   \;\log{\left(
       \frac{\left[M-3+\sqrt{M^2-3}^{^{}}\right]^2}{6(M-2)}
     \right)}
   \;\step(M-3)
   $
   \end{center}
 \end{minipage}
 \\
 \br
\end{tabular}
 }
\end{indented}
\end{table}
In \tref{SomeDensities} we list some examples of $3$-dimensional densities
whose expressions can be obtained analytically by integration of 
cut-diagrams.
Notice that due to its UV divergence the sunset diagram 
must be renormalized, and satisfies in our I-scheme 
a subtracted version of \eref{CauchyTheorem} which can be obtained 
from it by use of the replacement $1/(p^2 + M^2)\rightarrow 1/(p^2 + M^2)-1/M^2$.


A second possible direction to obtain effective vertices 
---which exploits the knowledge of densities
and the absence of renormalization of the involved amplitudes--- 
is the technique of \emph{line-dressing},
that we describe schematically below.

Let us suppose that the amplitude $\Amplitude_{\Diagram'}$ corresponding to a 
connected OPI diagram $\Diagram'$ with $\NLoops'$ loops is analytically known, 
with the condition that 
the mass ${m_j'}$ 
associated to the $j$-th propagator is different 
from all other masses in residual propagators,
\beq\label{dressingI}
\Amplitude_{\Diagram'}:=
\int \prod_{l'=1}^{\NLoops'}d^3\!\bell_{l'}'
\;\left[\mathrm{block}\right]
\;\left(\frac{1}{{k_j'}^2+{m_j'}^2}\right) 
;\eeq
let us suppose in addition that the amplitude $\AmplitudeD^{(2,0)}(p^2)$ 
associated to some two-point connected OPI diagram $\Diagram$ admits a spectral representation 
as the one given in \eref{CauchyTheorem}.
Then the amplitude of the diagram $\Diagram''$ 
obtained by replacing (or, familiarly, ``dressing'') the $j$-th line 
of diagram $\Diagram'$ with diagram $\Diagram$, 
\beq 
\Amplitude_{\Diagram''}:=
\int \prod_{l'=1}^{\NLoops'} d^3\!\bell_{l'}'
\;\left[\mathrm{block}\right]
\;\AmplitudeD^{(2,0)}({k_j'}^2)
\label{dressingIHalf}
,\eeq
can be written as
\beq
\Amplitude_{\Diagram''}=
\int_0^{\infty}d\!M^2
\AmplitudeDDensity\!\left(M\right)
\;\left(\Amplitude_{\Diagram'}\vert_{m'_j = M}^{^{^{}}}\right)
\label{dressingII}
.\eeq
The term $\left[\mathrm{block}\right]$ stands for the same 
${m_j'}^2$-independent  expression in both \eref{dressingI} and 
\eref{dressingIHalf}.

Please bear in mind that we used the ---implicit but essential--- 
hypothesis
that  $\Diagram$, $\Diagram'$ and $\Diagram''$ need not to be renormalized.
Examples of $\Amplitude_{\Diagram'}$ eligible to line-dressing 
are all the one-loop functions whose expressions
are known for non-equal masses (see \cite{MelroseReduction}, 
\cite{NickelDiagrams}),
in all kinematical configurations (see \cite{oneloop}). 

However, it must be noticed that the practical implementation 
of \eref{dressingII} can be very complicated; thus, 
we decided for the moment to make use of 
the technique of line-dressing only in the case of the first two-loop diagram 
on the left of \fref{OtherTwoLoopFunctions},
which can be seen as a triangle
with one line dressed by a bubble: for this effective vertex, the 
resulting cost turns out to be  $\TotalCost=1$.

\subsubsection{Low-cost subdiagrams}\label{LowCostSubdiagrams} 
In some cases, we can take advantage of 
particular ``hidden'' symmetries of the diagram we are dealing with 
to obtain low-cost integral representations for some higher-loop 
functions, with full dependence on external momenta. 

\begin{figure}[ht]
\begin{indented}
\item[]\begin{tabular}{@{}cc}
    \begin{minipage}[c]{0.25\textwidth}
      \includegraphics[width=\textwidth]{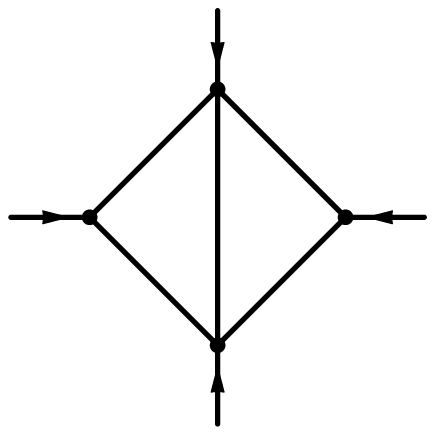}
    \end{minipage}
    &
    \begin{minipage}[c]{0.25\textwidth}
      \includegraphics[width=\textwidth]{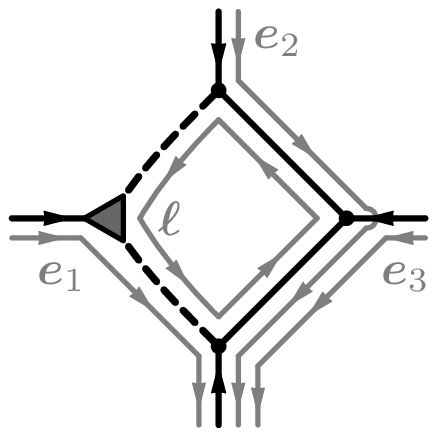}
    \end{minipage}
\end{tabular}
\end{indented}
\caption{\label{SwD} ``Square-with-diagonal'' diagram, and ---on the right---
the momentum assignement and the choice of a triangle effective vertex  
leading to its low-cost parametrization. 
Please remark that vectors ${\boldsymbol{e}}_i$ 
denote the incoming external momenta.}
\end{figure} 
An example of such a situation is given by the \emph{``square-with-diagonal''} diagram 
of \fref{SwD}, 
which ---by means of a clever assignement of loop momenta 
in the spirit of \sref{Loop Momenta}--- 
can be parametrized in terms of a one-loop triangular effective vertex and
only $2$ residual 
integrations. In fact, using the momentum assignement
described in \fref{SwD} and performing the integral over $\bell$ 
in spherical coordinates $(|\bell|,\cos\theta,\phi)$ 
with the $\mathbf{z}$-axis chosen to be parallel to the external momentum
$\boldsymbol{e}_1$, it turns out that the triangular effective vertex is $\phi$-independent; consequently, the integration over $\phi$ of the two residual propagators factors and can be performed analytically. 

The underlying symmetry leading to angular independence
is still present in the more general situations of diagrams
with two or three external legs, respectively, bordered by
an additional chain of propagators connecting two extremal vertices
---that is, vertices connected to external lines---;
in this sense, in fact, the ``square-with-diagonal'' diagram
can be considered as a triangle bordered by a chain of two propagators.
Using in such two general cases a parametrization analogous 
to that in \fref{SwD}, and spherical coordinates 
oriented as explained above, one readily obtains 
that the three-leg (resp.\ two-leg) subblock
is $\phi$-independent (resp.\ $(\theta,\phi)$-independent) 
and that these angular integrations 
involve only the propagators belonging to the bordering chain: 
consequently, such integrations may possibly be done analytically. 
(As usual, the absence of renormalization 
is an essential hypothesis for this property to safely hold.) 
In the case of a two-leg subblock bordered by a chain 
the considerations above imply that 
if the angular integral over $(\theta,\phi)$ of the chain of propagators 
is analytically known one can spare two integrations. 
The same result can be obtained by the technique of 
\emph{line-dressing} one-loop skeletons
with the spectral density of the two-leg subblock in question
(see \sref{spectral}), if this density is available:
this latter technique seems more adequate for 
the implementation of two-loop diagrams corresponding to
a one-loop bubble bordered by a chain of propagators.

\begin{figure}[ht]
\begin{indented}
\item[]\begin{tabular}{@{}cccc}
    \begin{minipage}[c]{0.135\textwidth}
      \includegraphics[width=\textwidth]{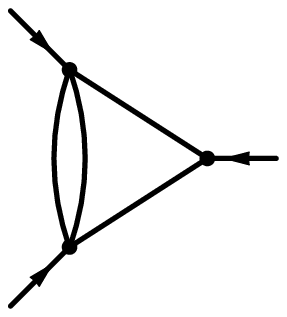}
    \end{minipage}
    &
    \begin{minipage}[c]{0.15\textwidth}
      \includegraphics[width=\textwidth]{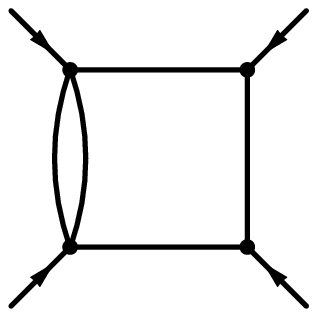} 
    \end{minipage}
    &
    \begin{minipage}[c]{0.15\textwidth}
      \includegraphics[width=\textwidth]{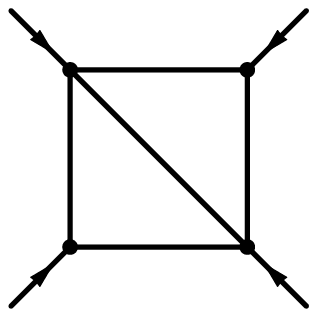} 
    \end{minipage}
    &
    \begin{minipage}[c]{0.225\textwidth}
      \includegraphics[width=\textwidth]{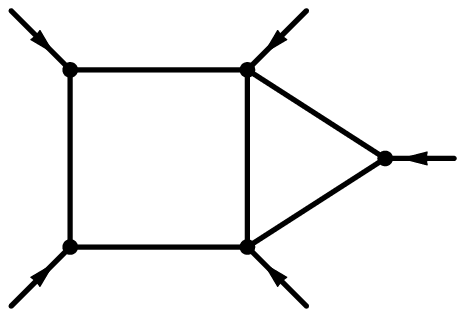} 
    \end{minipage}
\end{tabular}
\end{indented}
\caption{\label{OtherTwoLoopFunctions}Principal two-loop low-cost diagrams 
(no more than two integrations are required for their evaluation, see discussion
in \sref{LowCostSubdiagrams}).}
\end{figure}

The parametrization described above
turns out to be particularly useful when applied to  
the family of two-loop diagrams constructed by bordering with a chain two 
extrema of a triangle one-loop subdiagram; all such diagrams 
can be evaluated at cost 
$\TotalCost=2$ by performing the integration over $\phi$ 
of the corresponding chain of propagators, and using 
the known analytic expression for the triangle subdiagram.
The simplest graphs belonging to this family are shown in 
\fref{OtherTwoLoopFunctions} together with some useful 
bordered bubbles: employed as effective vertices they  
will prove important, for example, to reduce the complexity of some $7$-loop 
diagrams back to a more manageable size.
Following considerations exposed above, we will assign a cost $\TotalCost=1$ 
only to the first graph, which corresponds to a bubble bordered by a two-line
chain or, equivalently, to a line-dressing of a triangle 
with a bubble density ---see \sref{spectral}---; even if similar 
considerations could be applied to the second graph, which is as well a
bordered bubble, we will content ourselves to treat it here as an effective
vertex of cost $\TotalCost=2$, since the development of the code corresponding to the
$\TotalCost=1$ version would require a much larger programming effort.

\subsubsection{Full-cost subdiagrams}
Quite surprisingly, implementing effective vertices as standalone basic 
blocks can lead to important simplifications even if we are not able 
to compute the effective vertex in a cheap way and spare integrations;
the mere fact that more than one basic block can be replaced in one amplitude
is sometimes enough to give rise to reductions in its final cost.

\begin{figure}[ht]
\begin{indented}
\item[]\begin{tabular}{@{}cccc}
    \begin{minipage}[c]{0.18\textwidth} 
      \includegraphics[width=\textwidth]{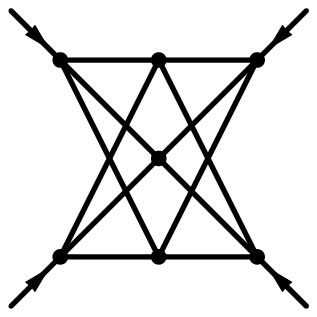}
    \end{minipage}
    &
    \begin{minipage}[c]{0.21\textwidth} 
      \includegraphics[width=\textwidth]{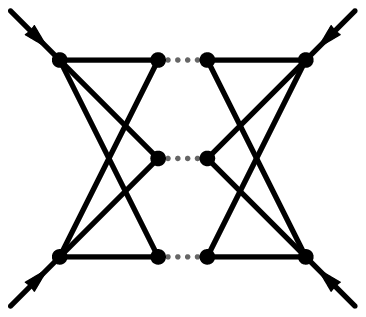}
    \end{minipage}
    &
    \begin{minipage}[c]{0.192\textwidth} 
      \includegraphics[width=\textwidth]{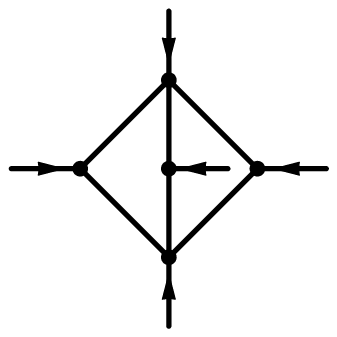}
    \end{minipage}
    &
    \begin{minipage}[c]{0.18\textwidth} 
      \includegraphics[width=\textwidth]{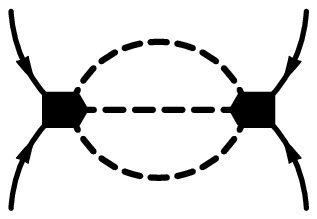}
    \end{minipage}
\\
(a)
&
(b)
&
(c)
&
(d)
\end{tabular}
\end{indented}
\caption{\label{K43} A difficult graph together with its treatment.}
\end{figure} 

An enlightening example of such a situation is offered by the diagram
shown in \fref{K43}.a; 
due to its complexity it looks formidable, until one 
realizes that it can be split as in \fref{K43}.b 
and expressed in terms of two identical blocks, each having the form 
illustrated in \fref{K43}.c; thus, the block of \fref{K43}.c 
can be considered as
a new kind of effective vertex with $5$ asymmetric legs, which can be 
identified and replaced into larger diagrams in the usual way  
(as illustrated in \fref{K43}.d for the case of 
diagram of \fref{K43}.a itself). 

From a more extensive analysis of \fref{K43}.d 
---using the definition of cost \eref{Cost} 
to deal with the nonzero-cost effective vertices,
and computing loop costs as described in \sref{autopar}--- we 
obtain for this diagram a final cost of $6$ ($3$ for the computation 
of loop integrals, plus $3$ for the computation of the effective vertices, 
which is in fact the full cost for computing such two-loop diagrams); 
on the other hand, it should be noticed that due to topological obstructions
only two one-loop ordinary ``low-cost'' functions 
could be replaced into the diagram of \fref{K43}.a
if we did not introduce the ``full-cost'' effective vertex of \fref{K43}.c, 
leading to a much larger final cost of $9$.

\subsection{Momentum representation: identities for vanishing external momenta}\label{ZeroExternalMomenta}
In this section we collect some simplifying 
diagrammatic identities, which take advantage 
of the fact that our amplitudes are always computed at zero external momenta
(and that renormalization is trivial in our scheme).

\subsubsection{Factorizable amplitudes}\label{FactAmpli}

Quite often  connected OPI diagrams in $\left( \phi^2 \right)^2$ theory 
show {\emph{cut-vertices}}, i.e. ---in this context--- vertices such that the separation of 
their four legs into two appropriate groups of two legs
disconnects the diagram in two parts.
\begin{figure}[ht]
\begin{indented}
\item[]\begin{tabular}{@{}ccc}
    \begin{minipage}{0.3\textwidth}
      \vspace*{1mm}
      \includegraphics[width=\textwidth]{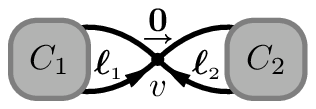}
    \end{minipage}
    &
    \begin{minipage}{2.7mm}\vspace*{3.25mm}$=$\end{minipage}
    &
    \begin{minipage}[c]{0.32\textwidth}
      \vspace*{2.2mm}
      \includegraphics[width=\textwidth]{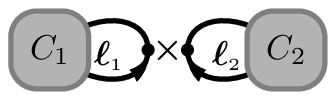} 
    \end{minipage}
\end{tabular}
\end{indented}
\caption{\label{FactorizationProperty}Factorization property of amplitudes of
  diagrams with cut-vertices when external momenta vanish. The vector $\mathbf{0}$
  refers to the vanishing of the momentum flow between the two blocks $C_1,C_2$.}
\end{figure} 
In \fref{FactorizationProperty} we depicted a cut-vertex 
$v$, which divides its graph in two subgraphs $C_1$ and $C_2$.
At vanishing external momenta, momentum conservation can be obeyed 
if and only if there is no total momentum flow from $C_1$ to $C_2$; 
using the locality property of momentum representation
and an appropriate choice of loop momenta
we can always separate the integrations relative to block $C_1$ 
from those relative to block $C_2$,
expressing the original integral as the product of two integrals, 
each one associated to a lower order connected OPI diagram evaluated at zero external momenta
and carrying  a $\int \phi^2$ insertion in place of the original 
$\left( \phi^2 \right)^2$ vertex $v$.
In the following we will refer to diagrams with one or more cut-vertices as 
to {\em factorizable} diagrams.

It turns out that a relevant fraction of the diagrams 
needed to compute RG-functions 
---about the $20\%$ up to the $7$-loop level--- 
is factorizable in terms of lower-order 
diagrams; these diagrams are in turn either easier to evaluate 
than the original one, or already known. This idea is fully implemented 
in our framework (the corresponding logical block has been labeled as 
{\em``filter for chain diagrams''}
in \fref{TheRoadmap}). 

\subsubsection{``Quail's leap''}

Other interesting relations can be obtained at zero external
momenta when additional symmetries are present in the expression 
of the amplitude for a given
diagram. In fact, it is sometimes possible 
to look at the diagram in
terms of a composition of blocks of propagators, with a momentum flow such
that blocks can be rearranged; one can in this way obtain other diagrams,
which differ in structure from the starting one but whose associated integrals have the same numerical 
value at zero external momenta.

\begin{figure}[t]
\begin{indented}
\item[]{\normalsize\begin{tabular}{@{}ccccc}
   $\AmplitudeD=$
   &
   \begin{minipage}[c]{0.2\textwidth}
   \includegraphics[width=\textwidth]{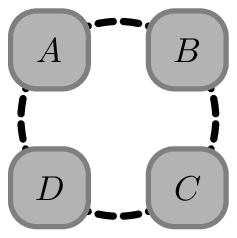}
   \end{minipage}
   &
   $=$
   &
   \begin{minipage}[c]{0.2\textwidth}
   \includegraphics[width=\textwidth]{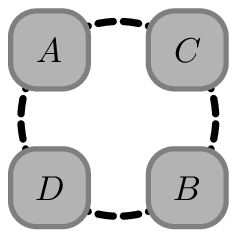}
   \end{minipage}
   &
   $=\Amplitude_{\Diagram'}$
\end{tabular}}
\end{indented}
\caption{\label{Quail}An example of {\em``quail's leap''} identity. 
The dashed lines express here momentum flow between blocks, 
and do not correspond to real propagators.}
\end{figure} 

In \fref{Quail} we give a pictorial illustration of an example of such 
set of identities; from now on, we will familiarly call it 
the {\em``quail's leap''}.
The analytic counterpart of \fref{Quail} follows straightforwardly 
if one writes down the explicit form of amplitudes $\AmplitudeD$ 
and $\Amplitude_{\Diagram'}$ as schematically defined by \fref{Quail}, 
and then relates them using the commutativity of the ordinary product 
in momentum space; as stated, this leads to the identity
\beq
\AmplitudeD=\int d\bell \; A(\bell)\,B(\bell)\,C(\bell)\,D(\bell)
=
\int d\bell \; A(\bell)\,C(\bell)\,B(\bell)\,D(\bell)=\Amplitude_{\Diagram'}
\;.\label{anaQuaglia}\eeq
(It should be noticed here the essential role played 
by the vanishing of external momenta; this condition automatically
enforces momentum conservation in the diagram ${\Diagram'}$ associated to
the amplitude with permuted blocks.)
In spite of the simplicity of its analytic derivation,
consequences of ``quail's leap" are non-trivial at all.  
Starting from the identifications
\begin{center}
\begin{tabular}{ll}
 $\begin{array}{c}\includegraphics[width=0.07\textwidth]{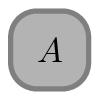}\end{array} 
  = 
  \begin{array}{c}\includegraphics[width=0.07\textwidth]{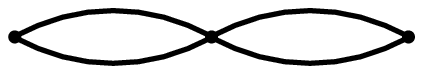}\end{array}
 $ 
& 
 $\begin{array}{c}\includegraphics[width=0.07\textwidth]{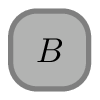}\end{array} 
  = 
  \begin{array}{c}\includegraphics[width=0.07\textwidth]{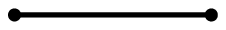}\end{array}$
\\
 $\begin{array}{c}\includegraphics[width=0.07\textwidth]{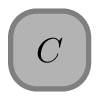}\end{array} 
  = 
  \begin{array}{c}\includegraphics[width=0.07\textwidth]{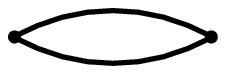}\end{array}
 $ 
& 
 $
 \begin{array}{c}\includegraphics[width=0.07\textwidth]{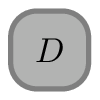}\end{array} 
 = 
 \begin{array}{c}\includegraphics[width=0.07\textwidth]{quailSPropagator.eps}\end{array}
 $
\end{tabular}
\end{center}
we can easily show an example of use of the relation in \fref{Quail}: 
substituting the indicated blocks into it
we readily get to the following non-trivial diagrammatic identity
\begin{eqnarray*}
\begin{array}{c}\includegraphics[width=0.12\textwidth]{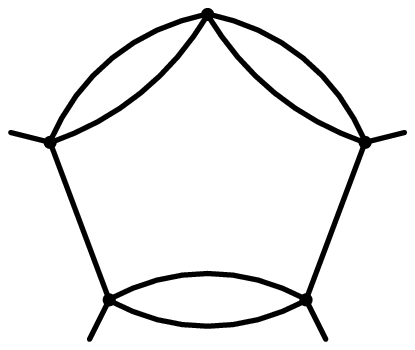}\end{array} 
  = 
 \begin{array}{c}\includegraphics[width=0.12\textwidth]{quail-one.eps}\end{array} 
  = 
  \begin{array}{c}\includegraphics[width=0.12\textwidth]{quail-two.eps}\end{array} 
  =
 \begin{array}{c}{\includegraphics[width=0.12\textwidth]{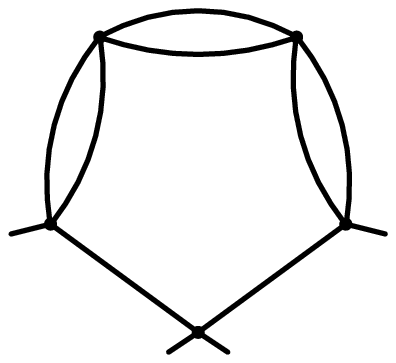}}\end{array} . 
\end{eqnarray*}
Strictly speaking, it is not guaranteed that the diagrams obtained applying
identities such as the one illustrated in \fref{Quail} will belong to the
same correlation function as the diagram one starts with; in practice,
it turns out that the number of such useful cases 
does not represent a large fraction of the diagrams required, 
and this trick is not implemented in the current form of our framework.

\subsection{Other aspects of momentum representation}
In this section we collect a couple of additional techniques, 
which, 
although not leading to a direct reduction in the dimensionality of the integrals 
to be performed, are nonetheless essential to complete our task of 
parametrizing amplitudes. In both cases, anyway, 
we take advantage of momentum representation of amplitudes 
to put the required manipulations in a neat diagrammatic form.
As usual, triviality of renormalization in the I-scheme 
is tacitly exploited.

\subsubsection{The derivative with respect to $p^2$} \label{DerivativeSection}

In this section we show how to compute 
$\left.{\de\Gamma^{(2,0)}_\I}/{\de p^2}\right\vert_{p=0}$
from the diagrammatic expansion of ${\Gamma^{(2,0)}_\I}$
--- for a better insight, we keep a generic dimension $d$ in the algebraic 
expressions presented below.

The derivative w.r.t.\ $p^2$
of a function $f(p^2)$ can be expressed in terms of derivatives w.r.t.\ 
momentum components $p_{\mu}$ by using the chain rule. 
In particular, the following relation
\[ \frac{\de}{\de p_{\mu}} \frac{\de}{\de p_{\mu}} f
   \left( p^2 \right) = 2 d \frac{\de}{\de p^2} f \left( p^2 \right)
   + 4 p^2 \frac{\de^2}{\de p^2 \de p^2} f \left( p^2 \right) 
\]
turns out to be useful. 
Assuming in addition $f(p^2)$ to be smooth at $p^2=0$, it follows:
\beq\label{dpsqZero}
\left( \frac{\de}{\de p^2}f(p^2)\right)_{p=0} = 
\frac{1}{2 d} \left( \frac{\de}{\de p_{\mu}} \frac{\de}{\de p_{\mu}} f(p^2)\right)_{p=0} .
\eeq 

The derivative w.r.t.\ $p^2$ of ${\Gamma^{(2,0)}_\I}$ at $p=0$ 
is obtained by distributing the linear differential operator ${\de}/{\de p^2}$
over each Feynman amplitude contributing to ${\Gamma^{(2,0)}_\I}$ at nonzero $p$, 
and then applying the identity \eref{dpsqZero} to each amplitude. 
(It should be noticed that each amplitude is a function 
of $p^2$ only, and is regular at $p=0$
 because all masses in propagators are nonvanishing.)
Commuting the derivatives with loop integrals leads us to the problem
of evaluating the action of differential operator   
$\left. {\de}/{\de p_{\mu}}\cdot{\de}/{\de p_{\mu}} \right\vert_{p=0}$
on a product of $p$-dependent propagators 
(obviously, $p$-independent propagators do not take part in the process, 
and stay unchanged).
In the following we will stick to the practically convenient case
when the external momentum $p$ flows without fractioning
from one external point of the diagram to the other 
through a chain of propagators
(the extension of given formulas to the general case is straightforward).

We label with the index $a$ the lines belonging to the chain, and with 
the index $b$ the lines not belonging to it; calling $k_a$ and $k_b$ 
the internal momenta associated to each line, and using for shortness 
the notations $\Propagator(k):=1/(1+k^2)$, 
$\Propagator_\mu(k):=k_\mu/(1+k^2)^2$, 
we easily obtain:   
\bea
\fl \left[ \frac{1}{2 d} \frac{\de}{\de p_{\mu}} \frac{\de}{\de p_{\mu}}
\left(  \prod_{a} \Propagator(p+k_a) \prod_b \Propagator(k_b)  \right) \right]_{p=0}
\nonumber
\\\lo=
\left( \prod_b \Propagator(k_b)  \right)
\sum_{a}\left[ 
\left( \frac{4-d}{d} \; \Propagator(k_a)^2 - \frac{4}{d}\; \Propagator(k_a)^3\right)\; \left( \prod_{a'\neq a} \Propagator(k_{a'})  \right)
\right]
\nonumber
\\
+
\left( \prod_b \Propagator(k_b)  \right)
\sum_{a'>a}\left[ 
\left( \frac{4}{d} \; \Propagator_\mu(k_a)\Propagator_\mu(k_{a'}) \right)\; \left( \prod_{a''\neq a,a'} \Propagator(k_{a''})  \right)
\right]
\label{chainD}.\eea
Remarkably, due to the freedom we have in choosing how to parametrize 
amplitudes the final integrated result will not depend on the choice 
of the chain; however, it is evident from \eref{chainD} 
that to compute the derivative of an amplitude using a chain of length $l$
one must deal with  ${l \left( l - 1\right)}/{2} + 2 l$ generalized diagrams 
--- involving, to worsen things, additional two-leg vertices 
or vector propagators.
As a consequence, to get an efficient automated framework for 
parametrizing amplitudes one must add a stage where all diagrams 
of ${\Gamma^{(2,0)}_\I}$ having a nontrivial dependence on external momentum 
are scanned; since each chain allows the computation of the derivative 
w.r.t.\ $p^2$ of the original diagram via the values of the set of graphs 
induced by \eref{chainD}, the goal will be to find the chain which is 
``minimal'' in terms of the set of integration costs of such induced 
set of graphs.

This processing stage has been indeed implemented in our code, 
and is in fact responsible for the 
majority of the time spent to parametrize the $2$-point function: the large 
number of chains to be examined, together with the large number of diagrams 
to be parametrized for each chain, give in general rise to an unpleasant 
explosion of the number of diagrams to be examined.

We notice as a final remark that a special prescription can be formulated 
to compute the derivative of factorizable amplitudes which we examined in 
\sref{FactAmpli}: in fact, in this case we can arbitrarily decide 
to confine the chain of propagators intervening in \eref{chainD} 
to the first block of the diagram, 
that is, to the factored block to which the two external legs of the diagram 
are connected; in this case, we will obtain the derivative of the whole 
diagram by computing the derivative (hopefully simpler) of the first 
block, and then multiplying each of the terms obtained from the r.h.s. of 
\eref{chainD} by all the remaining original factored blocks--- 
which will be left 
untouched by the action of the differential operator ${\de}/{\de p^2}$, 
since due to our particular choice of the chain they do not contain 
any of the propagators involved in the derivation. 
This consideration explains 
the intricated structure of the upper part of \fref{TheRoadmap}.

\subsubsection{Insertions of $\int\phi^2_\I$}\label{InsertionSection}
\begin{figure}[ht]
\begin{indented}
\item[]{\large\begin{tabular}{@{}ccc}
    $\begin{array}{c}\includegraphics[width=0.204\textwidth]{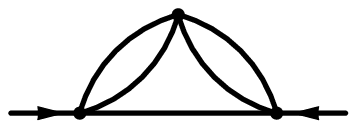}\end{array}$ 
    &
    $\longrightarrow$
    &
    $4 \times \begin{array}{c}\vspace*{-2mm}\includegraphics[width=0.174\textwidth]{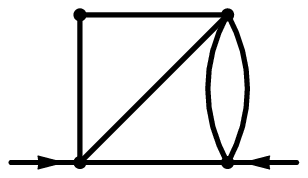}\end{array}+\begin{array}{c}\includegraphics[width=0.174\textwidth]{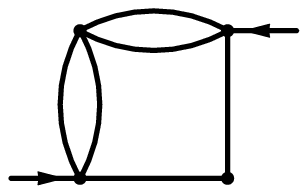}\end{array}$ 
\end{tabular}}
\end{indented}
\caption{\label{examplePhisqInsertion} A an example of diagram contributing to $\Gamma^{( 2,0 )}_\I$, 
and the corresponding groups of diagrams in $\Gamma^{( 2, 1 )}_\I$ generated by the insertion of one $\int \frac{\phi^2_\I}{2}$ operator.}
\end{figure}

The diagrammatic expansion for 
$\Gamma^{( 2, 1 )}_\I(\Many{\bo{p}}=\bo{0};\bo{q}=\bo{0})$ 
can be obtained from that of  $\Gamma^{( 2,0 )}_\I(\Many{\bo{p}}=\bo{0})$ 
by inserting $\int \frac{\phi^2_\I}{2}$ 
into each graph which belongs to the $2$-point function; 
from a diagrammatic point of view this operation 
corresponds to the insertion of a two-legs vertex 
(with no additional external momentum), and can be performed 
by replacing in all possible ways one propagator of the original diagram 
(that is, one line) with a chain of two propagators; when the same final
graph is generated in many different ways, its multiplicity has to be taken
into account (see \fref{examplePhisqInsertion} for an example).  

As observed in \cite{NickelPreprint}, 
it turns out from simple topological considerations that 
these diagrams correspond to a subset of the diagrams of 
$\Gamma^{(4,0)}_\I(\Many{\bo{p}}=\bo{0})$; this fact implies 
that we do not need to 
recompute the numerical values for the integrals associated to diagrams of 
$\Gamma^{( 2, 1 )}_\I(\Many{\bo{p}}=\bo{0};\bo{q}=\bo{0})$ at some loop order $L$, 
since such values can be deduced 
from a lookup in the list of values already computed during the evaluation 
of the $4$-point function at $L$ loops--- 
this nice property cannot be extended, however, to 
corresponding symmetry and $O(N)$ factors.

\section{Automatic parametrization of amplitudes}\label{autopar}
We are now ready to examine in full detail how an automatic framework 
for finding out minimal-cost parametrizations can be built.

\subsection{Overview}\label{autoparOverview}
Provided that the key concept of the method is the substitution of known
effective vertices into the analytic expression of the amplitudes, the problem
is how to fully exploit such an idea in a systematic, automated and optimal
way.

The starting point will be a choice of a set 
$\EffectiveVertices:=\{ \EffectiveVertexF_1,\,\ldots\, ,\,\EffectiveVertexF_{\NEffectiveVertices}\}$
of effective vertex functions. 
To each effective vertex $\EffectiveVertexF_e$ 
is associated a positive integer $\TotalCost_e$
called \emph{cost}.
When $\EffectiveVertexF_e$ is computed by use of an appropriate integral representation
the cost is defined to be the dimensionality of the integration, see \eref{Parametrization};
otherwise, if the vertex function is known analytically in terms of elementary functions, 
the cost $\TotalCost_e$ is zero. 

Formalizing now the ideas of \sref{EffectiveVerticesSection}, 
once $\NEffectiveVertices$ 
effective vertices have been replaced into the amplitude $\AmplitudeD$ 
corresponding to a given Feynman diagram $\Diagram$, with $\NLoopsResidual$ 
residual loops remaining, the new expression for that amplitude 
will be given by
\begin{eqnarray}
\AmplitudeD & \propto
\int \prod_{l=1}^{\NLoopsResidual} d^3\!\bell_{l} \;
\left(\prod_{j=1}^{\NLinesResidual}\frac{1}{1+k^2_j}\right)
\left(\prod_{e=1}^{\NEffectiveVertices}
  \EffectiveVertexF_e\!\left(\LoopScalarProducts_e\right)
\right)
\label{ScalarProducts}
\end{eqnarray}
where the effective vertices with non-zero cost are expressed as:
\beq
\EffectiveVertexF_e\!\left(\LoopScalarProducts_e\right)
:=
  \int d^{\TotalCost_e}\!\Many{\xi_e}\;\;
  {\EffectiveVertexK_e}\!\left(\LoopScalarProducts_e,\Many{\xi_e}\right)
\label{Parametrization}
.\eeq
$\NLinesResidual$ above is the number of residual propagators 
and $k_j$ the momentum flowing in each of them; 
the set $\LoopScalarProducts_e\subseteq\left\{\;\bell_{l_1}.\bell_{l_2}\;|\;l_1,l_2
\in 1\ldots\NLoopsResidual\mathrm{\ with\ }l_1\leq l_2^{^{}}\;\right\}$
encodes the dependence of the effective vertex $\EffectiveVertexF_e$
in terms of scalar products among loop momenta and plays a key role 
in the choice of the most appropriate angular measure, 
as will be shown in \sref{ChoiceOfAngularMeasure}.

The inclusion of effective vertices with non-zero cost in the set of possible
replacements imposes a refinement of the definition of the cost function 
itself.
The problem in question is how to extend 
the definition of total cost as dimensionality of the involved integration 
---which is appropriate when the amplitude is expressed in terms 
of one single multidimensional integral--- to situations
in which two or more effective vertices with non-zero cost ---and, thus, 
two or more blocks known only in terms of some integral representation---
are present in amplitude parametrization
and are themselves integrated as in equation \eref{ScalarProducts};
the solution to this problem is somewhat implementation-dependent. 
In the following we will refer to the concrete situation in which 
the numerical integration of the parametric integrals defining the blocks 
with non-zero cost via \eref{Parametrization} is performed
\emph{separately} for each block, and also separately 
from that of loop momenta.
In the presence of two or more blocks with non-zero cost it 
seems quite natural ---and convenient--- 
to prefer this type of implementation to that based on the opposite strategy 
of collecting and numerically evaluating all integrations together
in a single multidimensional integral.

Following these lines, we will assign to an amplitude parametrization like 
the one shown in \eref{ScalarProducts}-\eref{Parametrization} 
the refined total cost
\beq
\TotalCost:=\left(^{^{^{^{}}}}\mathrm{loop\ cost}^{^{^{^{}}}}\right)+\left(
\max_{1\leq e\leq\NEffectiveVertices}
\;\TotalCost_e\right)\label{Cost}
.\eeq
The ``loop cost'' appearing in this expression is given 
by the still unknown residual dimensionality of the integration of loop variables 
$\prod_{l=1}^{\NLoopsResidual} d^3\!\bell_{l}$. 
It will be the goal of \sref{ChoiceOfAngularMeasure} 
to compute, and possibly minimize, this number.

Some additional comments about formula \eref{Cost} are in order.
First of all, this refined definition of cost can be
easily justified 
by trusting equation \eref{IntegrationCost} and supposing that 
the number of operations of a sequence of integrations is dominated by the integration(s) of 
highest dimensionality.
Secondly, a finer structure exists: 
in essence, when a parametrization has a total cost $\TotalCost>0$ we can 
still say that to evaluate it we roughly need 
a number of operations proportional to those required by 
a $\TotalCost$-dimensional integral; however,
\begin{itemize}
\item if we adopt definition \eref{Cost} for the total cost, 
a constant proportionality factor in the number of operations is missed in all cases when 
a parametrization requires the evaluation of more than one single 
effective vertex having a cost exactly equal to 
$\max_{1\leq e\leq\NEffectiveVertices}\;\TotalCost_e$
\item even in the simple case when only one effective vertex needs to be 
integrated, and we are then free to choose whether or not to carry out 
the integration of the effective vertex along with that of loop variables, 
from the point of view of numerical integration the task of computing 
$\int_{\mathcal{L}}\left(\ldots\right)\cdot\int_{\Xi} {\EffectiveVertexK_e}$
 is not entirely 
equivalent to that of computing 
$\int_{\mathcal{L}\times \Xi}\left(\ldots\right) {\EffectiveVertexK_e}$ 
($\Xi$ being the integration domain of the effective vertex with non-zero cost,
and $\mathcal{L}$ the integration domain of loop variables); 
this happens  
first of all because more efficient non-separable integration rules can exist 
allowing to integrate the domain $\mathcal{L}\times \Xi$ as a whole, 
and then because a higher intermediate precision is in general needed 
for the evaluations of $\int_\mathcal{L}\left(\ldots\right)$ and 
$\int_{\Xi} {\EffectiveVertexK_e}$ if we want 
to obtain a given numerical precision for 
$\int_{\mathcal{L}\times \Xi}\left(\ldots\right){\EffectiveVertexK_e}$.
\end{itemize}
Nonetheless, since 
the task of specifying a precise formula for the computation of the cost 
is indeed a very difficult one 
---in the light of considerations carried out in \sref{WhichParametrization}--- 
we will stick to simpler formula 
\eref{Cost} throughout all this article.

Consequently, having picked up a set of known effective vertices $\EffectiveVertices$
with associated costs $\{\TotalCost_1,\ldots\, ,\TotalCost_\NEffectiveVertices\}$,
our strategy to reduce the total cost $\TotalCost$ 
for a given diagram $\Diagram$ will be as follows.

We start by considering the associated amplitude $\AmplitudeD$ parametrized in momentum space,
       \[
    \AmplitudeD \propto \int \prod_{l=1}^{\NLoops} d^3\!\bell_{l} \;\prod_{i=1}^{\NLines}\frac{1}{1+k^2_i}
      .\] 
Then, we try to reduce the cost \eref{Cost} by applying the following 
guidelines:
\begin{enumerate}
\item 
we look among all possible \emph{maximal substitutions} in the graph $\Diagram$
of subdiagrams from the given set of effective vertices $\EffectiveVertices$.
Each replacement is encoded via an \emph{effective graph} 
$\EffectiveDiagram$, which contains as many new vertices as possible 
describing the replaced effective vertices, and 
gives a new expression for the amplitude in terms of fewer integrals.
An algorithm to scan the space of all possible substitutions is presented in 
\sref{SubstitutingVertices}
\item we reduce the number of scalar products $\bell_{l_1}.\bell_{l_2}$ 
our integrand function depends on by means
  of a careful {\emph{choice of loop basis}} (see \sref{Loop Momenta})
\item we maximize the number of trivial angular integrations by {\em selecting the cheapest angular integration measure} 
which is allowed by our previous choices of substitutions and of loop basis (as explained in \sref{ChoiceOfAngularMeasure})
\item while picking up the best angular measure, we maximize at the same time 
  the possibilities of {\emph{factoring away simpler integrals}}, like for example 
\[
\int_{-1}^{1} d\!\left(\frac{\bell_{l_1}.\bell_{l_2}}{\ell_{l_1}\ell_{l_2}}\right)\;\left(\frac{1}{1+(\bell_{l_1}+\bell_{l_2})^2}\right)^n,\;\;\;\;\;n\in\mathbb{N}^{+}
\]
(see as before \sref{ChoiceOfAngularMeasure} for more details).
\end{enumerate}
It should be noticed that different (and perhaps more general) strategies 
can in principle be devised; of course, the list of tricks to 
minimize the number of residual integrations we just presented 
is only a faint approximation
of what a well-trained ``human neural network'' can do by examining all diagrams
one after the other, and finding out what is likely to be the {\em real}
optimal parametrization. However, since teaching mathematical intuition 
to a computer is clearly impossible, what we are interested in here is the 
much simpler task of creating a fast and robust algorithm relying on
a reasonably small number of simplifications, and working reasonably well 
for the majority of diagrams. 
A more in-depth discussion about this philosophy and the
optimality of the solutions it gives can be found in 
\sref{ImplementationAndResults}.
\subsection{Substitution of effective vertices}\label{SubstitutingVertices}
To fulfill our purpose of obtaining parametrizations of a given diagram
which have a minimal cost in the sense of equation (\ref{Cost}), 
we must as a first step find out all possible maximal substitutions 
of effective vertices for that diagram.

\begin{algorithm}\textbf{(Tree of Substitutions)}\label{TreeOfSubstitutions}
To find out all possible maximal substitutions
for a given Feynman diagram $\Diagram$,
\begin{enumerate}
\item {\textbf{read}} the set $\EffectiveVertices$ of known effective vertices, graphically expressed as pairs
\[
\fl p_e \equiv \left\{\; \mathrm{subgraph\ } \SubDiagram_e ,_{_{}} 
  \mathrm{effective\ graph\ }\EffectiveVertexG_e \;\mathrm{corresponding\ to\ replacement}\;\right\}
\]
\item {\textbf{define}}
  \begin{enumerate}
  \item the tree of substitutions: 
    to each node is associated a diagram which is obtained by replacing a
    collection of subdiagrams of $\Diagram$ ---non necessarily maximal---
    with corresponding elements of $\left\{\EffectiveVertexG_e\right\}$;
    the node also contains the list $\Replacements$ of all remaining 
    possible replacements,
    each one accompanied by a status flag describing 
    whether the related alternative has already been visited or not
  \item the set $\MaximalSubstitutions$ of maximal substitutions found so far
  \item the lookup node $\LookupNode$
  \end{enumerate}
\item {\textbf{initialize}}:
  \begin{enumerate}
  \item create the root node $\RootNode$ of the tree of substitutions; associate to it diagram $\Diagram$, and a list $\Replacements\!\left(\RootNode\right)$ appropriately computed
  \item $\MaximalSubstitutions\longleftarrow\emptyset$, $\LookupNode\longleftarrow\RootNode$
  \end{enumerate}
\item \label{Scanning Substitutions} 
  {{{\textbf{if}} the next unvisited replacement $R$ in 
the list of replacements $\Replacements\!\left(\LookupNode\right)$ exists,
      \begin{enumerate}
      \item update $\Replacements\!\left(\LookupNode\right)$
      \item apply $R$ to the diagram associated to lookup node 
        $\LookupNode$, thus obtaining a new diagram $\Diagram'$
      \item attach a new node $\Node'$ to $\LookupNode$; associate to it 
        diagram $\Diagram'$, and a list 
        $\Replacements\!\left(\Node'\right)$ appropriately computed
      \item update the lookup node: $\LookupNode\longleftarrow\Node'$
      \item {\textbf{go to}} step (\ref{Scanning Substitutions})
      \end{enumerate}}}

{\textbf{else}} 
  \begin{enumerate}
  \item put the diagram associated to lookup node $\LookupNode$ in the set 
    $\MaximalSubstitutions$ of maximal substitutions
  \item {\textbf{do backtrack}} up to the nearest node whose list of 
    unvisited replacements is not empty;

    {\textbf{if}} such a node exists,
      \begin{enumerate}
      \item set that node as the current lookup node
      \item {\textbf{go to}} step (\ref{Scanning Substitutions})
      \end{enumerate}

    {\textbf{else}} \textbf{end}.
  \end{enumerate}
\end{enumerate}
\end{algorithm}

\begin{figure}[ht]
\begin{indented}
\item[]\begin{minipage}{0.8\textwidth}   
\begin{eqnarray*}
\fl
 \Diagram 
 := 
 \begin{array}{l}\includegraphics[width=0.15\textwidth]{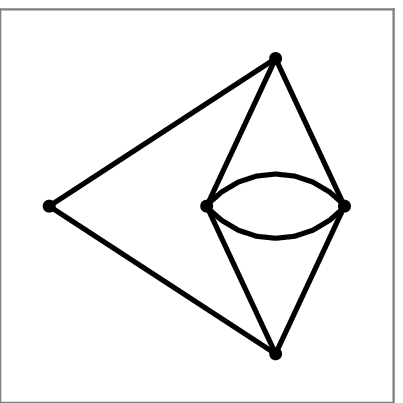}\end{array}
 \qquad
 \EffectiveVertices 
 :=
 \begin{array}{|l|l|l|l|l|}
 \hline
 \SubDiagram_e
 & \begin{minipage}[c]{0.12\textwidth}\vspace*{6.5mm}\includegraphics[width=\textwidth]{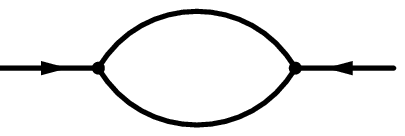}\vspace*{1.5mm}\end{minipage}
 & \begin{minipage}[c]{0.12\textwidth} \includegraphics[width=\textwidth]{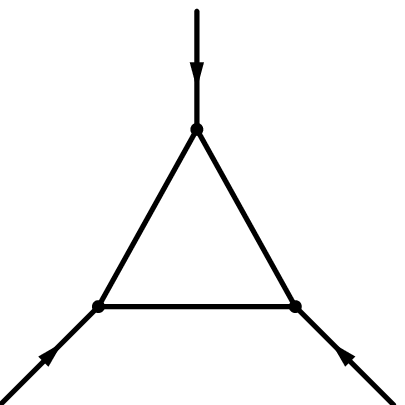}\end{minipage} 
 & \begin{minipage}[c]{0.12\textwidth} \includegraphics[width=\textwidth]{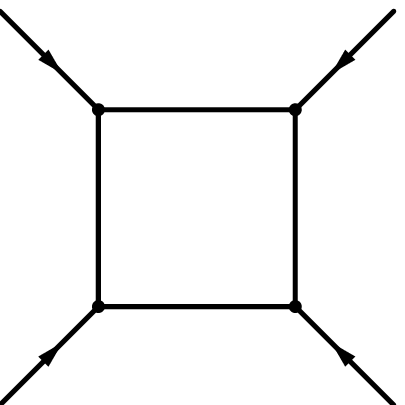}\end{minipage}
 & \begin{minipage}[c]{0.12\textwidth} \includegraphics[width=\textwidth]{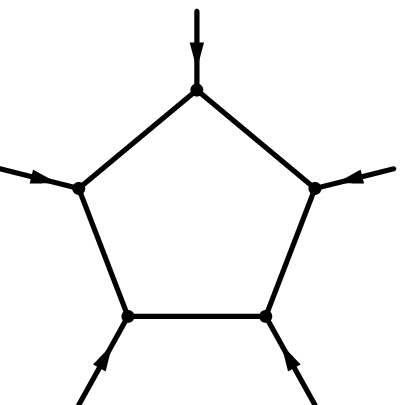}\end{minipage}
\\
 \hline
 \EffectiveVertexG_e 
 & \begin{minipage}[c]{0.12\textwidth}\vspace*{1.5mm}\includegraphics[width=\textwidth]{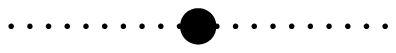}\vspace*{1.5mm}\end{minipage} 
 & \begin{minipage}[c]{0.12\textwidth} \includegraphics[width=\textwidth]{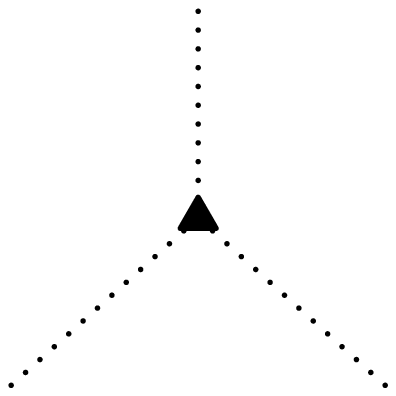}\end{minipage} 
 & \begin{minipage}[c]{0.12\textwidth} \includegraphics[width=\textwidth]{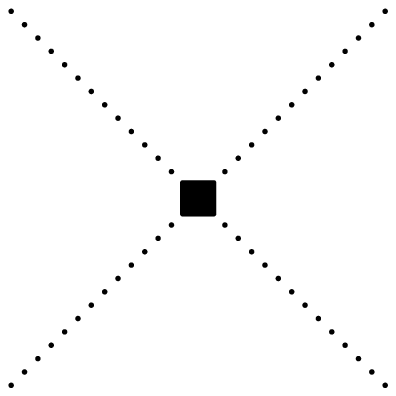} \end{minipage} 
 & \begin{minipage}[c]{0.12\textwidth} \includegraphics[width=\textwidth]{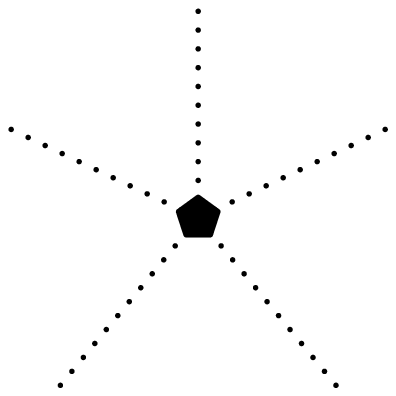}\end{minipage} 
 \\
 \hline
 \end{array} 
\end{eqnarray*}
\end{minipage}
\end{indented}
\caption{\label{ParametrizationExample} A test case for algorithm 
  (\ref{TreeOfSubstitutions}): a diagram $\Diagram$ and a set of effective 
  vertices $\EffectiveVertices$ to be replaced in it.
}
\end{figure}

As an example, we now illustrate how algorithm 1 
operates when it is applied to the diagram $\Diagram$ and the set 
of effective vertices $\EffectiveVertices$ which are presented 
in \fref{ParametrizationExample}.
\begin{figure}[ht]
\begin{center}   
\includegraphics[width=\textwidth]{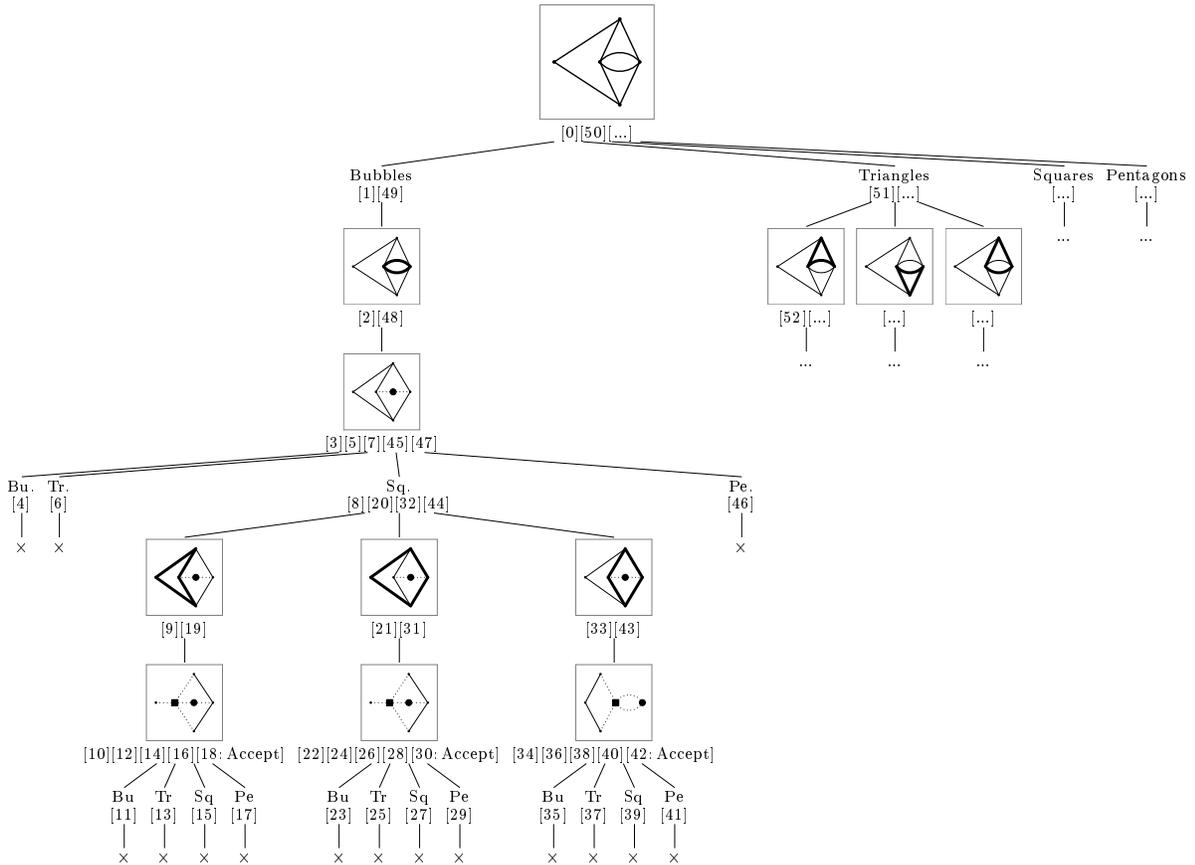}
\end{center}
 \caption{\label{Traversal}A partial tree of substitutions for
  the example of \fref{ParametrizationExample}. 
  The order of traversal of each node is explicitly specified between square parentheses. 
  A cross denotes the fact that the corresponding substitution is not possible.
  The label ``\emph{Accept}" means that the set of replacements associated to the node is maximal.}
\end{figure} 
\begin{figure}[ht]
\begin{center}   
\hfill\includegraphics[width=0.9\textwidth]{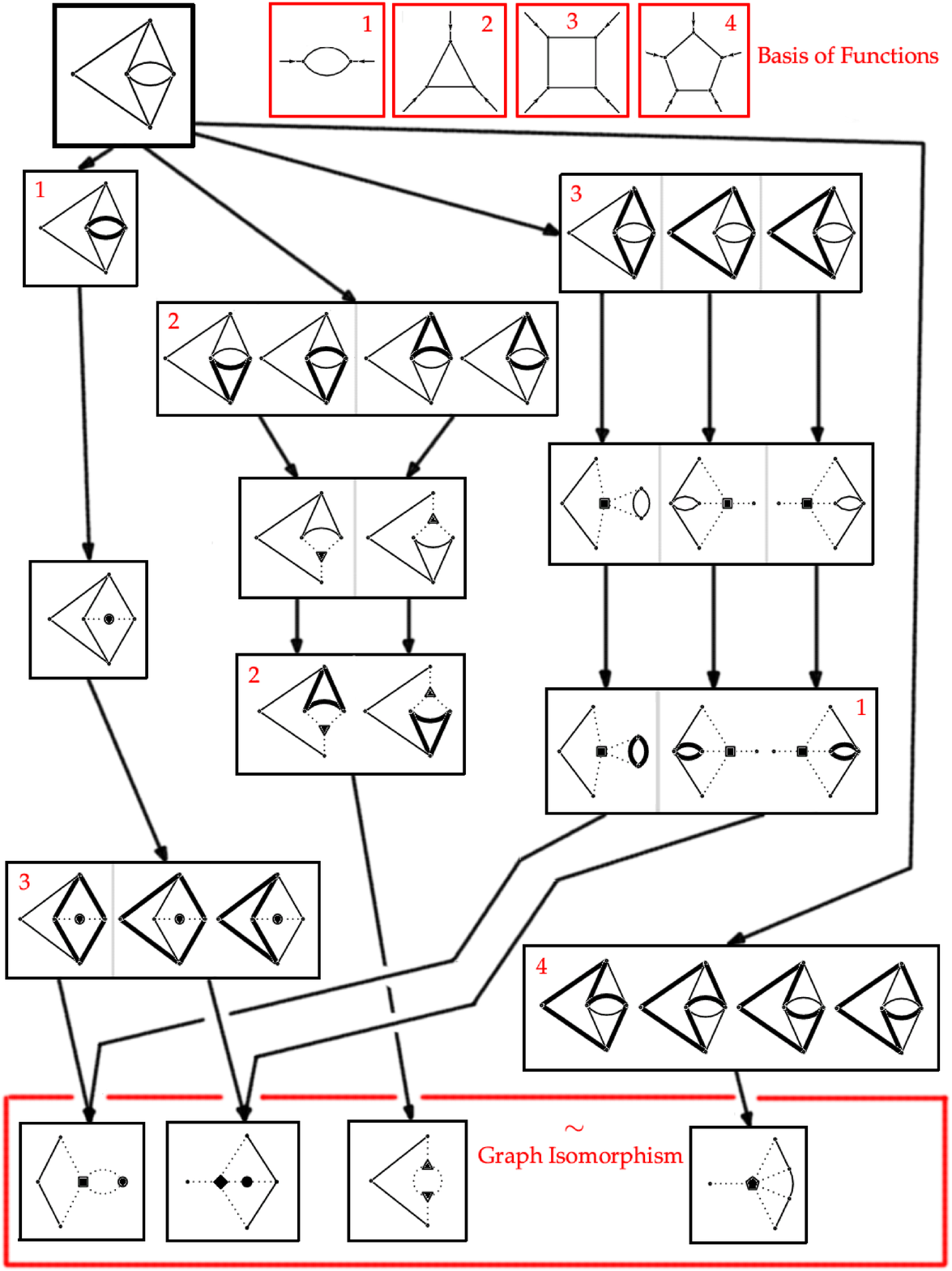}
\end{center}
 \caption{\label{SubstitutionsCompleteI}
 The complete tree of substitutions for the example of \fref{ParametrizationExample}. The numbers in frame corners indicate the element of the basis of effective vertices which is being replaced into the diagram in each case.}
\end{figure} 
In \fref{Traversal} we can observe how the traversal works: 
the algorithm always tries to push up the level of the tree, in the hope 
of finding out one effective vertex more in the same lookup diagram. 
A view of the entire tree
(more sketchy) can be found in \fref{SubstitutionsCompleteI}.
It turns out that for these choices of $\Diagram$ and $\EffectiveVertices$
there are exactly four possible different maximal substitutions, three with
a residual loop number $\NLoopsResidual = 2$ and one 
with $\NLoopsResidual = 3$.

Some general remarks are in order here:
\begin{enumerate}
\item we can easily convince ourselves that due to possible
topological obstructions the distance from the root of each terminal leaf of
the tree of substitutions is not the same. From a different perspective, 
we can say that 
{\em the final maximal 
substitutions will have in general a 
different number of residual loops, thus
leading to parametrizations with 
a different number of final integrations}. 
The importance of
such a conclusion in the perspective of the calculation of cost $\mathcal{C}$
should not be underestimated
\item by taking a closer look to \fref{SubstitutionsCompleteI}, it
turns out that the tree built up by this na\"\i ve formulation of the algorithm
must be pruned in order to be used in effective calculations. In fact, it is
clear that the same final parametrization can be in general reached following
a lot of different branches of the tree: being subdiagrams with a given form
but with a different embedding in the diagram
distinguishable from each other, the same final set of $n$ vertex functions
can be obtained from exactly $n!$ different tree paths.
Adding to this problem the observation that this algorithm too is
plagued by the problem of isomorphic copies, one can very well understand how
the size of such a na\"\i ve tree is soon pushed to astronomical values. 
Luckily, the use of {\emph{ordered searches}} for vertex
functions brings the complexity of the algorithm back to a manageable size.
\end{enumerate}

\subsection{Choice of loop momenta}\label{Loop Momenta}

After a choice of a maximal substitution of effective vertices 
in the original diagram $\Diagram$ has been performed, 
resulting in  an effective diagram $\EffectiveDiagram$, 
the second step in the generation of parametrizations is clearly the task of 
assigning loop momenta to $\EffectiveDiagram$.
In spite of the fact that the choice of loop momenta 
is often looked at as a trivial operation, a favourable assignement of loop momenta 
---if complemented by successive optimization of angular measure 
and maximization of simple angular integrals over residual propagators, 
see \sref{ChoiceOfAngularMeasure} and \sref{DisconnectibleCosineDiagrams}--- 
can in many cases provide further reductions in 
the number of final residual integrations.

The following facts ---which are common lore in graph theory--- 
will be very useful to get to algorithm \ref{LoopBases}, 
which allows the sequential generation of 
{\em all} possible loop assignments (further references 
and definitions adapted to more general contexts can be found 
for example in {\cite{Harary}}, {\cite{Skiena}} and {\cite{Rosen}}).
\begin{definition}
Given two subdiagrams $\SubDiagram_1$ and $\SubDiagram_2$ of a diagram
$\Diagram$, we define $\SubDiagram=\SubDiagram_1\oplus \SubDiagram_2$ 
as the subdiagram of $\Diagram$ whose lines are present 
in $\SubDiagram_1$ or in $\SubDiagram_2$, but not in both
(exclusive sum of lines).
\end{definition}
\begin{definition}
A $n$-{\em cycle} ---or $n$-{\em loop} in physicists' terminology--- 
is the diagram composed by the set of vertices
$\left\{v_1,\;\ldots\, ,\;v_n\right\}$
and the set of distinct lines
$\left\{v_1-v_2,\;\ldots\, ,\;v_n-v_1\right\}$.
A {\em tree} is a connected diagram with no cycles. 
A {\emph{spanning tree}} for a connected diagram is a 
tree subdiagram with the same vertices as the original diagram; by
topological count, the operation of adding a single line to a spanning
tree does always produce a subdiagram with loop number equal to $1$.
\end{definition}
\begin{theorem}
\label{VectorSpaceOfLoops}Given a connected diagram $\Diagram$, 
its loops span a vector space $\mathbb{V}\!_L$ in $\mathbb{Z}_2$ 
w.r.t.\ the operation $\oplus$.
The dimension of such a vector space is given by $\NLoops$,
the number of loops of the diagram familiar to physicists.
\end{theorem}

Not all the elements of $\mathbb{V}\!_L$ are loops, but all the loops of the 
diagram are contained in $\mathbb{V}\!_L$; thus, this theorem gives us a 
powerful tool to enumerate them all.

\begin{theorem}
\label{FirstLoopBasis}To get a {\em loop basis} $\LoopBasis$ 
for the vector space $\mathbb{V}\!_L$ associated to a connected diagram 
$\Diagram$,
\begin{enumerate}
\item produce a spanning tree $\Tree$ for the diagram
\item for each remaining line $\Line \in \Diagram$ such that
  $\Line \in \Diagram$ but $\Line \nin \Tree$,
  consider the subdiagram $\Tree_{\Line}$ obtained by adding
  the line $\Line$ to $\Tree$, that is
  $\Tree_{\Line} \equiv \Tree \cup \Line$;
  produce a cycle $\Cycle_i$ by identifying it as a subset of
  $\Tree_{\Line}$.
\end{enumerate}
The set $\LoopBasis \equiv \left\{ \Cycle_1, \ldots\, ,\Cycle_{\NLoops}
\right\}$ of cycles generated in this way is the desired basis.
\end{theorem}

As a consequence, we can now formulate the following
\begin{algorithm}\label{LoopBases}\textbf{(Loop bases).}
To produce all possible assignments of loop
momenta in a given connected diagram $\Diagram$
\begin{enumerate}
  \item build a spanning tree $\Tree$ for $\Diagram$
  
  \item deduce a first loop basis $\LoopBasis_1$ for the loops in the diagram
  $\Diagram$ by applying theorem \ref{FirstLoopBasis} to the couple
  $\left\{ \Diagram,\Tree \right\}$; 
   the property ${\NLoops} = \mathrm{card}
  \left( \LoopBasis_1 \right)$ will hold
  
  \item deduce the set $\Loops$ of all loops contained in the diagram by
  calculating all possible linear combinations over $\mathbb{Z}_2$ 
  of the elements of $\LoopBasis_1$ and by discarding combinations that are not loops
  
  \item {\textbf{for each}} extraction of ${\NLoops}$ loops $\Cycle_{i_1},
  \ldots\, ,\Cycle_{i_{\NLoops}} \in \Loops$
  \begin{description}
    {\textbf{if}} loops $\Cycle_{i_1}, \ldots\, ,\Cycle_{i_{\NLoops}}$ are
    linearly independent
    \begin{enumerate}
      \item take the set $\left\{ \Cycle_{i_1},
      \ldots\, ,\Cycle_{i_{\NLoops}} \right\}$ as a new possible choice of loop
      momenta
      \item {\textbf{for each}} $\Cycle_{i_n}$
      \begin{enumerate}
        \item assign an arbitrary orientation to it
        \item increment (according to the chosen loop orientation) the momenta associated to the lines of $\Cycle_{i_n}$ by a quantity $\alpha_n \bell_n$ --- where the $\alpha_n$'s are arbitrary non-null coefficients.
      \end{enumerate}
    \end{enumerate}
  \end{description}
\end{enumerate}
\end{algorithm}

\begin{figure}[htp]
  \begin{indented}
  \item[]
    \begin{minipage}{0.8\textwidth}
      \begin{eqnarray*}
        \fl
        \Tree 
        \left(\hspace*{-1.5mm}\begin{array}{c}\vspace*{-1.5mm}\includegraphics[width=0.2\textwidth]{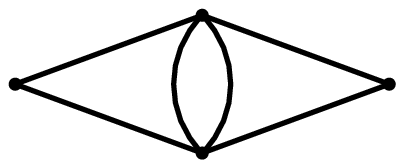}\end{array}\hspace*{-2mm}\right) 
        = \begin{array}{c}\vspace*{-1.5mm}\includegraphics[width=0.2\textwidth]{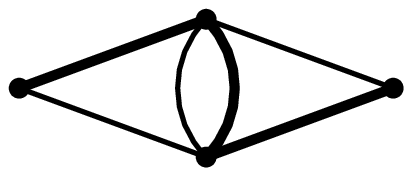}\end{array} 
        \\
        \fl
        \LoopBasis_1 = 
        \left\{\hspace*{-1mm}
          \begin{array}{ccccc}
            \vspace*{-3mm}\includegraphics[width=0.2\textwidth]{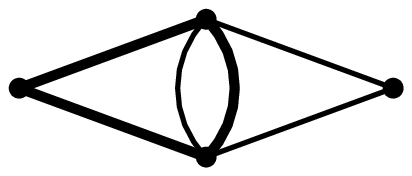}
            &\hspace*{-1mm}\begin{minipage}{1mm},\vspace*{7mm}\ \end{minipage}
            &\hspace*{-1mm}\includegraphics[width=0.2\textwidth]{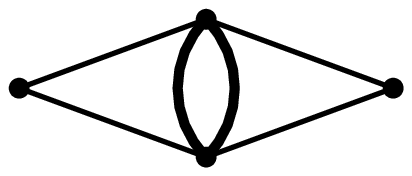}
            &\hspace*{-1mm}\begin{minipage}{1mm},\vspace*{7mm}\ \end{minipage}
            &\hspace*{-1mm}\includegraphics[width=0.2\textwidth]{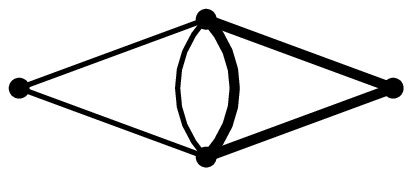}
          \end{array}
        \hspace*{-1mm}\right\} 
      \end{eqnarray*}
    \end{minipage}
    \vspace*{-2mm}
    \caption{\label{catsEyeI} ``Cat's eye'' diagram: a spanning tree $\Tree$ and the associated loop basis $\LoopBasis_1$.}   
    \vspace*{4mm}
  \item[]
    \begin{minipage}{0.6\textwidth}
      \begin{tabular}{cccc}
        \br
        $\left\{ 0, 0, 0 \right\}$ 
        & $\Cycle_1 \equiv \left\{ 0, 0, 1\right\}$ 
        & $\Cycle_2 \equiv \left\{ 0, 1, 0 \right\}$ 
        & $\Cycle_3 \equiv \left\{ 0, 1, 1 \right\}$
        \\
        \mr
        \textbf{$<1$ loops}
        &\begin{minipage}{0.25\textwidth}\includegraphics[width=\textwidth]{catsEyeLoopOne.eps}\end{minipage} 
        &\begin{minipage}{0.25\textwidth}\includegraphics[width=\textwidth]{catsEyeLoopTwo.eps}\end{minipage}
        &\begin{minipage}{0.25\textwidth}\includegraphics[width=\textwidth]{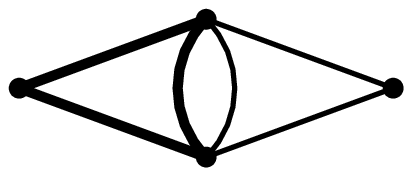}\end{minipage}
        \\
        \mr
        $\Cycle_4 \equiv \left\{ 1, 0, 0 \right\}$ 
        & $\Cycle_5 \equiv\left\{ 1, 0, 1 \right\}$ 
        & $\Cycle_6 \equiv \left\{ 1, 1, 0\right\}$ 
        & $\left\{ 1, 1, 1 \right\}$
        \\
        \mr
        \begin{minipage}{0.25\textwidth}\includegraphics[width=\textwidth]{catsEyeLoopThree.eps}\end{minipage}  
        &\begin{minipage}{0.25\textwidth}\includegraphics[width=\textwidth]{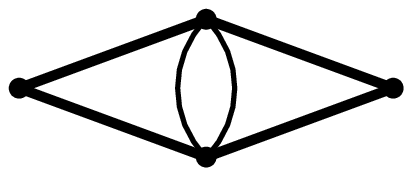} \end{minipage} 
        &\begin{minipage}{0.25\textwidth}\includegraphics[width=\textwidth]{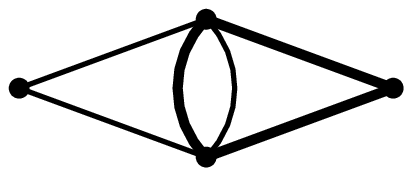} \end{minipage} 
        &\textbf{$>1$ loops}
        \\
        \br
      \end{tabular}
    \end{minipage}
    \vspace*{-2mm}
    \caption{\label{catsEyeII} ``Cat's eye'' diagram: all cycles.}   
    \vspace*{2mm}
  \item[]
    \begin{minipage}{0.36\textwidth}
      \includegraphics[width=\textwidth]{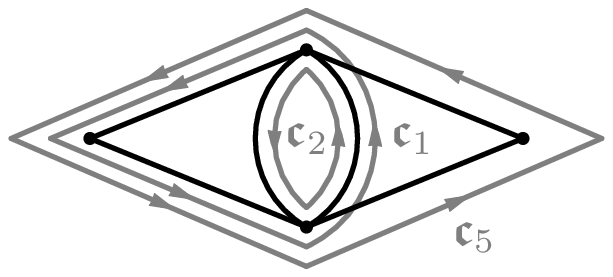}
    \end{minipage}
    \begin{minipage}{0.4\textwidth}
      \includegraphics[width=\textwidth]{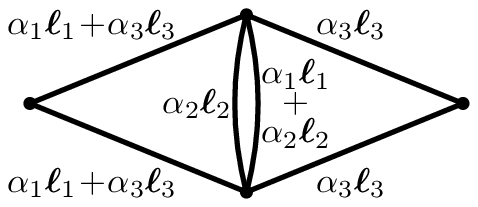}
    \end{minipage}
    \vspace*{-4mm}
    \caption{\label{catsEyeIII} ``Cat's eye'' diagram: 
      loop assignment deriving from loop basis 
      $\LoopBasis_2 \equiv \left\{\Cycle_1,\Cycle_2,\Cycle_5 \right\}$.
    }
  \end{indented}
\end{figure}

As an example of application of algorithm \ref{LoopBases} 
we illustrate in detail the loop parametrization of
the \emph{``cat's eye''} diagram; it is drawn 
in \fref{catsEyeI} together with a choice of a spanning tree, $\Tree$, 
and with the loop basis $\LoopBasis_1$ obtained by applying  
theorem~\ref{FirstLoopBasis} to $\Tree$. 

The three cycles which are elements of $\LoopBasis_1$ are identified (from left to right)  with the following 
basis vectors: $\left\{ 0, 0, 1\right\}$, $\left\{ 0, 1, 0 \right\}$ and $\left\{ 1, 0, 0 \right\}$. 
The set of all cycles contained in the original diagram is obtained by
 working out all possible linear combinations of such basis vectors in
$\mathbb{Z}_2$, and by discarding the resulting graphs which have more
 than, or less than, $1$ loop (see \fref{catsEyeII}).
All possible loop bases will then be given by all choices 
of three linearly independent loops among the six so far produced: 
for instance, 
$\LoopBasis_2 \equiv \left\{\Cycle_1,\Cycle_2,\Cycle_5 \right\}$ 
is a valid basis but  the same is not true for the choice 
$\left\{  \Cycle_1,\Cycle_2,\Cycle_3 \right\}$, 
due to the fact that $\Cycle_3 =\Cycle_1 \oplus \Cycle_2$.
  
Then, given a choice of loop basis and of suitable orientations, 
momentum assignements for the lines will follow by attributing momenta 
$\alpha_1\bell_1$, $\alpha_2\bell_2$ and $\alpha_3\bell_3$ (with $\alpha_i\neq0$ arbitrary) 
to the three cycles of the basis, respectively: in \fref{catsEyeIII} 
we depict the loop choice which is obtained for the ``cat's eye'' 
by choosing the $\LoopBasis_2$ defined above.

\begin{figure}[ht]
\begin{center}   
\hfill\includegraphics[width=0.9\textwidth]{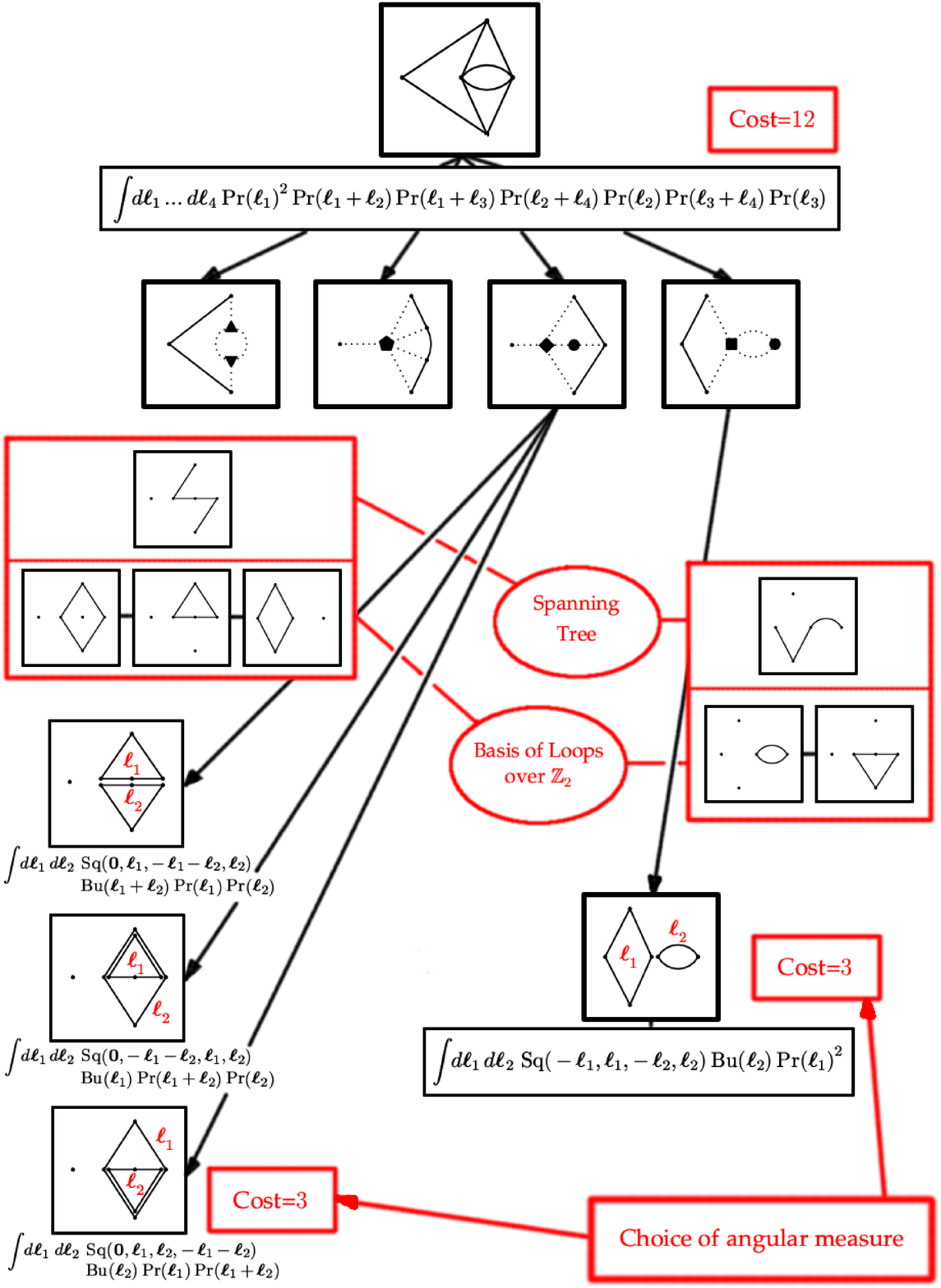}
\end{center}
 \caption{\label{SubstitutionsCompleteII}
 Example of \fref{ParametrizationExample} after various choices 
of loop momenta. The costs will be deduced only at a later stage, after 
angular measures for each diagram have been picked up 
(see \sref{ChoiceOfAngularMeasure}).}
\end{figure} 

It should be noticed that even the computation of loop bases can give rise 
to a combinatorial explosion of generated data, 
in particular when the number of residual loops 
$\NLoopsResidual$ of the effective diagram $\EffectiveDiagram$ is large: this fact can be easily understood 
by realizing that each loop basis is an extraction of $\NLoopsResidual$ 
linearly independent elements 
out of a set $\Loops$ composed by up to $2^\NLoopsResidual-1$ elements.

Finally, a summary of the entire parametrization procedure 
for the diagram introduced in \fref{ParametrizationExample} is presented
in  \fref{SubstitutionsCompleteII}. 
Please remark that the deduction of the costs shown in this figure  relies on the choice
of angular measures, which is the subject of the next \sref{ChoiceOfAngularMeasure}.

\subsection{Choice of angular measure}\label{ChoiceOfAngularMeasure}
In this section we deal with the problem of explicitly parametrizing the
integration measure for an integral expression of the form
\[
\Integral_F := \int \prod_{l=1}^{\NLoopsResidual} d^3\!\bell_{l}\; 
  F\!\left(\LoopScalarProducts\right)
\]
where, using notation previously introduced in \sref{autoparOverview}, 
$F$ depends\footnote{In reality, the fact that the $\NLoopsResidual\times \NLoopsResidual$ 
Gram matrix of scalar products $(\bell_i.\bell_j)$ has rank not greater than 
the integer space-time dimension $d$ implies ---when $\NLoopsResidual>d$--- the existence of $(\NLoopsResidual-d)(\NLoopsResidual-d+1)/2$ 
functional relations among such scalar products, and consequently that at most 
$d \NLoopsResidual-d(d-1)/2$ scalar products out of 
$\NLoopsResidual(\NLoopsResidual+1)/2$ can be functionally
independent. As a consequence, the same function $F$ might be defined 
in terms of different expressions, each involving an {\em a priori} 
different
subset of scalar products $(\bell_i.\bell_j)$. For shortness, in the following we use the words
``a function $F$'' meaning 
``an expression ---fixed throughout the analysis--- which gives a realization of the function $F$''; 
however, notice that the minimality of the 
set of scalar products on which the chosen representation depends, $\LoopScalarProducts$, is never 
assumed in our approach. In particular, eventual functional dependencies 
among the scalar products are \emph{automatically} taken into account 
in our framework, being the parametrizations of scalar products obtained from
parametrizations of vectors ---see eq. \eref{versorViaAxes}.} 
on a 
subset $\LoopScalarProducts$ of all scalar products 
$\{\;\bell_{l_1}.\bell_{l_2}\;|\;l_1,l_2\in 1\ldots\NLoopsResidual\mathrm{\ with\ }l_1\leq l_2\;_{_{}}\}$ among integration vectors.

Since an hyperrectangular domain seems to be the most convenient set-up when
we are to perform multidimensional numerical integration, in the following we will
restrict ourselves to the particular choice of spherical coordinates,
and will accordingly decompose loop vectors as the product of their 
norm times a versor: $\bell_l=\ell_l \;\vell_l $. 
Consequently, given an integral 
\beq
\Integral_F = \int\prod_{l=1}^{\NLoopsResidual} 
 d\ell_{l}\;\ell_{l}^{2}\;d\Omega_{l}\;F\!\left(
\Many{\ell},\ScalarProducts
\right) \label{Spheres}
,\eeq
where
\bea
\ScalarProducts\subseteq
\{\;\vell_{l_1}.\vell_{l_2}=:\cos\theta_{l_1 l_2}\;
   |\;l_1,l_2\in 1\ldots\NLoopsResidual\mathrm{\ with\ }l_1< l_2\;_{_{}}\}
\\
\Many{\ell}=(\ell_1,\,\ldots\, ,\,\ell_\NLoopsResidual),   
\eea
we examine here the problem of explicitly writing down an angular
measure $\prod_{l} d\Omega_{l}$ which is minimal in the number of required 
residual integrations, that is,
which allows a maximal number of angular integrations to be carried out 
analytically. Although the spirit of this idea can be easily generalized 
to a different number of dimensions $d$, our explicit formulas 
will be specialized to the case $d=3$.

This investigation is motivated by equation \eref{ScalarProducts}, which
shows that our most general amplitude is precisely of the form just 
mentioned;
the cost \eref{Cost} associated to a  parametrization of 
a given diagram cannot be fully evaluated without knowing
the cost of the integration over loop momenta, and due to this fact 
one is forced to explicitly determine such a ``minimal'' angular measure.

It should be noticed that parametrizations of angular measures 
completely different from the spherical ones can of course be considered 
---we think  e.g. to the techniques used in \cite{David:1992vv}---,
possibly leading as well to a significant reduction in the number
of residual integrations; however, we have not considered them in this work,
due mainly to the fact that high-performance numerical integrators 
seem to get along best with simple hyperrectangular domains.

\subsubsection{Cosine diagrams}\label{cosDiagrams}
As we will see later on,  $\ScalarProducts$,
the subset of scalar products on which
function $F$ in equation
\eref{Spheres} depends, plays an essential role in determining the optimal
choice of angular integration measure. This information is encoded in the
so-called {\em cosine diagram associated to function $F$}, $\CosDiagram$,
defined as follows:
\begin{enumerate}
\item for each integration variable $\vell_l$, add a vertex $v_l$
\item if function $F$ carries an explicit dependence on 
$\vell_{l_1}.\vell_{l_2}$ (with $l_1<l_2$), draw a line connecting 
vertex $v_{l_1}$ to vertex $v_{l_2}$.
\end{enumerate}
At this level the knowledge of $\ScalarProducts$ 
is equivalent to that of $\CosDiagram$, but
further information ---encoding additional properties of $F$---
can be attached to the cosine diagram by allowing 
different kinds of lines in $\CosDiagram$.
A fruitful application of this technique to the specific class of function we are interested in
is postponed to section \ref{DisconnectibleCosineDiagrams}.
In the following section \ref{ChoosingAxes} we will discuss 
at a more general level the problem of a ``good choice'' 
of parametrization of the angular measure.

\subsubsection{Choosing axes}\label{ChoosingAxes}
The abitual way of parametrizing the angular measure $\prod_{l} d\Omega_{l}$ 
is to decompose each $d\Omega_l$ as
\[
d\Omega_l=-d\!\left(\cos\theta_l\right)d\phi_l
;\]
however, this decomposition requires a previous choice of an axis 
$\ZVersor_l$ 
---in terms of which angle $\theta_l$ is measured--- and of an axis 
$\XVersor_l$ 
---such that angle $\phi_l$ can be defined as in usual $3$-dimensional 
spherical coordinates. 
(Please remark that in the following $\XVersor_l$, $\ZVersor_l$ will denote 
a choice of non-degenerate
---but \emph{not necessarily orthogonal}--- axis versors.)
Once a choice of axes has been picked up, one can explicitly parametrize 
all the domain of variation of versor $\vell_l$ in the standard way as 
\bea
\vell_l&=&\cos\theta_l\ZVersor_l
+\sin\theta_l\left(
   \cos\phi_l\;\widehat{\left(\ZVersor_l\wedge\XVersor_l\right)\wedge\ZVersor_l}
  +\sin\phi_l\;\widehat{\ZVersor_l\wedge\XVersor_l}
  \right)
\label{versorViaAxes}
\\
\nonumber&&\mathrm{with\ }\theta_l\in\left[0,\pi\right]\mathrm{\ and\ }\phi_l\in\left[0,2\pi\right)
\eea
(our standard notation $\widehat{\mathbf{v}}:=\mathbf{v}/|\mathbf{v}|$
has been used here).
In addition, 
after a parametrization is given for all versors $\vell_l$
one can usefully encode this information  in  
the {\em matrix of scalar products} $\MatrixScalarProducts$, 
which is defined as
\beq
\MatrixScalarProducts_{l_1 l_2}\!\!\left(\Many{\theta},\Many{\phi}\right):=
\cases{
  \vell_{l_1}.\vell_{l_2}\!\!\left(\Many{\theta},\Many{\phi}\right)&if $1\leq l_1<l_2\leq\NLoopsResidual$\\ 
  0&if $1\leq l_2\leq l_1\leq\NLoopsResidual$.\\
}
\label{MSP}\eeq
(Please remark that $\Many{\theta}$ means ``an appropriate subset of $\theta_1,\,\ldots,\,\theta_\NLoopsResidual$'',
and the same holds for $\Many{\phi}$; refer to \ref{renormalization} 
for an explanation of our notations.)

Now, supposing we have formulated a choice of axes $\XVersor_l,\ZVersor_l$
for each  $\vell_l$ and prepared the matrix of scalar products
$\MatrixScalarProducts$ associated to this parametrization,
the question arises:
how many nontrivial angular integrations are necessary to parametrize
the angular measure $\prod_{l} d\Omega_{l}$ for   
a given function $F$ which depends on a (sub-)set of scalar products 
$\ScalarProducts$ via a specific cosine diagram $\CosDiagram$?

In present framework the answer follows straightforwardly:
just use the informations encoded in the cosine diagram $\CosDiagram$
and in the matrix  $\MatrixScalarProducts$ of parametrized 
scalar products to  build up the set $\AngularVariables$ 
containing the ``nontrivial'' residual angular variables 
which $F$ really depends on:
\beq
\AngularVariables:=  
\bigcup_{\mathrm{line}\; l_1-l_2\; \in\CosDiagram}
\left\{\;\Many{\theta},\Many{\phi}\;|
\MatrixScalarProducts_{l_1 l_2} \mathrm{depends \ on\ variables\ } \Many{\theta},\Many{\phi}
\;\right\}
\label{AngularVariables}
.\eeq
The relevance of the set $\AngularVariables$ is immediately apparent:
given an angular measure $\prod_{l} d\Omega_{l}$ induced by a specific choice
for a parametrization,
all angular variables $\Many{\theta}$, $\Many{\phi}$ involved 
in its realization which \emph{are not} elements of $\AngularVariables$
correspond to flat directions, and can in accordance 
be trivially integrated out.

Coming back to the issue of parametrization via \eref{versorViaAxes},
it clearly appears that a great arbitrariness is still left over:
in fact, given a function $F$ with cosine diagram $\CosDiagram$,
many choices can be imagined for versors $\ZVersor_l$ and $\XVersor_l$, 
each one leading to an \emph{a priori} different number of final angular integrations,
depending on the particular form of $\CosDiagram$.

One result of the present work is the classification of 
\emph{all} possible cosine diagrams relevant for integrations up to $\NLoopsResidual=4$ 
and the systematic development of appropriate choices of axes 
to lower as much as possible
the number of final angular integrations.  
In the following we will clarify the theoretical framework, and introduce 
the reader to our parametrization strategy by giving some examples.

Let us suppose of having to cope with two integrand functions,
$F_1,F_2$ depending respectively on four loop versors 
$\vell_1,\,\ldots\, ,\,\vell_4$ 
via the set of scalar products $\ScalarProducts_1$ and $\ScalarProducts_2$
defined as:
\begin{eqnarray}
\ScalarProducts_1:=
\left\{\;\vell_1.\vell_2,\;\vell_1.\vell_3,\;\vell_1.\vell_4
      ,\;\vell_2.\vell_3,\;\vell_2.\vell_4,\;\vell_3.\vell_4\;\right\}\\
\ScalarProducts_2:=\left\{\;\vell_1.\vell_2,\;\vell_1.\vell_3\;\right\}
\end{eqnarray} 

As a warm-up we consider the natural choice of axes parametrization 
consisting in picking up two
{\em fixed} nondegenerate external vectors $\ZCapitalVersor$ and $\XCapitalVersor$, 
and in parametrizing  all the versors $\vell_l$
w.r.t.\ these axes via \eref{versorViaAxes}; 
the following table encodes this possibility.
\beq
\fl
\begin{minipage}{0.8\textwidth}
\begin{indented}
\item[]\begin{tabular}[b]{@{}cccc}
\br
Axis choices & $\vell_1$ & $\ldots$ & $\vell_\NLoopsResidual$\\
\mr
$\ZVersor_l$ & $\ZCapitalVersor$ & $\ldots$ & $\ZCapitalVersor$ \\
$\XVersor_l$ & $\XCapitalVersor$ & $\ldots$ & $\XCapitalVersor$ \\
\br
\end{tabular}
\end{indented}
\end{minipage}
\label{trivialChoice}\eeq
The choice in \eref{trivialChoice} would lead in the  $\NLoopsResidual=4$ case 
to the 
particular form of $\MatrixScalarProducts$ given by 
\beq
\fl
\pmatrix{
0&\Cos\theta_1\Cos\theta_2+\Sin\theta_1\Sin\theta_2\Cos\!\left(\phi_1-\phi_2\right)&\Cos\theta_1\Cos\theta_3+\Sin\theta_1\Sin\theta_3\Cos\!\left(\phi_1-\phi_3\right)&\Cos\theta_1\Cos\theta_4+\Sin\theta_1\Sin\theta_4\Cos\!\left(\phi_1-\phi_4\right)\cr
&0&\Cos\theta_2\Cos\theta_3+\Sin\theta_2\Sin\theta_3\Cos\!\left(\phi_2-\phi_3\right)&\Cos\theta_2\Cos\theta_4+\Sin\theta_2\Sin\theta_4\Cos\!\left(\phi_2-\phi_4\right)\cr
&&0&\Cos\theta_3\Cos\theta_4+\Sin\theta_3\Sin\theta_4\Cos\!\left(\phi_3-\phi_4\right)\cr
0&&&0
}
.
\label{trivialChoiceM}
\eeq
Following equation \eref{AngularVariables} we readily obtain the sets 
$\AngularVariables_{1,\mathrm{fixed}}$ and
$\AngularVariables_{2,\mathrm{fixed}}$ of nontrivial 
angular variables induced by the parametrization \eref{trivialChoice} for the
two cases $F_1$, $F_2$:
\bea
\AngularVariables_{1,\mathrm{fixed}}=
\left\{\;\theta_1,\theta_2,\theta_3,\theta_4,
         \phi_1,\phi_2,\phi_3,\phi_4\;\right\}\\
\AngularVariables_{2,\mathrm{fixed}}=
\left\{\;\theta_1,\theta_2,\theta_3,
         \phi_1,\phi_2,\phi_3\;\right\}
.\eea
According to the above results
we are now able to estimate the cost of residual angular integrations 
\emph{as induced by the parametrization \eref{trivialChoice}} 
to $8$ for $F_1$ and to $6$ in the case of $F_2$.
Impressively enough, the developed formalism reveals quite immediately that 
due to the simpler angular dependences of $F_2$ 
$2$ angular integrations can be spared w.r.t.\ the
``full cost'' case of $F_1$.
(In fact, a deeper analysis of \eref{trivialChoiceM} would show that 
an additional integration could be
spared for both $F_1$ and $F_2$ since $\MatrixScalarProducts$ depends only on 
differences of $\phi$ variables; however, this point is inessential in the present discussion because 
---as shown below---
one can  do much better than that.) 

The key that opens the door to much more efficient parametrizations 
is the possibility to choose  
as parametrizing axes $\ZVersor_l$ and $\XVersor_l$ for $\vell_l$
appropriate {\em ``moving''} reference vectors.

A fruitful example of such a choice is the  parametrization defined below,
which we call \emph{``standard''};
\beq
\label{Standard}
\fl
\begin{minipage}{0.8\textwidth}
\begin{indented}
\item[]\begin{tabular}[b]{@{}cccccc}
\br
Axis choices & $\vell_1$ & $\vell_2$& $\vell_3$& $\ldots$ & $\vell_{\NLoopsResidual}$\\
\mr
$\ZVersor_l$ & -         & ${\vell_1}$ & ${\vell_1}$ & ${\vell_1}$& ${\vell_1}$\\
$\XVersor_l$ & -         & -               & ${\vell_2}$ & ${\vell_2}$& ${\vell_2}$\\
\br
\end{tabular}
\end{indented}
\end{minipage}
\eeq
in the case of $\NLoopsResidual=4$ our ``standard'' choice 
leads to a cosine matrix $\MatrixScalarProducts$ of the form
\beq
\label{StandardMatrix}
\pmatrix{
0&\Cos\theta_2&\Cos\theta_3&\Cos\theta_4\cr
&0&\Cos\theta_2\Cos\theta_3+\Sin\theta_2\Sin\theta_3\Cos\phi_3&\Cos\theta_2\Cos\theta_4+\Sin\theta_2\Sin\theta_4\Cos\phi_4\cr
&&0&\Cos\theta_3\Cos\theta_4+\Sin\theta_3\Sin\theta_4\Cos\!\left(\phi_3-\phi_4\right)\cr
0&&&0
}
.
\eeq
Please remark that in the standard parametrization the expression of 
$\MatrixScalarProducts$ in \eref{StandardMatrix} is left unchanged by
\emph{any} legal choice of $\XVersor_1$, $\ZVersor_1$ and $\XVersor_2$: 
that is why no explicit choice is given for such versors in definition 
\eref{Standard}.

The sets 
$\AngularVariables_{1,\mathrm{standard}}$, 
$\AngularVariables_{2,\mathrm{standard}}$
induced by the standard parametrization \eref{Standard} for our examples 
$F_1$, $F_2$ follow readily:
\bea
\AngularVariables_{1,\mathrm{standard}}=
\left\{\;\theta_2,\theta_3,\theta_4,\phi_3,\phi_4\;\right\}
\label{AVOneStd}
\\
\AngularVariables_{2,\mathrm{standard}}=
\left\{\;\theta_2,\theta_3\;\right\}
\label{AvTwoStd}
,\eea
resulting in $5$ residual angular integrations for the full-cost case $F_1$ and 
---quite spectacularly--- in only $2$ residual angular integrations for the simpler 
case of $F_2$.

From a closer inspection of  \eref{Standard},
and of its implications on the matrix of scalar products $\MatrixScalarProducts$
---think to  \eref{StandardMatrix} and its generalizations--- 
it should appear evident that standard parametrization implements in an explicit way 
the possibility of making use of the overall $SO\left(3\right)$-symmetry 
of the integrand to drop the trivial integration 
over a global $SO\left(3\right)$ rotation of the $\vell_l$. 
In the full-cost case of a function depending on $\NLoopsResidual$ 
vectors $\bell_l$
with complete cosine diagram $\CosDiagram$, 
the standard parametrization would give 
\beq
\AngularVariables_\mathrm{standard}=
\left\{\;\theta_2,\,\ldots\, ,\,\theta_{\NLoopsResidual},\phi_3,\,\ldots\, ,\,\phi_{\NLoopsResidual}\;\right\},
\label{worst}\eeq 
i.e. $2\NLoopsResidual-3$ residual angular integrations to which 
one must add $\NLoopsResidual$ integrations over the vector norms,
obtaining the total cost $3\NLoopsResidual-3$\setcounter{footnote}{0}\footnote{It is remarkable that 
$3\NLoopsResidual-3$ is equal to the maximal number of 
functionally independent scalar products which can be obtained 
from $\NLoopsResidual$ $3$-dimensional vectors, see footnote at 
the beginning of \sref{ChoiceOfAngularMeasure}.}, in agreement with the 
$d$-dimensional formula introduced in \sref{WhichParametrization}.
(By the way, a $d$-dimensional generalization of \eref{Standard} 
can be straightforwardly derived. 
See \cite{Cicuta:2001ke} for a similar result from a quite different parametrization.) 

The example of function $F_2$ above teaches us that
when the angular dependence of the function is simple, 
additional integrations
can be spared 
---within the framework of standard parametrization---
compared to the worst-case equation \eref{worst};
however, to reach this goal, the properly \emph{permuted} 
version of the standard parametization
 \eref{Standard} must be used, think e.g. to a function $F_3$ with 
 $\ScalarProducts_3:=\left\{\;\vell_1.\vell_3,\;\vell_2.\vell_3\;\right\}$.

All in all standard parametrization \eref{Standard} 
proves to be useful to treat a large variety of angular 
dependencies; however, it should be emphasized that 
considering other choices of moving axes $\XVersor_l$, $\ZVersor_l$ 
for specific cosine diagrams can sometimes result in parametrizations 
even cheaper than the standard one. 

Let us examine, for instance, the example of an integrand function $F_4$
depending on $\vell_1,\,\ldots\, ,\, \vell_4$ via a set of scalar products 
$\ScalarProducts_4:=\left\{\;\vell_1.\vell_2,\;\vell_1.\vell_3,\;\vell_2.\vell_4\;\right\}$.   
In this case the standard parametrization  \eref{Standard} 
would give a set of residual variables
\beq
\AngularVariables_{4,\mathrm{standard}}=
\left\{\;\theta_2,\theta_3,\theta_4,\phi_4\;\right\},
\eeq 
corresponding to $4$ angular integrations; this result could not be improved 
by permutations of \eref{Standard}.

On the other hand, the ``non-standard'' choice
\[
\fl
\begin{minipage}{0.8\textwidth}
\begin{indented}
\item[]\begin{tabular}[b]{@{}ccccc}
\br
Axis choices & $\vell_1$ & $\vell_2$      & $\vell_3$      & $\vell_4$\\
\mr
$\ZVersor_l$ & -         & ${\vell_1}$ & ${\vell_1}$ & ${\vell_2}$\\
$\XVersor_l$ & -         & -               & ${\vell_2}$ & ${\vell_3}$\\
\br
\end{tabular}
\end{indented}
\end{minipage}
\]
leads us to the minimal set of angular 
variables 
\beq
\AngularVariables_{4,\mathrm{non-standard}}
=\left\{\;\theta_{12},\;\theta_{13},\;\theta_{24}\;\right\}
,\label{aNonStandard}\eeq
corresponding to an angular cost of $3$: two angular integrations are spared
w.r.t.\ the standard parametrization (please remark that in 
\eref{aNonStandard} we defined as usual 
$\theta_{l_1 l_2}:=\arccos \left(\vell_{l_1}.\vell_{l_2}\right)$).
 
The variety of parametrizations based on moving axes 
is enormous, and limited only by the imagination of the user. 
The only important constraint when giving a choice of axes 
$\XVersor_l$, $\ZVersor_l$ in terms of integration variables $\vell_l$
is that ---to avoid inconsistences--- 
one must fix an order of integration over the variables $\vell_l$ 
and coherently respect it,
by defining $\XVersor_l$, $\ZVersor_l$ only in terms of those $\vell_{l'}$
which are integrated before $\vell_l$ in the chosen integration order.

Given a \emph{generic} cosine diagram $\CosDiagram$,
 the problem of finding for each $\vell_l$ 
 a choice of axes $\XVersor_l$, $\ZVersor_l$
giving rise to an angular measure with the
\emph{minimal} number of residual integrations over $\theta$, $\phi$ 
is ---in our knowledge--- still unsolved.
A partial answer can be found in the following proposition, which deals with
cosine diagrams of tree type.

\begin{proposition}\label{treecos} 
Given a cosine diagram $\CosDiagram$ 
with $\NLoopsResidual$ vertices and vanishing loop number,
a choice of axes $\XVersor_l$, $\ZVersor_l$ can be given for each $\vell_l$ 
---in the sense of \eref{versorViaAxes}--- such that 
the angular measure is expressed in terms of $\NLoopsResidual$ 
residual integrations of $\cos\theta$-type (or, equivalently, such 
that $\AngularVariables$ has $\NLoopsResidual$ elements, 
all of $\theta$-type).
\end{proposition}

\noindent\textbf{Proof} 
Suppose first that $\CosDiagram$ is a tree (hence connected by definition).
Choose a root vertex $v_{l_0}$ on $\CosDiagram$.
Fix arbitrarily a couple 
of nondegenerate versors $(\XCapitalVersor,\ZCapitalVersor)$ 
and assign 
$(\XVersor_{l_0},\ZVersor_{l_0})=(\XCapitalVersor,\ZCapitalVersor)$
for the root vertex and $\XVersor_l=\XCapitalVersor$ 
for all other vertices $v_l$ of $\CosDiagram$.
A choice of axis $\ZVersor_l$ 
for all vertices $v_l\neq v_{l_0}$ will now be given,
by iterating over their distance $n$ from the root vertex $v_{l_0}$.
First define $\ZVersor_{l_1}=\vell_{l_0}$ 
for all vertices $v_{l_1}$ at distance $1$ from the root
--- $\vell_{l_0}$ being the integration versor associated by definition to the root vertex. 
Then iterate over increasing $n\geq 1$ as following: 
for each vertex $v_{l_n}$ at  distance $n$ 
define $\ZVersor_{l_{n+1}}=\vell_{l_n}$ for all vertices $v_{l_{n+1}}$
departing from $v_{l_n}$ and having distance $n+1$ from the root
--- $\vell_{l_n}$ being the integration versor associated by definition to $v_{l_n}$. 
If $\CosDiagram$ is not connected 
repeat the above reasoning for each connected component of $\CosDiagram$,
using the fact that with this parametrization the angular measure factors 
in term of angular measures over the connected components 
of the cosine diagram.\hfill$\Box$\medskip

In table \ref{ThreeLoopsAngular}
we will give the complete classification of cosine diagrams for the case 
$\NLoopsResidual=3$, as well as the best choice of parametrizations which we 
were able to find according to the lines of reasoning presented 
in this section (including also some additional tricks which will be
introduced in the following \sref{DisconnectibleCosineDiagrams}).

\subsubsection{Cosine diagrams with disconnectible lines}\label{DisconnectibleCosineDiagrams}
As mentioned in section \ref{cosDiagrams}, 
it is sometimes useful to encode additional properties of the integrand function $F$
in the cosine diagram $\CosDiagram$, by distinguishing different types of lines. 
This technique comes particularly at hand when parametrizing 
loop integrations for the amplitudes described 
in equation \eref{ScalarProducts}.
Below we identify a particular situation when the integrand function
depends on its arguments in a simpler way:
an appropriate codification of this information in the cosine diagram
and a careful analysis may sometimes lead to 
the detection of new types of factored analytically-known angular integrals
and, consequently, 
to  an additional  decrease of the number of residual angular integrations.

The situation of interest is   
when the integrand function $F$ can be factored as follows:
\beq\fl
F=\left(\mathrm{block\ with\ no\ dependence\ on\ }\vell_{l_1}.\vell_{l_2}^{^{^{^{}}}}\right) \left(\frac{1}{1+\ell_{l_1}^2+\ell_{l_2}^2+2\;\bell_{l_1}.\bell_{l_2}}\right)^n
,\label{disconnectibleI}\eeq
$n$ being a strictly positive integer;
in this case the line joining vertices $v_{l_1}$ and $v_{l_2}$ in $\CosDiagram$ will be 
graphically denoted with a dashed style instead of a continuous one, 
and we will say that such a line 
---together with the associated cosine $\cos\theta_{l_1 l_2}:={\vell_{l_1}.\vell_{l_2}}$ 
between loop momenta $\bell_{l_1}$ and $\bell_{l_2}$--- 
is {\em disconnectible}.
 
The motivation for introducing the notion of disconnectible lines 
derives from the following observation: 
supposing that we have been able to pick up a parametrization of angular measure
$\prod_{l} d\!\left(\cos \theta_l\right) d\phi_l$ such that
\begin{enumerate}
\item $\cos\theta_{l_1 l_2}$ belongs to the set of angular 
integration variables $(\cos \theta_1, \cdots ,\cos \theta_{\NLoopsResidual})$ 
\item 
the only dependence of the integrand on  $\cos\theta_{l_1 l_2}$ 
is via the scalar product $\vell_{l_1}.\vell_{l_2}$, 
i.e. no spurious dependencies on  
$\cos\theta_{l_1 l_2}$ of other scalar products are re-introduced by the parametrization
 ---see \sref{ChoosingAxes}, 
\end{enumerate}
then the integration
\beq
\int_{-1}^{1} d\!\left(\cos\theta_{l_1 l_2}\right)\;
\left(\frac{1}{1+\ell_{l_1}^2+\ell_{l_2}^2
     +2\;\ell_{l_1}\ell_{l_2}\;\cos\theta_{l_1 l_2}}\right)^n
\label{CosIntegral}
\eeq
can be factored away  
and carried out analytically, thus leading to a lower number of required 
residual integrations.

However, this is not the actual end of the history: 
a weaker situation can sometimes happen when it is
not possible to integrate directly over 
a disconnectible cosine $\cos\theta_{l_1 l_2}$,
but such a cosine can nonetheless be expressed 
in terms of some of the variables that parametrize the angular measure,
with the condition that one of these variables must appear \emph{only} 
in the expression of the scalar product $\vell_{l_1}.\vell_{l_2}$.
In this case, again, the integral over such distinguished angular variable 
of the power of propagators containing $\vell_{l_1}.\vell_{l_2}$ 
---see \eref{disconnectibleI}--- 
can be factored and, possibly, performed analitically.
A typical example of such a case is given 
by the combination
\bea
\nonumber
\int_0^{2\pi} d\phi_3\;\left(\frac{1}{1+\ell_{2}^2+\ell_{3}^2+2\;\ell_2\ell_3\;\vell_2.\vell_3}\right)^n
\\
\vell_{2}.\vell_{3}=\cos\theta_2\cos\theta_3+\sin\theta_2\sin\theta_3\cos\phi_3
\label{PhiIntegral}
,\eea
originating for example in some cases from the ``standard'' angular parametrization 
defined in \eref{Standard}; 
it should be noticed that integrals in the form of \eref{PhiIntegral} 
can always be performed by means of the technique of complex residues.

Additional complications arise when we want to minimize the number of angular 
integrations in presence of an integrand function $F$ whose associated 
cosine diagram $\CosDiagram$ possesses disconnectible lines. 
As a matter of fact, in this case we must face two different requirements at 
the same time: as in previous section,
we must minimize the number of angular integration variables on which $F$
depends; \emph{in addition}, we must now maximize as well 
the number of occurrences of integrals in the form of 
\eref{CosIntegral} and \eref{PhiIntegral}.

Such a problem is quite complicated, and finding a solution implies the 
enumeration and the inspection of all possible cosine diagrams with disconnectible lines 
which can arise up to the maximal number of residual loops $\NLoopsResidual$ 
we are interested in.

\begin{table}[ht]
\caption{\label{ThreeLoopsAngular}
  Our choice of ``cheapest'' angular measures for the case 
  $\NLoopsResidual=3$. See the text for all the explanations about 
  graphical conventions used in this table.}
\begin{indented}
\item[]\begin{tabular}[t]{@{}clcc}
\br
Initial cosine diagram & Final cosine diagram & Variables lines depend on & Cost\\
\mr
\begin{minipage}{0.11\textwidth}
  \includegraphics[width=\textwidth]{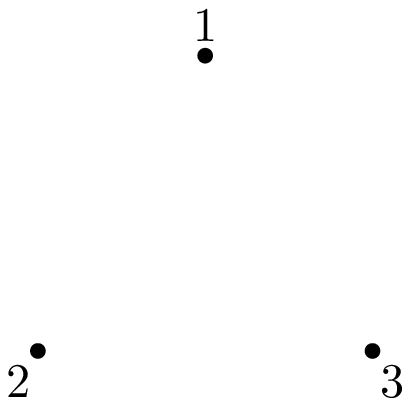}
\end{minipage}
&
\begin{minipage}{0.11\textwidth}
  \includegraphics[width=\textwidth]{000.eps}
\end{minipage}
&
&
\begin{minipage}{0.05\textwidth}$(0,0)$\end{minipage}\vspace*{1mm}\\
\begin{minipage}{0.11\textwidth}
  \includegraphics[width=\textwidth]{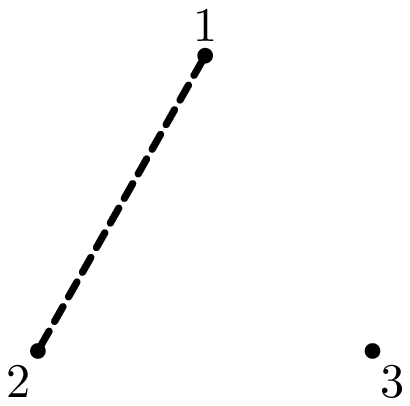}
\end{minipage}
&
\begin{minipage}{0.11\textwidth}
  \includegraphics[width=\textwidth]{200.eps}
\end{minipage}
&
&
\begin{minipage}{0.05\textwidth}$(0,0)$\end{minipage}\vspace*{1mm}\\
\begin{minipage}{0.11\textwidth}
  \includegraphics[width=\textwidth]{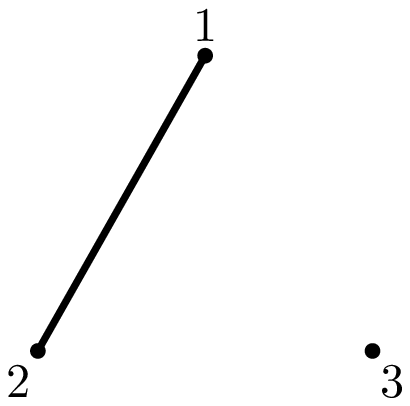}
\end{minipage}
&
\begin{minipage}{0.11\textwidth}
  \includegraphics[width=\textwidth]{100.eps}
\end{minipage}
&
\begin{minipage}[c]{0.23\textwidth}
line $1$--$2$ $\longrightarrow$ $\left\{_{_{}}\theta_2\right\}$
\end{minipage}
&
\begin{minipage}{0.05\textwidth}$(1,0)$\end{minipage}\vspace*{1mm}\\
\begin{minipage}{0.11\textwidth}
  \includegraphics[width=\textwidth]{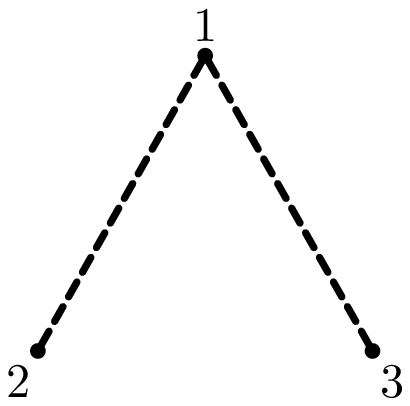}
\end{minipage}
&
\begin{minipage}{0.11\textwidth}
  \includegraphics[width=\textwidth]{220.eps}
\end{minipage}
&
&
\begin{minipage}{0.05\textwidth}$(0,0)$\end{minipage}\vspace*{1mm}\\
\begin{minipage}{0.11\textwidth}
  \includegraphics[width=\textwidth]{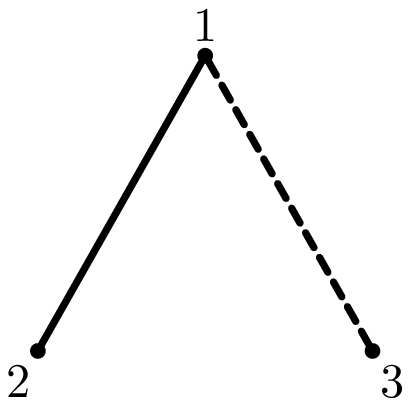}
\end{minipage}
&
\begin{minipage}{0.11\textwidth}
  \includegraphics[width=\textwidth]{120.eps}
\end{minipage}
&
\begin{minipage}[c]{0.23\textwidth}
line $1$--$2$ $\longrightarrow$ $\left\{_{_{}}\theta_2\right\}$
\end{minipage}
&
\begin{minipage}{0.05\textwidth}$(1,0)$\end{minipage}\vspace*{1mm}\\
\begin{minipage}{0.11\textwidth}
  \includegraphics[width=\textwidth]{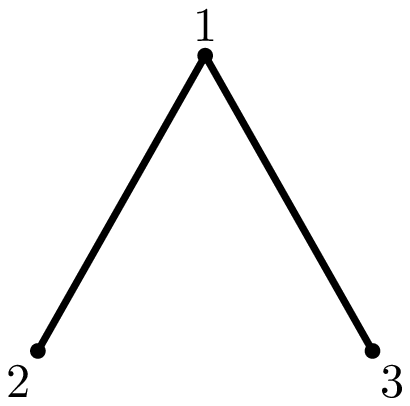}
\end{minipage}
&
\begin{minipage}{0.11\textwidth}
  \includegraphics[width=\textwidth]{110.eps}
\end{minipage}
&
\begin{minipage}[c]{0.23\textwidth}
line $1$--$2$ $\longrightarrow$ $\left\{_{_{}}\theta_2\right\}$\\
line $1$--$3$ $\longrightarrow$ $\left\{_{_{}}\theta_3\right\}$
\end{minipage}
&
\begin{minipage}{0.05\textwidth}$(2,0)$\end{minipage}\vspace*{1mm}\\
\begin{minipage}{0.11\textwidth}
  \includegraphics[width=\textwidth]{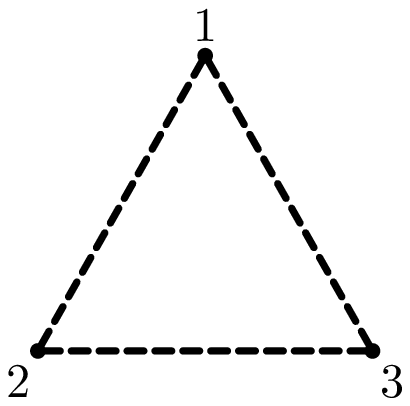}
\end{minipage}
&
\begin{minipage}{0.11\textwidth}
  \includegraphics[width=\textwidth]{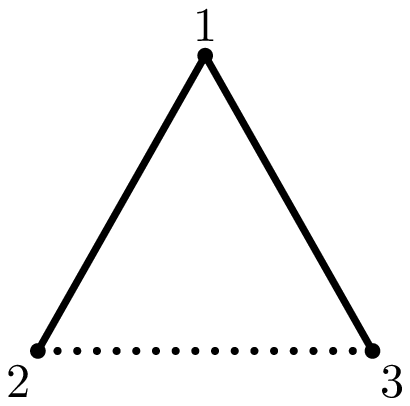}
\end{minipage}
&
\begin{minipage}[c]{0.23\textwidth}
line $1$--$2$ $\longrightarrow$ $\left\{_{_{}}\theta_2\right\}$\\
line $1$--$3$ $\longrightarrow$ $\left\{_{_{}}\theta_3\right\}$
\end{minipage}
&
\begin{minipage}{0.05\textwidth}$(2,0)$\end{minipage}\vspace*{1mm}\\
\begin{minipage}{0.11\textwidth}
  \includegraphics[width=\textwidth]{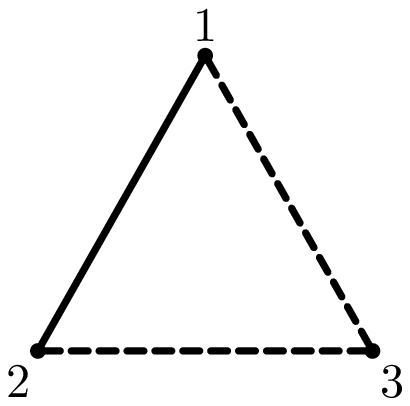}
\end{minipage}
&
\begin{minipage}{0.11\textwidth}
  \includegraphics[width=\textwidth]{113.eps}
\end{minipage}
&
\begin{minipage}[c]{0.23\textwidth}
line $1$--$2$ $\longrightarrow$ $\left\{_{_{}}\theta_2\right\}$\\
line $1$--$3$ $\longrightarrow$ $\left\{_{_{}}\theta_3\right\}$
\end{minipage}
&
\begin{minipage}{0.05\textwidth}$(2,0)$\end{minipage}\vspace*{1mm}\\
\begin{minipage}{0.11\textwidth}
  \includegraphics[width=\textwidth]{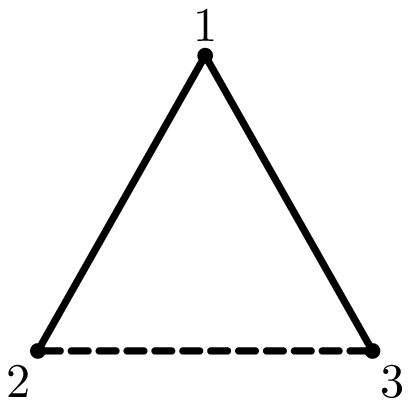}
\end{minipage}
&
\begin{minipage}{0.11\textwidth}
  \includegraphics[width=\textwidth]{113.eps}
\end{minipage}
&
\begin{minipage}[c]{0.23\textwidth}
line $1$--$2$ $\longrightarrow$ $\left\{_{_{}}\theta_2\right\}$\\
line $1$--$3$ $\longrightarrow$ $\left\{_{_{}}\theta_3\right\}$
\end{minipage}
&
\begin{minipage}{0.05\textwidth}$(2,0)$\end{minipage}\vspace*{1mm}\\
\begin{minipage}{0.11\textwidth}
  \includegraphics[width=\textwidth]{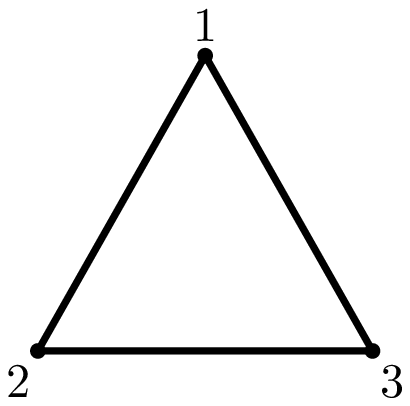}
\end{minipage}
&
\begin{minipage}{0.11\textwidth}
  \includegraphics[width=\textwidth]{111.eps}
\end{minipage}
&
\begin{minipage}[c]{0.23\textwidth}
line $1$--$2$ $\longrightarrow$ $\left\{_{_{}}\theta_2\right\}$\\
line $1$--$3$ $\longrightarrow$ $\left\{_{_{}}\theta_3\right\}$\\
line $2$--$3$ $\longrightarrow$ $\left\{_{_{}}\theta_2,\theta_3,\phi_3\right\}$
\end{minipage}
&
\begin{minipage}{0.05\textwidth}$(2,1)$\end{minipage}\vspace*{1mm}\\
\br
\end{tabular}
\end{indented}
\end{table}

To give a taste of how things look like, we illustrate in 
\tref{ThreeLoopsAngular} the ``simple'' case $\NLoopsResidual=3$.
In the first column, all possible cosine diagrams are listed; 
a dashed line means that the associated cosine is disconnectible 
in the sense defined by equation \eref{disconnectibleI}.  
As explained above, 
in some cases the corresponding angular measures can be simplified 
by analytically integrating over appropriate angles; thus, 
in the second column ``improved'' cosine diagrams can be found, describing 
the cheapest measures which will be used in the final parametrization; 
in this column dashes  
mean {\em ``a $\theta$ has been integrated''} 
---see \eref{CosIntegral}---, while 
dots mean {\em ``a $\phi$ has been integrated''} 
---see \eref{PhiIntegral}---.
In the third column one can see the contribution to the set 
$\AngularVariables$ coming from each line of the improved cosine diagram 
--- see \eref{AngularVariables}. 
Finally, in the last column we indicate the cost of the final angular measure 
as a couple 
$($number of integrations on $\Many{\theta}$, number of integrations on $\Many{\phi}$$)$. 

We just mention that in the more realistic case of $\NLoopsResidual=4$ 
we have $65$ different possible cosine diagrams, which have been analyzed 
by hand and, successively, incorporated in our code. In addition, 
the program takes automatically care of the fact that at $\NLoopsResidual=4$ 
each basic cosine diagram can manifest 
in up to $4!=24$ different possible permutations. 

\section{Implementation, and results}\label{ImplementationAndResults}
\begin{figure}[t]
\begin{indented}
\item[]\begin{minipage}[c]{0.8\textwidth}
    \includegraphics[width=\textwidth]{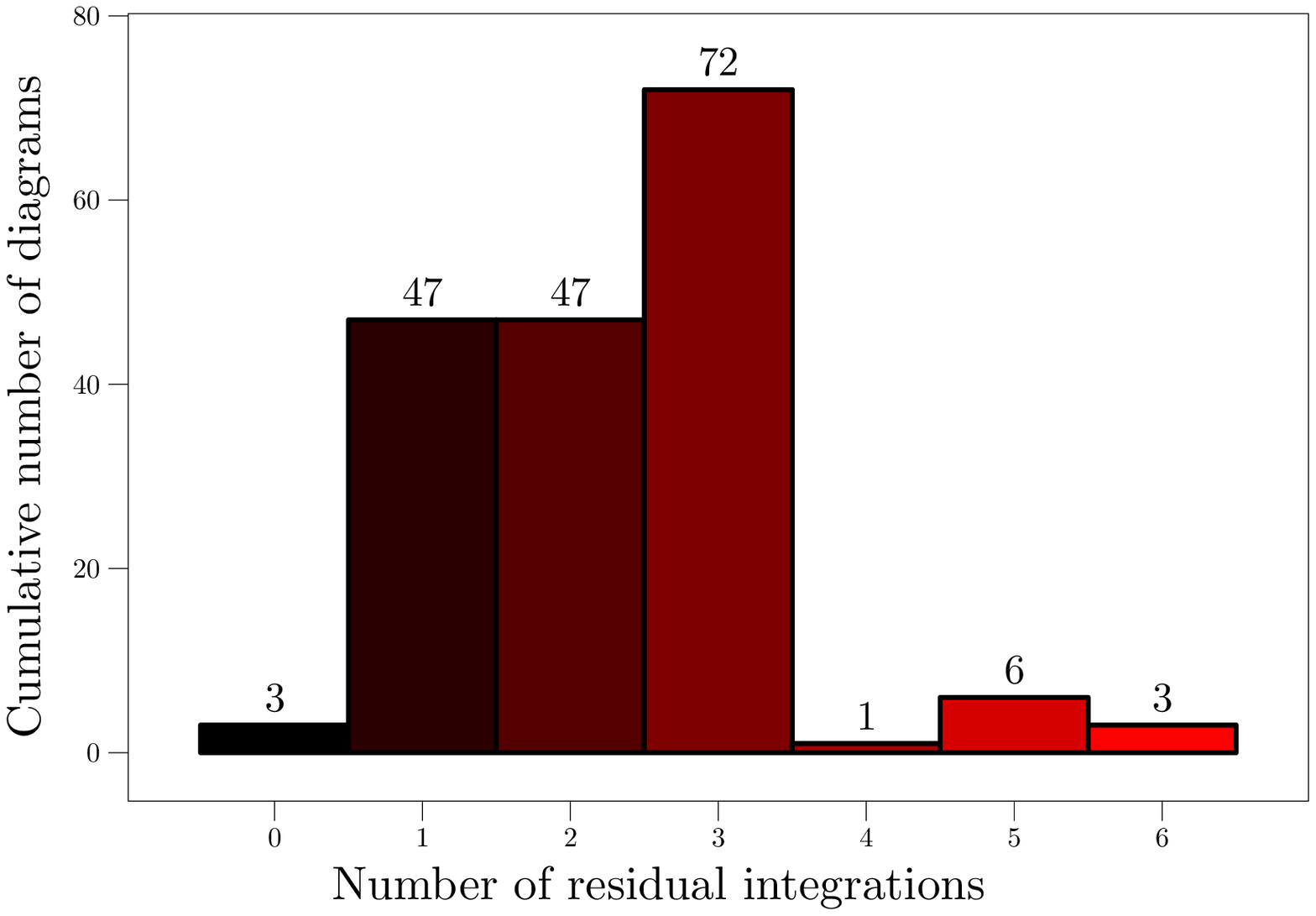}
  \end{minipage}
\end{indented}
\caption{
Total cost of loops from $1$ to $5$.
The histogram accounts for amplitudes contributing to 
$\Gamma^{(2,0)}_\I$, $\de\Gamma^{(2,0)}_\I/\de p^2$ and  $\Gamma^{(4,0)}_\I$,
including higher-points amplitudes obtained from factorization --- see 
\sref{InsertionSection} and \sref{FactAmpli}.  
Used basis of effective vertices are $1$-loop functions up to the pentagon 
(see \fref{ParametrizationExample}) 
plus $2$-loop functions of \fref{OtherTwoLoopFunctions}, 
plus the renormalized sunset. 
}
\label{ResultsI-V}
\end{figure}

\begin{figure}[t]
\begin{indented}
\item[]\begin{minipage}[c]{0.8\textwidth}
    \includegraphics[width=\textwidth]{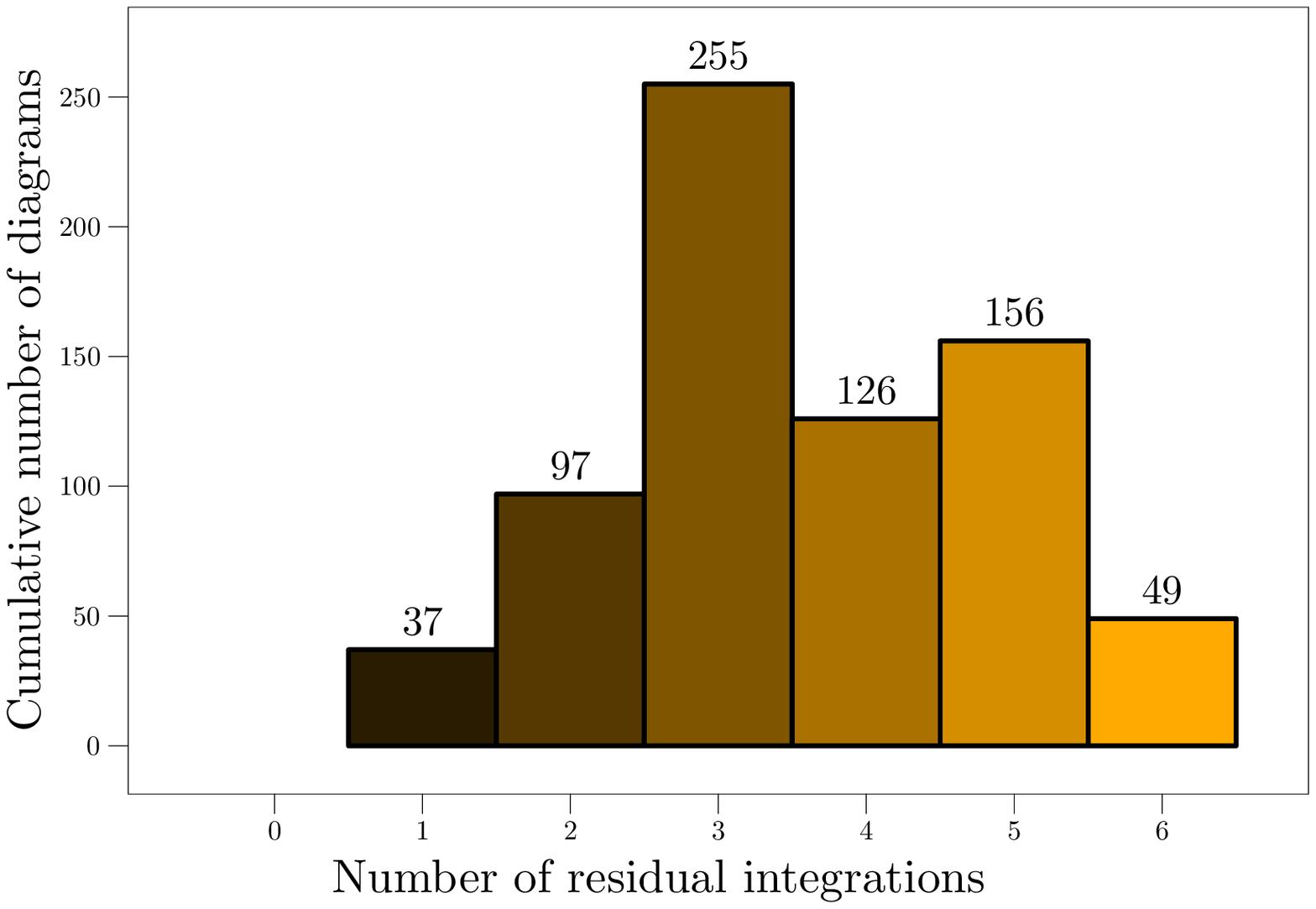}
  \end{minipage}
\end{indented}
\caption{
Total cost of the $6$-th loop alone.
The histogram accounts for amplitudes contributing to 
$\Gamma^{(2,0)}_\I$, $\de\Gamma^{(2,0)}_\I/\de p^2$ and  $\Gamma^{(4,0)}_\I$,
including higher-points amplitudes obtained from factorization --- see 
\sref{InsertionSection} and \sref{FactAmpli}.  
Used basis 
of effective vertices are $1$-loop functions up to the pentagon 
(see \fref{ParametrizationExample}) 
plus $2$-loop functions of \fref{OtherTwoLoopFunctions}, 
plus diagram of \fref{K43}.c, 
plus the renormalized sunset. 
}
\label{ResultsVI}
\end{figure}

\begin{figure}[t]
\begin{indented}
\item[]\begin{minipage}[c]{0.8\textwidth}
    \includegraphics[width=\textwidth]{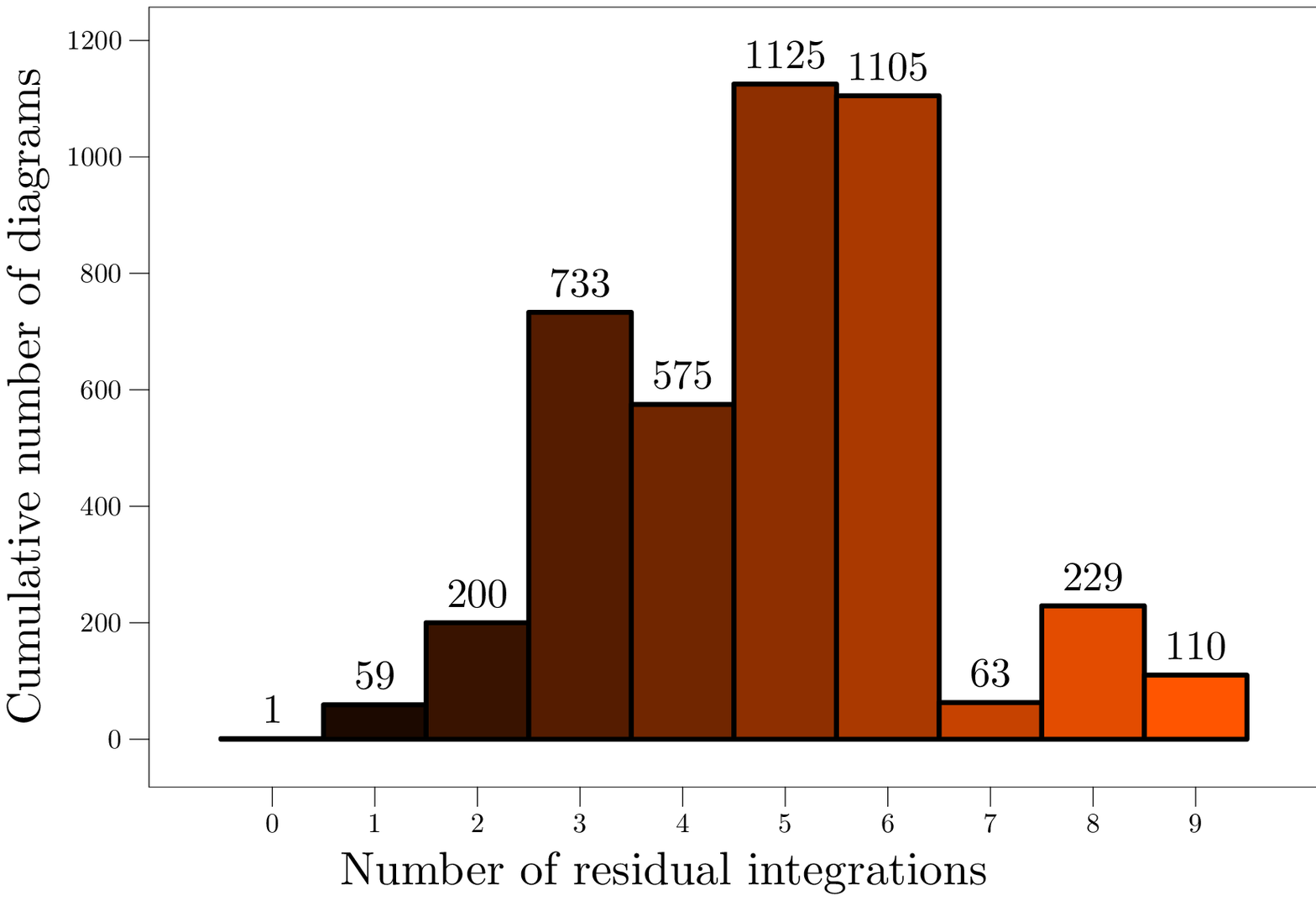}
  \end{minipage}
\item[]\begin{minipage}[c]{0.8\textwidth}
    \includegraphics[width=\textwidth]{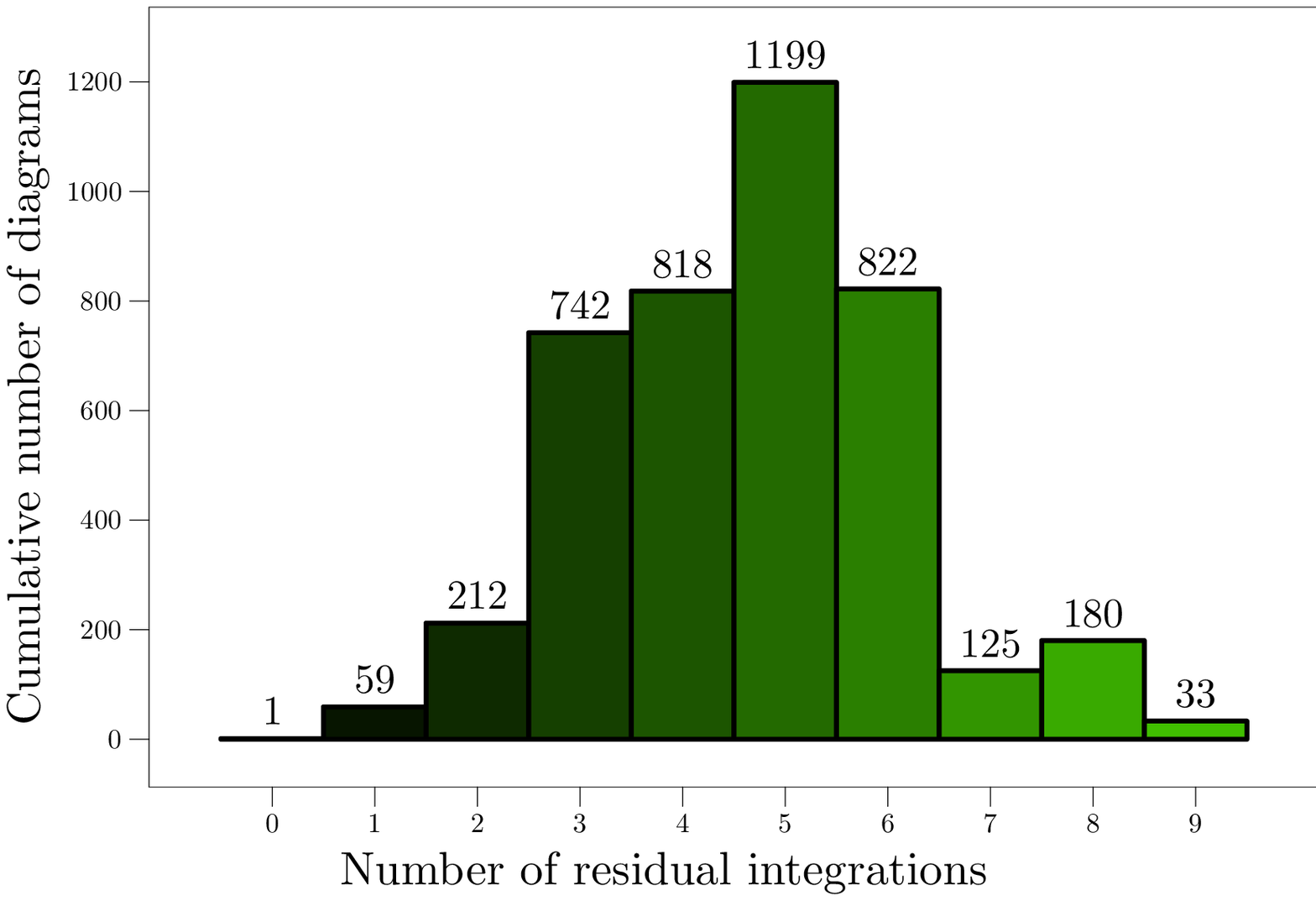}
  \end{minipage}
\end{indented}
\caption{
Total cost of the $7$-th loop alone.
The histogram accounts for amplitudes contributing to 
$\Gamma^{(2,0)}_\I$, $\de\Gamma^{(2,0)}_\I/\de p^2$ and $\Gamma^{(4,0)}_\I$,
including higher-points amplitudes obtained from factorization --- see 
\sref{InsertionSection} and \sref{FactAmpli}.  
Used bases of effective vertices are: above, 
$1$-loop functions up to the pentagon 
(see \fref{ParametrizationExample}) plus the renormalized sunset;
below, $1$-loop functions up to the pentagon, 
plus $2$-loop functions of \fref{OtherTwoLoopFunctions}, 
plus diagram of \fref{K43}.c, 
plus the renormalized sunset.
}
\label{ResultsVII}
\end{figure}

The techniques described in \sref{autopar} 
were actually used to produce a \texttt{C++} program; 
given a graphical description of the basis of effective vertices one intends 
to use ---as described in \sref{SubstitutingVertices}---, 
and a specification of the set of diagrams 
to be parametrized, our code produces
\begin{itemize}
\item an histogram of the distribution of the cost $\TotalCost$ ---as defined 
in \sref{autoparOverview}--- 
for the specified set of diagrams whose parametrization has been 
required. Examples of such distributions for some different choices of 
diagrams and bases of effective vertices can be found in \fref{ResultsI-V},
\fref{ResultsVI} and \fref{ResultsVII}
\item the \texttt{C++} functions needed to perform the numerical integration 
of amplitudes, ready to be compiled if an actual numerical implementation 
of each of the effective vertices which have been used 
during the parametrization stage is given; the set of amplitudes is then 
placed in a dynamical library, with each subroutine ready to be retrieved 
by the integration program. We do not enter here into the very complicated 
details of numerical evaluation of amplitudes, which will be described 
in future publications
\item the specifications needed to compute the RG functions 
$\beta_\M (g_\M),\eta_\M(g_\M),\eta_{2\M}(g_\M)$; 
these informations consist in a list of all required 
combinatorial and $O\left(N\right)$- factors, plus ---after a suitable file 
containing the results of numerical integration has been provided--- the 
actual value of each integral in question, with an estimation of its 
numerical error.
\end{itemize}
In general, the program presents a lot of useful switches to precisely tune 
its behaviour: for example, subsets of all the needed graphs ---or even 
single graphs--- can be selected to be parametrized, and a graphical 
automated output containing all the details of the process is provided 
for debugging purposes if requested; 
more than one parametrization for the same amplitude 
can be produced and stored when a comparison is desired.

The code consists of about $14.000$ lines of \texttt{C++} ---about $7.000$
for graph library and other auxiliary libraries, the rest in higher-level 
routines for the manipulation of scalar amplitudes--- which have been 
written and tested from scratch.
Being modular in the spirit of \texttt{C++} classes (see \cite{Stroustrup}) 
---with a different class corresponding to each different concept exposed in 
\sref{autopar}, like substitutions of effective vertices, loop bases, 
angular measures and so on--- the code can be quite easily modified to 
include new computational simplifications or new strategies, 
as discussed in \sref{autoparOverview}.
On a typical machine of the \textsf{GHz}-era, the $7$ loops can be entirely 
parametrized ---and the corresponding \texttt{C++} code produced--- 
in some hours; peak memory occupation is entirely acceptable. 

Some additional details about the structure, the design and the results 
of the program which we had to omit here due to the lack of space can be 
found in \cite{RibecaPhD} (although it documents a slightly outdated 
version of our framework): for example, the explicit parametrization 
of all amplitudes up to the $4$-loop level with ``simple'' one-loop 
effective vertices can be found there.


We would now like to discuss in detail some of the typical results 
produced by the program.

Two histograms detailing the costs of the parametrization 
of amplitudes for loops $1$ to $5$ all together, and for loop $6$ alone%
\footnote{
By \emph{``amplitudes contributing at loop $L$ alone 
to some collection of quantities $\mathbb{X}$''},  
we mean the set of all \emph{new} amplitudes which have to be evaluated in our I-scheme
to obtain all the quantities in $\mathbb{X}$ at $L$ loop level, 
assuming that all the amplitudes needed for the evaluation at loop level $L-1$ 
of $\Gamma^{(2,0)}_\I$, $\de\Gamma^{(2,0)}_\I/\de p^2$ and $\Gamma^{(4,0)}_\I$
at vanishing external momenta,
are already known.}, respectively, are shown in \fref{ResultsI-V} and \fref{ResultsVI}; 
the basis of functions employed here is a ``standard'' one, composed by the
polygons up to the pentagon, the renormalized ``sunset'' effective vertex,
all other low-cost $2$-loop functions shown in 
\fref{OtherTwoLoopFunctions} (see \sref{LowCostSubdiagrams}) 
and ---from $6$ loops and higher--- 
the effective vertex corresponding to the $2$-loop diagram of \fref{K43}.c.
All resulting costs are not more than $6$; this fact is in agreement 
with what has been previously found by Nickel et al., who declare 
in \cite{NickelPreprint} that they were able to obtain all their 
parametrizations of $6$-loop diagrams with integrations in not more than $6$ 
residual dimensions.

In spite of the fact that we are working in different renormalization schemes
and that, consequently, we have a different number of diagrams to evaluate,
it would be nevertheless quite interesting to be able to perform a more precise quantitative 
comparison between our histograms and the corresponding distributions
obtained by Nickel and coworkers; 
however, this information has never been published. Our general impression 
is that our results could be slightly sub-optimal, mainly due to the fact
that for some ``difficult'' graphs it is probably possible to work out
cheaper parametrizations by hand, in the spirit of \fref{K43}. All in all, the 
results we obtained for the $6$-loop case can be also interpreted as being 
very encouraging, if one bears in mind that
\begin{itemize}
\item the simplifications implemented in our code (see the list in 
\sref{autoparOverview}) are just a few, and not very complicated;
nonetheless, they are enough to obtain a set of parametrizations 
which would seem comparable to that derived in the original work of 
Nickel and coworkers
\item our program is very general, so that other techniques can be easily 
added if needed
\item a separate treatment of a few particularly expensive graphs 
is always possible
\item if we want to tackle the computation of the $7$-th loop, we must be 
anyway prepared to the appearance of a consistent number of integrals 
with a residual dimensionality much larger than $6$.
\end{itemize}

Going on to the results relative to the parametrization of the $7$-th loop 
shown in \fref{ResultsVII} 
---which are the most important since their knowledge represents 
the motivation for this work, in view of their possible use to improve 
available estimates of critical quantities--- the situation looks even more 
promising. The larger residual dimensionality is here $9$, still low enough 
to make possible a high-precision numerical evaluation of the amplitudes if 
enough computer power is available. By comparing the two profiles shown 
in the figure one can notice how the two-loop effective functions play 
an essential role in lowering the overall complexity of the computation 
of a factor of $\sim3$ (this rough estimate can immediately be 
obtained by remembering, as explained in \sref{WhichParametrization}, 
that the main part of computer time will be spent in the evaluation 
of the integrals of highest dimensionality). 
Such results bring us to the remarkable conclusion 
that \emph{even at a $7$-loop level the approach of Baker, Nickel, Green 
and Meiron remains fruitful, leading ---within the present framework--- 
to an acceptable set of parametrizations for all the required diagrams,
including those contributing to the $4$-point function}.

As a final remark, some preliminary looks at the corresponding results 
for the $2$-point functions at $8$ loops (not reported in this article) 
seem to indicate that the set of effective vertices which allows a
successful parametrization of the $7$ loops would not be adequate to
conveniently
parametrize the $8$ loops; anyway, due to the number and the complexity 
of required integrals the possibility of carrying out the computation 
of the $8$-th loop is purely academic at the moment.

\section{Epilogue}\label{epilogue}
In this work we present a solution 
to the problem of building an automated framework for the 
parametrization of Feynman amplitudes needed to compute 
---at the level of $7$ loops and higher---
the Renormalization-Group $\beta$-function 
and the anomalous dimensions $\eta,\eta_2$ 
for the $3$-dimensional $O(N)$ vector model. 

The core of the employed computational method  ---in the spirit
of \cite{NickelI}-\cite{NickelII}--- consists in replacing 
effective-vertex functions and applying other analytic simplifications
to the original expressions for amplitudes in momentum space. 

Some of the key features of our own approach, as compared to that in \cite{NickelI}-\cite{NickelII},
are the following:
\begin{itemize}
\item 
 a different intermediate renormalization scheme ---the I-scheme--- has been chosen,
 which has simpler renormalization properties as well as a 
 Callan-Symanzik equation with exactly-known coefficients, see \sref{IScheme}
 \item 
 a systematic treatement of the parametrization of loop integrations 
 in momentum space
 (leading in our framework to a ``good choice of angular measures'') 
 has been developed in \sref{autopar}
  \item 
 an extended set of effective vertices, 
 including some convenient $2$-loop functions, 
 has been used to perform the analysis.   
\end{itemize}
The employed technique can eventually allow to reformulate the needed amplitudes in terms of 
relatively low-dimensional integral representations; however,
a general proof of the effectiveness of the method is missing, 
and its practical applicability to the full $7$-loop case had not been demonstrated so far.

In spite of the theoretical possibility that some topological obstructions 
could show up, leading to some very difficult graph with a very 
high number of residual integrations, the automated parametrization 
of all amplitudes needed to compute RG-functions revealed that 
---luckily--- at the level of $7$ loops nothing similar happens:
this work shows that a reasonably small set of computational 
simplifications is still sufficient to obtain for all the needed amplitudes 
efficient parametrizations with a residual dimensionality 
of not more than $9$, thus opening the way to a full $7$-loop evaluation of 
RG $\beta$-function.

On the other hand, the large number of needed diagrams ($\sim 4000$) 
and the necessity of carrying out each integration with $7$-$8$ digits of 
precision in dimensionalities up to $9$ nonetheless classify the problem as a 
formidable one in terms of required computational resources.
How to solve it is an open issue, and possible solutions 
will be discussed in an incoming publication.

\ack
P.R. is deeply indebted to all the people at the Service de 
Physique Th\'eorique in Saclay, Paris 
(where the main part of this work has been worked out) 
for the constant and invaluable support, both material 
and moral. 
R.G. wants to thank M. Bauer, M. Berg{\`e}re and F. David for 
many enlightening discussions.
Both authors wish to thank J. Zinn-Justin for his continuous 
encouragement 
and many fruitful discussions. This work was partly supported by 
Sonderforschungsbereich-Transregio ``Computational Particle Physics'' 
(SFB-TR9).

\appendix

\section{Renormalization of  the $O(N)$ vector model and Callan-Symanzik equations}
\label{renormalization}
After a brief review of our notations, 
we give in this section a concise
presentation of the $O(N)$ vector model, of its renormalization and 
of Callan-Symanzik equations. 
The reader looking for a more detailed and pedagogical introduction 
should refer e.g. to 
\cite{LeBellac}, \cite{Zinn}, \cite{AmitMayorBook},  
\cite{Kleinert}.

We will work in the Euclidean space $\mathbb{R}^3$, and
consider an $O(N)$-invariant quantum field theory
built out of scalar fields $\phi^i(\bo{x})$ in a fundamental representation of $O(N)$ --- where $i=1\,\ldots\, ,\, N$.

For shortness we will omit field arguments, $O(N)$ or $\mathbb{R}^3$ indices 
unless they play a crucial role;
we  will write $\Many{x}$ instead of a collection of objects 
 $(x_j\in \mathbb{S})_{j=1}^n$ if no ambiguity arises (that is, 
if the set $\mathbb{S}$ and the value $n$ can be clearly deduced 
from the context).
In this notation, the Fourier transform of a function 
$f(\bo{x}_1,\,\ldots\, ,\,\bo{x}_n)$ 
invariant under global space translations is written
\beq
 f(\Many{\bo{p}}) \;(2\pi)^3\;\delta^3(\sum \Many{\bo{p}})
 :=
 \int d\Many{\bo{x}}\;\exp\!\left(\rmi\,\Many{\bo{p}}.\Many{\bo{x}}\right) f(\Many{\bo{x}})
\eeq
meaning as usual
\beq
\fl
 f(\bo{p}_1,\,\ldots\, ,\,\bo{p}_n) \;(2\pi)^3\;\delta^3(\sum_{j=1}^n \bo{p}_j)
 :=
\int \left( \prod_{j=1}^n d^3\bo{x}_j\;\exp\!\left(\rmi\,\bo{p}_j.\bo{x}_j\right)\right) f(\bo{x}_1,\,\ldots\, ,\,\bo{x}_n)
.\eeq

The bare action with sources $J_\B, K_\B$ is defined by:
\beq\fl
S_\B(\phi_\B, J_\B, K_\B) :=\int d^3x 
\left(
  \frac{1}{2}(\nabla \phi_\B)^2+{m^2_\B \over 2} \phi_\B^2
  +{\lambda_\B \over 4!}(\phi_\B^2)^2 
  -J_\B \phi_\B +\frac{1}{2} K_\B \phi_\B^2
\right).
\eeq
It should be noticed that 
\beq
\fl
(\phi_\B^2)^2 \equiv \delta_{i_{1}\,i_{2}\,i_{3}\,i_{4}} 
\phi_\B^{i_1}\phi_\B^{i_2}\phi_\B^{i_3}\phi_\B^{i_4}
\qquad
\delta_{i_{1}\,i_{2}\,i_{3}\,i_{4}}:= 
\frac{1}{3}\left(
 \delta_{i_{1}\,i_{2}}\delta_{i_{3}\,i_{4}}
+\delta_{i_{1}\,i_{3}}\delta_{i_{2}\,i_{4}}
+\delta_{i_{1}\,i_{4}}\delta_{i_{2}\,i_{3}}
\right)
.
\eeq 
The generating functional for the bare theory is defined by: 
\beq
Z_\B (J_\B, K_\B):=
\int [d\phi_\B] \exp\!\left(-S_\B(\phi_\B,J_\B,K_\B)_{_{_{}}}\right) 
\eeq
(an appropriate regularization parametrized by an ultraviolet cutoff 
$\Lambda$ is assumed, to give meaning to functional integration).
The effective action $\Gamma_\B$ is defined via Legendre transform
of the generator of connected diagrams, $\log Z_\B$:
\beq
\fl
\log Z_\B (J_\B,K_\B)+\Gamma_\B(\phi_\B,K_\B):=\int d^3x\;J_\B(x)\,\phi_\B(x)
\qquad
\phi_\B:=\frac{\delta \log Z_\B}{\delta J_\B}
.
\eeq
\emph{One-particle irreducible} (OPI) correlators are defined as usual 
from the expansion of  the effective action
 w.r.t.\ its sources:
\[\fl   
\Gamma_{\B}(\phi_{\B},K_{\B})= 
\sum_{\NExternal,\NVerticesII=0}^{\infty} {1\over \NExternal!\NVerticesII!} \int d\Many{\bo{x}}\,d\Many{\bo{y}}
 \;\Gamma_{\B}^{(\NExternal,\NVerticesII)}(\Many{\bo{x}};\Many{\bo{y}}) 
  \; \phi_{\B}(\bo{x}_1)\,\ldots\,\phi_{\B}(\bo{x}_\NExternal)
\; {K_{\B}}(\bo{y}_1)\,\ldots\,{K_{\B}}(\bo{y}_{\NVerticesII})
.\]
Exploiting $O(N)$-invariance of the theory,
we factor some  correlators 
evaluated at zero momenta into a $O(N)$-invariant tensor 
times a $O(N)$-scalar function:
\bea\label{on-fact-I}
{\Gamma_{\B}^{(2,0)}}_{i_1\,i_2}(\bo{p}_1,\bo{p}_2;\cdots)\vert_{\Many{\bo{p}}=\bo{0}}
=:
\delta_{i_1\,i_2}\,{\Gamma_{\B}^{(2,0)}}(\cdots)
\\\label{on-fact-II}
{\Gamma_{\B}^{(2,1)}}_{i_1\,i_2}(\bo{p}_1,\bo{p}_2;\bo{q};\cdots)\vert_{\Many{\bo{p}}=\bo{q}=\bo{0}}
=:
\delta_{i_1\,i_2}\,{\Gamma_{\B}^{(2,1)}}(\cdots)
\\\label{on-fact-III}
{\Gamma_{\B}^{(4,0)}}_{i_1\,i_2\,i_3\,i_4}(\bo{p}_1,\bo{p}_2,\bo{p}_3,\bo{p}_4;\cdots)\vert_{\Many{\bo{p}}=\bo{0}}
=:
\delta_{i_1\,i_2\,i_3\,i_4}\,{\Gamma_{\B}^{(4,0)}}(\cdots)
\eea
(the dots $\cdots$ in expressions above denote the dependence on other parameters).

In a generic renormalization scheme, bare quantities are related to 
corresponding renormalized ones 
according to the following transformations:
\bea
\phi_\B   =: \sqrt{\zphi{\R}} \;\phi_{\R}  \label{BR-phi} 
\\
m^2_\B    =:\left(m_{\R}^2 +\delta m_{\R}^2\right)/ \zphi{\R} 
\label{BR-m}
\\
\lambda_\B =: \frac{\zlambda{\R}}{\left(\zphi{\R}\right)^2} \;\lambda_{\R}
\label{BR-lambda}
\\
K_{\R}\!\left[\phi^2\right]_{\R}
:= K_{\R}\left(
       \frac{\zphiphi{\R}}{\zphi{\R}}\;\phi_\B^2
       +\zphione
       \right) 
\label{BR-phisq}.
\eea
The transformation from bare to renormalized quantities
is performed at fixed regularization parameter $\Lambda$  
and is parametrized by the scheme-dependent quantities $\delta m^2_\R$ and $Z$'s.
The dependence on the renormalization scheme 
will be emphasized by capitalized suffixes such as $\M$ or $\I$.
The special suffix $\R$ is used to label scheme-dependent quantities 
in relations holding for a generic renormalization scheme 
with non-zero renormalized mass.
The same conventions used in the case of  
bare quantities and their Legendre and Fourier transforms 
will be adopted also 
for renormalized quantities (apart from a straightforward change of suffix).

The renormalized (sourceless) action is obtained 
from the bare one after expressing it
in terms of renormalized quantities: 
\bea
\fl
S_{\R}(\phi_{\R})&:=&S_\B\!\left(\phi_\B=\sqrt{\zphi{\R}}\,\phi_{\R},J_\B=0,K_\B=0\right) 
\\
\fl
&\;=&
\int d^3x \left(\frac{\zphi{\R}}{2} (\nabla \phi_{\R})^2+ 
\frac{1}{2}\left(m_{\R}^2 +\delta m_{\R}^2\right)\phi_{\R}^2
+\zlambda{\R} {\lambda_{\R}\over 4!} (\phi_{\R}^2)^2\right).
\eea
The renormalized functional generator $Z_\R$,
the renormalized effective action $\Gamma_\R$
and OPI correlators are expressed 
in terms of bare quantities as follows:
\bea
\fl
Z_{\R}(J_{\R},K_{\R}) & := &
   \int [d\phi_{\R}]\; 
      \exp\!{\left(
            -S_{\R}(\phi_{\R})
            +\int J_{\R} \phi_{\R} -\int \frac{1}{2}K_{\R} [\phi^2]_{\R}
            \right)}
\label{ZR}
\\
\fl 
& = &
Z_\B\!\!\left(\!J_\B=\frac{J_{\R}}{\sqrt{\zphi{\R}}}
    ,K_{\B}=\frac{\zphiphi{\R}}{\zphi{\R}}K_{\R}\!\right)
\exp\!{\left(-\frac{1}{2} \int K_{\R}\zphione\right)}\nonumber
\\
\fl
\Gamma_{\R}(\phi_{\R},K_{\R}) & = &
\Gamma_{\B}\!\!\left(\!\phi_{\B}=\sqrt{\zphi{\R}}\phi_{\R}
            ,K_{\B}=\frac{\zphiphi{\R}}{\zphi{\R}}K_{\R}\!\right)
+\frac{1}{2}\int K_{\R}\zphione
\label{GAR-GAB}
\\
\fl
\Gamma_{\R}^{(\NExternal,\NVerticesII)}\!(\Many{\bo{x}};\Many{\bo{y}})
& = &
\left(\zphi{\R}\right)^{\NExternal/2} \left(\frac{\zphiphi{\R}}{\zphi{\R}}\right)^{\!\!\NVerticesII}
 \Gamma_\B^{(\NExternal,\NVerticesII)}\!(\Many{\bo{x}};\Many{\bo{y}})
+\frac{1}{2}\;\delta_{\NExternal\,0}\;\delta_{\NVerticesII\,1}\; 
\zphione
\label{GAR-GAB-II}.
\eea
In the rest of this article we will only deal with correlators
with $\NExternal>0$, so we do not consider anymore 
the local contact term 
in the right hand side of \eref{GAR-GAB-II}; however, it should be noticed
that in the case $(\NExternal,\NVerticesII)=(0,1)$ 
the contact term \emph{does} modify
Callan-Symanzik equations \eref{RCallanSymanzik}.

The Callan-Symanzik differential operator in
a generic scheme is defined as
\beq
\mathcal{D}_\R := 
\left. m_\R \frac{d}{d m_\R} \right\vert_{\lambda_\B,\Lambda};
\label{RCallanSymanzikD}
\eeq
Callan-Symanzik equations
are obtained by
application of the Callan-Symanzik operator on both sides of
(Fourier transform) of \eref{GAR-GAB-II} and 
by use of identities
\beq 
\fl\mathcal{D}_\R\Gamma_{\B}^{(\NExternal,\NVerticesII)}\left(\Many{\bo{p}};\Many{\bo{q}}\right)
=\;\mathcal{D}_\R[m^2_\B]\;
\frac{\de\Gamma_{\B}^{(\NExternal,\NVerticesII)}}{\de m^2_\B}\left(\Many{\bo{p}};\Many{\bo{q}}\right)
=\;\mathcal{D}_\R[m^2_\B]\;
\Gamma_{\B}^{(\NExternal,\NVerticesII+1)}\left(\Many{\bo{p}};\Many{\bo{q}},\bo{q}_{\NVerticesII+1}=\bo{0}\right)
\eeq
which follow from the chain rule and 
from the fact that a derivative of bare OPI correlators 
w.r.t.\ the bare mass 
is equivalent ---in non-pathological cases---
to an insertion of $\phi^2_\B/2$ at zero momentum.

The resulting Callan-Symanzik equations in a generic renormalization scheme
are (assuming $\NExternal>0$):
\bea
\fl
\left[
 m_\R\frac{\de}{\de m_\R}
 +\beta_\R(g_\R)\frac{\de}{\de g_\R}
 -\frac{\NExternal}{2}\eta_\R(g_\R)
 -{\NVerticesII}\Bigl(\eta_{2\R}(g_\R)-\eta_{\R}(g_\R)\Bigr)
\right]
\Gamma_{\R}^{(\NExternal,\NVerticesII)}\left(\Many{\bo{p}};\Many{\bo{q}}\right)
\nonumber\\
\lo{=}
m_\R^2 \;\sigma_\R(g_\R)\;
\Gamma_{\R}^{(\NExternal,\NVerticesII+1)}\left(\Many{\bo{p}};\Many{\bo{q}},\bo{q}_{\NVerticesII+1}=\bo{0}\right)
\label{RCallanSymanzik}
\eea
where we introduced the adimensional renormalized coupling 
\beq
g_\R:=\lambda_\R/m_\R
\eeq 
and the RG-functions:
\bea
\beta_\R(g_\R):=\mathcal{D}_\R[g_\R]
\label{RBeta}
\\
\eta_\R(g_\R):=\mathcal{D}_\R[\log \zphi{\R}]
\label{REta}
\\
\eta_{2\R}(g_\R):=\mathcal{D}_\R[\log \zphiphi{\R}]
\label{REtaTwo}
\\
\sigma_\R(g_\R):=
 \left(\frac{\zphiphi{\R}}{\zphi{\R}}\right)^{-1}
 \frac{\mathcal{D}_\R[m^2_\B]}{m^2_\R}
 \label{RSigma}.
\eea

\section*{References}
\bibliographystyle{unsrt}
\bibliography{autopar}

\begin{thebibliography}{10}

\bibitem{NickelI}
G.A. Baker, B.G. Nickel, M.S. Green, and D.I. Meiron.
\newblock {I}sing-model critical indices in $3$ dimensions from the
  {C}allan-{S}ymanzik equation.
\newblock {\em Physical Review Letters}, 36(23):1351--1354, June 1976.

\bibitem{NickelII}
G.A. Baker, B.G. Nickel, and D.I. Meiron.
\newblock Critical indices from perturbation analysis of the
  {C}allan-{S}ymanzik equation.
\newblock {\em Physical Review B}, 17(3):1365--1374, February 1978.

\bibitem{Kadanoff:1966wm}
L.~P. Kadanoff.
\newblock Scaling laws for {I}sing models near {$T_c$}.
\newblock {\em Physics}, 2:263--272, 1966.

\bibitem{Wilson:1971bg}
K.~G. Wilson.
\newblock Renormalization {G}roup and critical phenomena. 1. {R}enormalization
  {G}roup and the {K}adanoff scaling picture.
\newblock {\em Phys. Rev.}, B4:3174--3183, 1971.

\bibitem{Wilson:1971dh}
K.~G. Wilson.
\newblock Renormalization {G}roup and critical phenomena. 2. {P}hase space cell
  analysis of critical behavior.
\newblock {\em Phys. Rev.}, B4:3184--3205, 1971.

\bibitem{Wegner:1972my}
F.~J. Wegner.
\newblock Corrections to scaling laws.
\newblock {\em Phys. Rev.}, B5:4529--4536, 1972.

\bibitem{Wegner:1972xxx}
F.~J. Wegner.
\newblock Critical exponents in isotropic spin systems.
\newblock {\em Phys. Rev.}, B6:1891--1893, 1972.

\bibitem{Wilson:1974jj}
K.~G. Wilson and J.~B. Kogut.
\newblock The {R}enormalization {G}roup and the {$\epsilon$}-expansion.
\newblock {\em Phys. Rept.}, 12:75--200, 1974.

\bibitem{WilsonFisher}
K.G. Wilson and M.E. Fisher.
\newblock Critical exponents in $3.99$ dimensions.
\newblock {\em Physical Review Letters}, 28:240, 1972.

\bibitem{ParisiTwoLoops}
G.~Parisi.
\newblock Cargese Lecture Notes, Columbia University preprint 1973.

\bibitem{Parisi}
G.~Parisi.
\newblock Field-theoretical approach to second-order phase transitions.
\newblock {\em Journal of Statistical Physics}, 23:49, 1980.

\bibitem{Magnen:1977ha}
J.~Magnen and R.~Seneor.
\newblock Phase space cell expansion and borel summability for the euclidean
  $\phi^4$ in $3$-dimensions theory.
\newblock {\em Commun. Math. Phys.}, 56:237, 1977.

\bibitem{LeBellac}
M.~Le Bellac.
\newblock {\em Des Ph{\'e}nom{\`e}nes Critiques aux Champs de Jauge}.
\newblock Savoirs Actuels. InterEditions and Editions du CNRS, deuxi{\`e}me
  edition, 1988.

\bibitem{Zinn}
J.~Zinn-Justin.
\newblock {\em Quantum Field Theory and Critical Phenomena}.
\newblock Oxford Science Publications, fourth edition, 2002.

\bibitem{ParisiBook}
G.~Parisi.
\newblock {\em Statistical Field Theory}.
\newblock Perseus Books Group, 1998.

\bibitem{AmitMayorBook}
D.~J. Amit and V.~M. Mayor.
\newblock {\em Field Theory, the {R}enormalization {G}roup and Critical
  Phenomena Graphs to Computers}.
\newblock World Scientific Publishing, third edition, 2005.

\bibitem{Kleinert}
H.~Kleinert and V.~Schulte-Frohlinde.
\newblock {\em Critical Properties of {$\phi^4$}-theories}.
\newblock World Scientific, 2001.

\bibitem{BaBe}
C.~Bagnuls and C.~Bervillier.
\newblock Exact {R}enormalization {G}roup equations: an introductory review.
\newblock {\em Physics Reports}, 348:91, 2001.

\bibitem{PelissettoVicari}
A.~Pelissetto and E.~Vicari.
\newblock Critical phenomena and {R}enormalization-{G}roup theory.
\newblock {\em Physics Reports}, 368:549--727, 2002.
\newblock {\tt cond-mat/0012164}.

\bibitem{NickelDiagrams}
B.G. Nickel.
\newblock Evaluation of simple {F}eynman graphs.
\newblock {\em Journal of Mathematical Physics}, 19(3):542--548, 1978.

\bibitem{ZinnTwoLoops}
E.~Br{\'e}zin, J.C.~Le Guillou, and J.~Zinn-Justin.
\newblock {\em Physical Review D}, 8:434, 1973.

\bibitem{NickelPreprint}
B.G. Nickel, D.I. Meiron, and G.A. Baker.
\newblock Compilation of {$2$}-pt. and {$4$}-pt. graphs for continuous spin
  models.
\newblock University of Guelph Report, 1977.

\bibitem{MuNi84}
M.~Muthukumar and B.~G. Nickel.
\newblock Perturbation theory for a polymer chain with excluded volume
  interaction.
\newblock {\em J. Chem. Phys.}, 80:5839, 1984.

\bibitem{NickelPreprintII}
D.B. Murray and B.G. Nickel.
\newblock Revised estimates for critical exponents for the continuum $n$-vector
  model in $3$ dimensions.
\newblock University of Guelph Report, 1991.

\bibitem{Guida:1998bx}
R.~Guida and J.~Zinn-Justin.
\newblock Critical exponents of the {$N$}-vector model.
\newblock {\em J. of Phys. A}, 31:8103, 1998.

\bibitem{Calabrese:2002qi}
P.~Calabrese, A.~Pelissetto, and E.~Vicari.
\newblock Critical structure factors of bilinear fields in {$O(N)$}- vector
  models.
\newblock {\em Phys. Rev.}, E65:046115, 2002.

\bibitem{Calabrese:2002sz}
P.~Calabrese, A.~Pelissetto, and E.~Vicari.
\newblock Randomly dilute spin models with cubic symmetry.
\newblock {\em Phys. Rev.}, B67:024418, 2003.

\bibitem{Kastening:2003iu}
B.~M. Kastening.
\newblock {B}ose-{E}instein condensation temperature of homogenous weakly
  interacting {B}ose gas in variational perturbation theory through seven
  loops.
\newblock {\em Phys. Rev.}, A69:043613, 2004.

\bibitem{ItzyksonZuber}
C.~Itzykson and J-B. Zuber.
\newblock {\em Quantum Field Theory}.
\newblock International Series in Pure and Applied Physics. McGraw-Hill, 1980.

\bibitem{Collins}
J.~C. Collins.
\newblock {\em Renormalization. {A}n introduction to renormalization, the
  {R}enormalization {G}roup, and the operator product expansion}.
\newblock Cambridge University Press, 1984.

\bibitem{GAP}
{\tt http://www-gap.mcs.st-and.ac.uk}.

\bibitem{FeynARTS}
{\tt http://www.feynarts.de}.

\bibitem{Stroustrup}
B.~Stroustrup.
\newblock {\em The {\texttt{C++}} programming language}.
\newblock Addison-Wesley, third edition, 1997.

\bibitem{SchmidtDruffel}
D.C. Schmidt and L.E. Druffel.
\newblock A fast backtracking algorithm to test directed graphs for isomorphism
  using distance matrices.
\newblock {\em Journal of ACM}, 23:433--445, 1976.

\bibitem{McKay}
B.~McKay.
\newblock Practical graph isomorphism.
\newblock {\em Congressus Numerantium}, 21:45--87, 1981.

\bibitem{Rosen}
K.H. Rosen, editor.
\newblock {\em Handbook of Discrete and Combinatorial Mathematics}.
\newblock CRC Press, 2000.

\bibitem{qgraf}
{\tt http://cfif.ist.utl.pt}.

\bibitem{Heap}
B.R. Heap.
\newblock The enumeration of homeomorphically-irreducible star graphs.
\newblock {\em Journal of Mathematical Physics}, 7:1582, 1966.

\bibitem{Nagle}
J.F. Nagle.
\newblock On ordering and identifying undirected linear graphs.
\newblock {\em Journal of Mathematical Physics}, 7:1588, 1966.

\bibitem{RibecaPhD}
P.~Ribeca.
\newblock {\em Precise evaluation of universal quantities in second-order phase
  transitions}.
\newblock PhD thesis, Universit\`a di {Genova} and Universit\'e de Paris 11,
  2003.

\bibitem{MelroseReduction}
D.B. Melrose.
\newblock Reduction of {F}eynman diagrams.
\newblock {\em Il Nuovo Cimento}, 1965.

\bibitem{oneloop}
R.~Guida and P.~Ribeca.
\newblock Kinematically safe reductions of one-loop correlators.
\newblock preprint HU-EP-05/40, 2005.

\bibitem{Remiddi:1982hn}
E.~Remiddi.
\newblock Dispersion relations for {F}eynman graphs.
\newblock {\em Helv. Phys. Acta}, 54:364, 1982.

\bibitem{Harary}
F.~Harary.
\newblock {\em Graph Theory}.
\newblock Addison-Wesley, Reading, Massachusetts, 1969.

\bibitem{Skiena}
S.~Skiena.
\newblock {\em Implementing Discrete Mathematics}.
\newblock The Advanced Book Program. Addison-Wesley, 1990.

\bibitem{David:1992vv}
F.~David, B.~Duplantier, and E.~Guitter.
\newblock Renormalization theory for interacting crumpled manifolds.
\newblock {\em Nucl. Phys.}, B394:555--664, 1993.

\bibitem{Cicuta:2001ke}
G.~M. Cicuta, L.~Molinari, and G.~Vernizzi.
\newblock Yang-{M}ills integrals.
\newblock {\em J. Phys.}, A35:L51--L59, 2002.

\end{thebibliography}

\end{document}